%% file: DESI2024_kp7_MG_value_added_paper.tex
\documentclass[a4paper,11pt]{article}
\usepackage[mathlines]{lineno}
\usepackage{jcappub}

\usepackage[table,svgnames,dvipsnames]{xcolor}

\usepackage[normalem]{ulem}
\usepackage{aas_macros}
\usepackage{subfigure}
\usepackage{graphicx}
\usepackage{graphics}
\usepackage{dcolumn}
\usepackage{bm}
\usepackage{orcidlink}
\usepackage{cases}
\usepackage{booktabs}
\usepackage{comment}
\usepackage{multirow}
\usepackage{makecell}
\usepackage{siunitx}
\usepackage{tabularx}
\usepackage{xspace}
\usepackage{soul} 
\graphicspath{{figs/}}
\usepackage{tensor}

\usepackage{listings}
\usepackage{caption}
\lstdefinelanguage{yaml}{
  keywords={true, false, null, yes, no, y, n},
  keywordstyle=\color{blue},
  basicstyle=\ttfamily\scriptsize,
  comment=[l]{\#},
  commentstyle=\color{gray},
  stringstyle=\color{orange},
  moredelim=[l][\color{brown}]{\&},
  moredelim=[l][\color{magenta}]{*},
  moredelim=[s][\color{gray}]{:}{\ },
  sensitive=false
}

\usepackage{orcidlink} 
\usepackage{hanging} 
\usepackage{arydshln}
\def\tpdf#1{\texorpdfstring{#1}{Lg}}

\renewcommand{\arraystretch}{1.4}
\def\be{\begin{equation}}
\def\ee{\end{equation}}

\def\ba#1\ea{\begin{align*}#1\end{align*}}

\renewcommand{\emph}[1]{\textit{#1}}
\definecolor{RoyalBlue}{rgb}{0.25,.41,.88}
\definecolor{WildStrawberry}{HTML}{EE2967}
\definecolor{RedWine}{rgb}{0.743,0,0}
\definecolor{bittersweet}{rgb}{1.0, 0.44, 0.37}
\definecolor{burntorange}{rgb}{0.8, 0.33, 0.0}
\definecolor{midnightgreen}{rgb}{0.0, 0.29, 0.33}
\definecolor{otherblue}{rgb}{0.20, 0.73, 0.92}

\usepackage[nameinlink,noabbrev]{cleveref}
\crefname{equation}{Eq.}{Eqs.}
\crefname{section}{Section}{Sections}
\crefname{figure}{Figure}{Figures}
\crefname{table}{Table}{Tables}
\crefname{appendix}{Appendix}{Appendices}
\Crefname{figure}{Figure}{Figures}
\Crefname{equation}{Equation}{Equations}
\Crefname{section}{Section}{Sections}
\Crefname{table}{Table}{Tables}

\newcommand{\mksym}[1]{\ifmmode {\rm #1}\else #1\fi}
\newcommand{\dataplus}{\allowbreak+}

\newcommand{\leftparbox}[2]{\parbox{#1}{\begin{flushleft} #2 \end{flushleft}}}

\newcommand{\twoonesig}[4][\pbwidth]{
\begin{equation}
\left.
 \begin{aligned}
#2 \\ #3
 \end{aligned}
\ \right\} \ \ \mbox{\text{\leftparbox{#1}{#4}}}
\end{equation}
}

\newcommand{\threeonesig}[5][\pbwidth]{
\begin{equation}
\left.
 \begin{aligned}
#2 \\ #3 \\ #4
 \end{aligned}
\ \right\} \ \ \mbox{\text{\leftparbox{#1}{#5}}}
\end{equation}
}

\newcommand{\fouronesig}[6][\pbwidth]{
\begin{equation}
\left.
 \begin{aligned}
#2 \\ #3 \\ #4 \\ #5
 \end{aligned}
\ \right\} \ \ \mbox{\text{\leftparbox{#1}{#6}}}
\end{equation}
}

\newcommand{\Om}{\Omega_\mathrm{m}}
\newcommand{\Ocdm}{\Omega_\mathrm{cdm}}
\newcommand{\Ob}{\Omega_\mathrm{b}}

\newcommand{\ob}{\omega_\mathrm{b}}
\newcommand{\ocdm}{\omega_\mathrm{cdm}}

\newcommand{\Neff}{N_{\mathrm{eff}}}

\newcommand{\lcdm}{$\Lambda$CDM} 
 
\newcommand{\wowacdm}{$w_0w_a$CDM} 
\newcommand{\lya}{Ly$\alpha$\xspace}

\newcommand{\Planck}{\emph{Planck}}

\newcommand{\hinvmpc}{\,h^{-1}{\rm Mpc}}
\newcommand{\hmpcinv}{\,h\,{\rm Mpc^{-1}}}
\newcommand{\kmsMpc}{\,{\rm km\,s^{-1}\,Mpc^{-1}}}

\input{MG.authorlist.nov18}

\emailAdd{mishak@utdallas.edu}

\title{Modified Gravity Constraints from the Full Shape Modeling of Clustering Measurements from DESI 2024}

\date{\today}

\abstract{
We present cosmological constraints on deviations from general relativity (GR) from the first-year of clustering observations from the Dark Energy Spectroscopic Instrument (DESI) in combination with other available datasets including the CMB data from Planck with CMB-lensing from Planck and ACT, BBN constraints on the physical baryon density, the galaxy weak lensing and clustering from DESY3 and supernova data from DESY5. 
We first consider the $\mu(a,k)$--$\Sigma(a,k)$ modified gravity (MG) parameterization (as well as $\eta(a,k)$) in a \lcdm\ and a \wowacdm\ cosmological backgrounds. 
Using a functional form for time-only evolution gives $\mu_0= 0.11^{+0.44}_{-0.54}$ from DESI(FS+BAO)+BBN and a wide prior on $n_{s}$. 
Using DESI(FS+BAO)+CMB+DESY3+DESY5-SN, we obtain $\mu_0  =  0.05\pm 0.22$ and $\Sigma_0 =  0.008\pm 0.045$ and similarly $\mu_0 = 0.02^{+0.19}_{-0.24}$ and $\eta_0 = 0.09^{+0.36}_{-0.60}$, in an \lcdm\ background. In \wowacdm\, we obtain $\mu_0  =-0.24^{+0.32}_{-0.28}$ and $\Sigma_0 = 0.006\pm 0.043$, consistent with GR, and we still find a preference of the data for a dynamical dark energy with $w_0>-1$ and $w_a<0$.
Using functional dependencies in both time and scale gives $\mu_0$ and $\Sigma_0$ with a same level of precision as above but other scale MG parameters remain hard to constrain. 
We then move to binned parameterizations in a \lcdm\ background starting with two bins in redshift and obtain, 
 $\mu_1=1.02\pm 0.13$, $\mu_2=1.04\pm 0.11$, $\Sigma_1=1.021\pm 0.029$ and $\Sigma_2=1.022^{+0.027}_{-0.023}$, all consistent with the unity value of GR in the binning formalism. 
 We then extend the analysis to combine two bins in redshift and two in scale giving 8 MG parameters that we find all consistent with GR. 
We note that we find here that the tension reported in previous studies about $\Sigma_0$ being inconsistent with GR when using Planck PR3 data goes away when we use the recent \texttt{LoLLiPoP}+\texttt{HiLLiPoP} likelihoods. As noted in previous studies, this seems to indicate that the tension is indeed related to the CMB lensing anomaly in PR3 which is also resolved when using the recent likelihoods.
We then constrain the class of Horndeski theory in the effective field theory of dark energy approach. We consider both EFT-basis and $\alpha$-basis in the analysis. Assuming a power law parameterization for the EFT function $\Omega$, which controls non-minimal coupling, we obtain $\Omega_0 = 0.012^{+0.001}_{-0.012}$ and $s_0 = 0.996^{+0.54}_{-0.20}$ from the combination of DESI(FS+BAO)+DESY5SN+CMB in a $\Lambda$CDM background, which are consistent with GR.
Similar results are obtained when using the $\alpha$-basis and assuming no-braiding ($\alpha_B=0$) giving $c_M<1.14$ at $95\%$ CL in a $\Lambda\rm CDM$ background, also in agreement with GR. However, we see a mild yet consistent indication for $c_B > 0$ when $\alpha_B$ is allowed to vary which will require further study to determine whether this is due to systematics or new physics.
}
\begin{document}
\maketitle
\flushbottom

\section{Introduction}
\label{sec:intro}
A pillar of our standard model of cosmology is the general relativity theory of gravity that was proposed over a century ago by Einstein \cite{Einstein1915}. The model and its foundational theory have since flourished through a lot of successes in making predictions that continue to be confirmed by various astronomical observations \cite{Will2018,BertiElAlReview2015,Will:2014kxa}.  However, the problem of cosmic acceleration and the dark energy associated with it, as well as the tedious problems of the cosmological constant, have motivated further the question of testing general relativity at cosmological scales \cite{Weinberg1988CC,Padmanabhan2003,Sahni2000,Carroll2001CC,Peebles2003,Copeland2006,Ishak:2005xp}. Is cosmic acceleration the symptom of an extension or deviation from general relativity? Or is it a hint for a new understanding of our notion of space and time?  Moreover, it is worth noting that testing general relativity on cosmological scales is an appealing endeavor on its own right with all the high precision data that is accumulating from many surveys \cite{DES:2022ccp,Blake:2020mzy,Joudaki:2016kym,Planck:2015bue,Jullo:2019lgq,Linder:2019khz}. Testing general relativity on cosmological scales has been the subject of a tremendous effort within the cosmology community and we refer the reader to multiple reviews on the subject and the references therein \cite{Clifton:2011jh,Koyama:2015vza,Joyce:2016vqv,Ishak:2018his,Martinelli:2021hir}. 

From the multitude of works done on this subject in the last two decades, three major approaches to testing general relativity in cosmology have emerged. The first approach is to add to the perturbed Einstein equations some physically motivated phenomenological parameters or functions that are free to vary and are constrained from fits to the data. Such parameters are predicted by general relativity to take some specific known values. The goal then becomes to try to measure if these parameters take values that are different from those predicted by the theory of general relativity. Then, significant deviations may either point out to systematics in the data that have not been accounted for yet, or indicate a deviation in the underlying theory of gravity away from general relativity. An advantage of this approach is that we do not need to know in advance what the potential of the exact modified gravity model is. In this sense, this approach is more general and rather than looking for any deviations from general relativity as a first step and then if such a departure is found, one can use such new values to investigate what models could be associated with such modified gravity signatures. Additionally, there are systematic approaches for constructing parameterized forms, such as those used in effective field theories of dark energy and modified gravity (see e.g., \citep{Gubitosi:2012hu, Hu:2013twa}). Despite the advantages of this approach, some modified gravity models feature degrees of freedom that cannot be captured by such parameterizations. Another promising possibility is to employ non-parametric methods to reconstruct the time evolution of the modified gravity functions directly from the data \citep[e.g.,][and references therein]{Pinho:2018unz, Raveri:2019mxg, Pogosian:2021mcs, Ruiz-Zapatero:2022xbv, Alestas:2022gcg, Calderon:2022cfj,Calderon:2023msm,Mu:2023zct}.

The second approach is to analyze specific modified gravity models that are consistent theories on their own, or the low energy limit of what may be conceived as a more fundamental description of nature. Among theories which have driven interest in cosmology in recent years are those that exhibit a screening of these modifications on certain environments or scales, such as the popular $f(R)$ or nDGP. The following reviews describe in more detail the different categories behind some of these gravity theories, see e.g.\ \cite{Clifton:2011jh,Langlois:2018dxi}. A known difficulty in this second type of analyses is that they require to derive non trivial specific cosmological observables and functions to be able to confront them to observations.  Also, when it comes to available cosmological simulations, such modified gravity models are far behind $\Lambda$CDM simulations \cite{Alam:2020jdv} and computational frameworks have also not been raised to the same sophistication level or code-running speed as in $\Lambda$CDM. For some of these models, it remains a challenge to compare them to the full CMB data or the  weak lensing and galaxy clustering data and their cross-correlation. Nevertheless, such specific models can allow one to study gravity beyond the limited orders of phenomenological parameterization approaches.  

A third method that could be looked at as being an indirect method is that of analyzing and quantifying tensions and discordance between cosmological parameters within the standard model as determined by different datasets, see e.g. \cite{DES:2020hen,Raveri:2018wln,Andrade:2021njl,Linder2005,DES:2020iqt,2006-Ishak-splitting,Lin:2017ikq}. The detection and quantification of such a significant inconsistency would signal a problem with the standard model used and its underlying theory of gravity. Studies for this third approach are complementary and usually try to motivate the need for going to extended models to alleviate discordance such as the Hubble tension or the amplitude of matter fluctuation (see e.g \cite{Abdalla:2022yfr,MGH01_SolaPeracaula:2019zsl,MGH010_Rossi:2019lgt,MGH02_Ballardini:2016cvy,MGH03_Umilta:2015cta,MGH06_Yan:2019gbw,MGH07_Frusciante:2019puu,MGH08_Braglia:2020iik} for modified gravity models that aim to alleviate cosmological tensions).

In this paper, we focus on the first approach, i.e. parameterizations of deviations from general relativity to test gravity using clustering data from the Dark Energy Spectroscopic Instrument (DESI), in combination with other datasets. 

{DESI is an instrument on the Mayall telescope capable of capturing thousands of simultaneous spectra during each sky exposure, aided by 5000 positioners on the focal plane \cite{DESI2016b.Instr, DESI2022.KP1.Instr, FocalPlane.Silber.2023, Corrector.Miller.2023,SurveyOps.Schlafly.2023} and 9 spectrographs with a spectral resolution ranging from 2000 to 5000 over the wavelength range of 3600 to 9800 Angstroms \cite{Spectro.Pipeline.Guy.2023}.  Among many interesting scientific cases (see, for example, \cite{Snowmass2013.Levi, DESI2016a.Science, ChaussidonY1fnl}), DESI has the potential to test gravity using various estimators across targets spanning multiple redshift bins \cite{DESI2016a.Science, Alam:2020jdv}. In particular, DESI's} full shape clustering measurements trace the growth rate of large scale structure in the universe which is sensitive to the underlying theory of gravity.  The growth of large-scale structure is able to distinguish between gravity theories even if they have exactly the same expansion history and can thus act as a discriminant between such theories. 

We explore various combinations of DESI with other datasets such as redshifts and distances to supernovae, the cosmic microwave background (CMB) radiation temperature and polarization data as well as CMB lensing, constraints on the baryons physical density from the Big Bang Nucleosynthesis (BBN), the weak lensing and clustering data and their cross-correlation from the Dark Energy Survey (DES).        

The paper outline is as follows. In \cref{sec:data_methodology}, we describe the datasets and the inference methodology we use. In \cref{sec:phenomenological_parameters}, we constrain physically motivated phenomenological parameters that are added to the perturbed Einstein gravitational field equations including functional and binned forms for time and scale evolution. 
In the following \cref{sec:EFT_alpha_parameterization}, we constrain the parameters of the EFT parameterization of modified gravity. We conclude in \cref{sec:conclusions}. 

\section{Data and Methodology}
\label{sec:data_methodology}

{This work uses the observations of DESI's Data Release 1 (DR1, \cite{DESI2024.I.DR1}),  which covers the first year of observations in the main survey. The main survey began in May 2021, following a successful survey validation campaign with its associated data release \cite{DESI2023a.KP1.SV, DESI2023b.KP1.EDR}, which included a detailed characterization of the extra-galactic target selection \cite{TS.Pipeline.Myers.2023} and visual inspections \cite{VIGalaxies.Lan.2023,VIQSO.Alexander.2023}. }

{The DESI 2024 results using DR1 focus on the separate two point statistics measurements: Baryonic Acoustic Oscillations (BAO) and the full-shape of the power spectrum. These two point measurements were extensively validated, as described in \cite{DESI2024.II.KP3} for the galaxy and quasar sample, and in \cite{KP6s6-Cuceu,DESI2024.IV.KP6} for the Lyman alpha forest (Lya). A first group of publications centers around galaxy \cite{DESI2024.III.KP4} and Lya \cite{DESI2024.IV.KP6} BAO measurements, along with its cosmological parameter inference \cite{DESI2024.VI.KP7A}. Accompanying this work, a second set of publications explores the full-shape using galaxies and quasars \cite{DESI2024.V.KP5} and the corresponding impact on cosmological models \cite{DESI2024.VII.KP7B}.}

\subsection{DESI DR1 data and full-shape measurements}

The measurements adopted by DESI's DR1 are derived from the redshifts and positions of over 4.7 million unique galaxies and QSOs observed over a $\sim$$7{,}500$ square degree footprint covering the redshift range $0.1 < z < 2.1$.  These discrete tracers, described in detail in \cite{DESI2024.V.KP5}, are broken into four target classes:  300,017 galaxies from the magnitude-limited bright galaxy survey (BGS, \cite{BGS.TS.Hahn.2023}); 2,138,600 luminous red galaxies (LRG, \cite{LRG.TS.Zhou.2023}); 1,415,707 emission line galaxies (ELG, \cite{ELG.TS.Raichoor.2023}) and 856,652 quasars (QSO, \cite{QSO.TS.Chaussidon.2023}) (see Table 1 of \cite{DESI2024.V.KP5}). These four tracers are split into six redshift bins: one bin with the BGS ($0.1<z<0.4$), three bins with the LRGs ($0.4<z<0.6$, $0.6<z<0.8$, and $0.8<z<1.1$), one bin with the ELGs ($1.1<z<1.6$), and one redshift bin with the QSOs ($0.8<z<2.1$). The power spectrum in each redshift bin is subsequently computed as described below.

We complement the information from discrete tracers with that from the \lya\ forest --- features in the spectra of distant QSOs that are sensitive to the large-scale structure in the intergalactic medium.  Measurements of the 3D correlation function of the DR1 \lya\ forest data are presented in \cite{DESI2024.IV.KP6}. We currently only use the baryon acoustic oscillation information in the large-scale clustering of the \lya\ forest to constrain the background geometry, and do not provide a measurement of growth. 

Power-spectrum measurements, including the treatment and control of systematic errors that include sky masks, fiber-assignment completeness, imaging systematics, redshift systematics, and other sources of error are all described in \cite{DESI2024.II.KP3}. In brief, we measure the monopole, quadrupole and hexadecapole of the Fourier-space power spectra, as they quantify the information imprinted by redshift-space distortions. These measurements are obtained with the Yamamoto estimator \cite{yamamoto2006} that was implemented in \textsc{pypower}.\footnote{\url{https://github.com/cosmodesi/pypower}} The power-spectrum measurements are obtained from the galaxy catalogs (``data") and from synthetically generated catalogs with random distribution of points (``randoms") to which we assign the same selection as for the data. We also use the random catalog to compute the window matrix \cite{Beutler:2021eqq,KP3s5-Pinon} that relates the measured power spectrum multipoles to the theory power spectrum prediction. We make use of the power-spectrum measurements in wavenumber range $0.02 < k / (\hinvmpc) < 0.2$, adopting the binning width of $\Delta k = 0.005 \; \hmpcinv$. We only use the monopole and quadrupole of the power spectrum for our cosmological tests as the hexadecapole did not pass the systematics tests \cite{DESI2024.V.KP5}.

To obtain the cosmological tests, we employ the methodology that was described in \cite{DESI2024.V.KP5,DESI2024.VII.KP7B}, and we summarize the approach here. The essential element are the ``full-shape" measurements, that is, measurements of the monopole and quadrupole in redshift bins as a function of scale. The full-shape measurements rely on a perturbation-theory model that directly fits to power-spectrum multipoles. In the perturbative expansion, the growth of structure is treated systematically by expanding order-by-order in the amplitude of the initial fluctuations, with nonlinearities at small scales encoded using a series of ``counterterms'' that are constrained by the symmetries of the equations of motion. This so-called ``effective-field theory" approach also consistently treats the fact that the objects that we utilize (galaxies, quasars, and the \lya\ forest) are biased tracers of the large-scale structure. Our approach has been described and developed in some detail in \cite{KP5s2-Maus,KP5s3-Noriega,KP5s4-Lai,KP5s5-Ramirez} and references therein, and validated via the comparison of several perturbation theory codes, and to a series of simulations, in \cite{KP5s1-Maus,DESI2024.V.KP5}. As a default, we use the Eulerian perturbation theory implementation in \texttt{velocileptors} \cite{Chen20}.

We combine these full-shape measurements with BAO measurements obtained from post-reconstruction correlation functions \cite{DESI2024.III.KP4} for all six redshift bins. To combine power spectrum and BAO measurements, we compute the complete covariance matrix that covers the power spectrum measurements, the post-reconstruction BAO parameters, and their mutual correlation (for more details, refer to Sec 2.3.1 of \cite{DESI2024.VII.KP7B}). 

The total log-likelihood is sum of log-likelihood for the six redshift bins and the Ly$\alpha$ BAO log-likelihood \cite{DESI2024.IV.KP6}, closely following the methodology outlined in our DR1 Full-Shape analysis \cite{DESI2024.VII.KP7B}.

\subsection{External datasets}
We use the same external datasets as the key DESI DR1 cosmology papers \cite{DESI2024.VI.KP7A,DESI2024.VII.KP7B}, and summarize them as follows. 
\begin{itemize}
    \item \textbf{Cosmic Microwave Background (CMB)}: We employ the official Planck 2018 high-$\ell$ TTTEEE (\texttt{plik}) likelihood, supplemented with the low-$\ell$ TT (\texttt{Commander}) and EE (\texttt{SimAll}) likelihoods. In addition to the temperature and polarisation anisotropies, we also include measurements of the lensing potential auto-spectrum $C_\ell^{\phi\phi}$ from Planck's \texttt{NPIPE} PR4 CMB maps \cite{Carron:2022eyg}, in combination with lensing measurements from the Atacama Cosmology Telescope (ACT) Data Release 6 (DR6) \cite{ACT:2023kun,Qu:2023}. Specifically, we use \texttt{cobaya}'s public implementation, using the \texttt{actplanck\_baseline} option\footnote{For details, visit \url{https://github.com/ACTCollaboration/act_dr6_lenslike}.}. Finally, whenever relevant, we will also report the constraints using two updated likelihood releases of the Planck data Camspec \cite{Efstathiou:2021,Rosenberg:2022} and \texttt{LoLLiPoP}-\texttt{HiLLiPoP} \cite{Tristram:2021,Tristram:2023}, based on the PR4 CMB maps.

    \item \textbf{Type Ia supernovae (SN~Ia)}:     We use the DES-SN5YR dataset, a compilation of 194 low-redshift SN~Ia ($0.025<z<0.1$) and 1635 photometrically classified SN~Ia covering the range $0.1<z<1.3$ \cite{DES:2024tys}. Since supernovae are not expected to significantly affect the constraints on modified gravity models when combined with DESI and CMB data, we have opted to utilize just this one SN~Ia dataset, and not three datasets (i.e.\ we do not use PantheonPlus \cite{Scolnic:2021amr} or  Union 3 compilation \cite{Rubin:2023}) which we employ when we constrain dark energy (in general relativity) in our companion paper \cite{DESI2024.VII.KP7B}. We chose this sample just for simplicity and to avoid unnecessary repetitions of results in the case of modified gravity.  
    
    \item \textbf{Weak Gravitational Lensing (WL)}:  We follow the MG analysis by DES \cite{DES:2022ccp}, and utilize the results from DES Year 3, which include the combined measurements of cosmic shear, galaxy-galaxy lensing, and galaxy clustering, which we refer to as the ``DESY3 (3$\times$2-pt)" analysis. In line with this, we do not apply the Limber approximation for galaxy clustering on large angular scales, but rather follow the exact method described in \cite{Fang:2019xat}. The DESY3 (3$\times$2-pt) analysis was conducted using source galaxies in four redshift bins [0, 0.36, 0.63, 0.87, 2.0] and lens galaxies from the Maglim sample in the first four redshift bins [0.20, 0.40, 0.55, 0.70, 0.85, 0.95, 1.05]. To confine the 3$\times$2-pt data to linear scales and improve constraints on MG parameters, we apply conservative cuts consistent with the DES linear scale cuts and set \texttt{use\_Weyl=true}.

    \item \textbf{Big Bang Nucleosynthesis (BBN)} We add the prior on the physical baryon density, $\Ob h^2$, coming from  Big Bang Nucleosynthesis in the dataset combinations that do not include the CMB information. Measurements of light elements abundances from BBN, specially deuterium (D) and Helium ($^4$He), constrain the baryon density. However, this depends on details of the theoretical framework, particularly the crucial input of nuclear interaction cross-sections. We adopt the results obtained  in a recent analysis \cite{Schoeneberg:2024} that made use of the \texttt{PRyMordial} code \cite{Burns:2024} to recompute the predictions while marginalizing over uncertainties in the reaction rates. We utilize the joint constraint on $\Ob h^2$ and the number of relativistic species $\Neff$,\footnote{The $2\times 2$ covariance matrix in $\Ob h^2$ and $\Neff$, and their respective central values, are available at \url{https://tinyurl.com/29vzc592}.} and fix the latter parameter to its fiducial value of 3.044; the resulting projected constraint on the physical baryon density is $\ob\equiv \Ob h^2= 0.02198 \pm 0.00053$. When we combine DESI with the CMB data, however, we do not use the BBN prior on $\Omega_b h^2$ as this parameter is tightly constrained by the CMB.
    
  \end{itemize}

\subsection{Cosmological inference, likelihoods and modeling}
Our inference approach follows the methodology described in the two DESI DR1 cosmology papers \cite{DESI2024.VI.KP7A} but as in \cite{DESI2024.VII.KP7B}, we include a larger number of nuisance parameters specific to the full-shape analysis. We employ the cosmological inference software \texttt{cobaya} \cite{Torrado:2019, Torrado:2021}, incorporating the DES-SN5YR likelihood along with our DESI likelihood, \texttt{desilike}. For CMB likelihoods, we utilize public packages that are either part of the \texttt{cobaya} distribution or available from the respective research teams.

For our modified gravity parameterizations including functional and binning methods, we use the \texttt{ISiTGR} (Integrated Software in Testing General Relativity) code \cite{Dossett:2011tn, Garcia-Quintero:2019xal}, which is built on \texttt{CAMB}~\cite{LewisCAMB:2000, HowlettCAMB:2012}, and is also integrated within \texttt{cobaya} through a Python-wrapper described in \cite{Garcia-Quintero:2020mja}. 
\texttt{ISiTGR} can run on a \lcdm\ or dynamical dark energy background and allows for time-dependent equation of state as well as a spatially flat or curved background. It has been tested to provide a consistent implementation of anisotropic shear to model massive neutrinos throughout the full formalism. It allows to use functional, binned and hybrid time- and scale-dependencies for MG parameters. 

We perform Bayesian inference using the Metropolis-Hastings MCMC\footnote{
We require a convergence of $R-1\leq0.01$ for MCMC chains for 
the large majority of the $\mu-\Sigma$ and $\mu-\eta$ parameterizations. However, for a few combinations, and in particular with the new Planck likelihood \texttt{LoLLiPoP}-\texttt{HiLLiPoP}, the chains converge at a slower rate than those using PR3. We required for those $R-1\leq 0.03$.  Also, for the case of EFT parameterization, the chains run at a much slower rate and we required $R-1\leq0.03$ for the EFT-basis, but for the $\alpha$-basis we accepted $R-1 \le 0.1$ , similar to other works, see e.g. Ref. \cite{Chudaykin:2024gol}.} sampler \cite{LewisMCMC:2002, LewisMCMC:2013} within \texttt{cobaya}.

\Cref{tab:priors} provides a summary of the cosmological parameters sampled in different runs and the priors applied to them. For the basic DESI (FS+BAO) analysis in the \lcdm\ background model, we vary five key cosmological parameters: Hubble constant ($H_0$), physical densities of baryons ($\Ob h^2$) and cold dark matter ($\Ocdm h^2$), and the amplitude ($A_s$) and spectral index ($n_s$) of the primordial density perturbations. When incorporating the CMB likelihood, we replace $H_0$ with $\theta_{\rm MC}$, an approximation of the acoustic angular scale $\theta_\ast$, and include the optical depth to reionization parameter ($\tau$). In the dynamical dark-energy background model (\wowacdm), we introduce two additional parameters: $w_0$ and $w_a$. Additionally, we account for a set of nuisance parameters necessary to describe the full-shape clustering signal. Detailed descriptions of the DESI FS nuisance parameters can be found in \cite{DESI2024.V.KP5}, and they are listed at the bottom of \Cref{tab:priors}. 
In the same table we also summarize the modified gravity parameters and their priors; we introduce the definitions of some of these parameters in the respective sections and subsections. 

{As a test of the effect of the priors on the full shape EFT modeling parameters, we increased the prior range on these parameter by a factor of 3 for some test runs. We find only a small shift in the mean on the MG parameter $\mu_0$ (well-below the 1$\sigma$ uncertainty)  for the baseline $\mu$-$\Sigma$ analysis and for the less constraining combination of data DESI+BBN+n$_{s10}$.  The shift goes away when using the next constraining combination  DESI+CMBL with no-shift of $\mu_0$. A shift is even less expected for the more constraining dataset combinations like when adding 3x2pts and/or supernova.}

Although we provide results for DESI (FS+BAO)+BBN+$n_{s10}$ data for completeness, we provide various combinations of datasets where projection effects are expected to be 
 effectively mitigated (see \cite{DESI2024.V.KP5} and  \cite{DESI2024.VII.KP7B}). Projection effects have been extensively explored in the \cite{DESI2024.V.KP5} analysis and then further discussed for beyond LCDM (i.e. \wowacdm) in 
 \cite{DESI2024.VII.KP7B}. 
The reason that the projection effects are much more pronounced in the FS+BAO constraints in \wowacdm\ than in the equivalent BAO-alone analysis \cite{DESI2024.VI.KP7A} is the presence of many additional nuisance parameters in the full-shape analysis which allow additional freedom and open new degeneracy directions. 
In models beyond \lcdm, when using combinations of datasets that included both CMB and Supernova data sets with DESI (and also DES Y3 data here), such projection effects were found to be effectively mitigated.  We mainly focus here on results where we have used the combination DESI+CMB+DESy3+DESY5SN, but for less constraining combinations like DESI+BBN+ns10 or DESI+CMB results should be taken with some caution. {This is illustrated in Appendix \ref{sec:projection} where we show in Figure~\ref{fig:mg_projection} that the maximum a posteriori and mean values for the MG parameters $\mu_0$ and $\Sigma_0$ lie well within 1$\sigma$. This indicates that projection-induced biases are minimal when those datasets are combined.}

For the EFT-basis modified gravity inference, we utilize the \texttt{EFTCAMB} \cite{Hu:2013twa,Raveri2014} code which implements the EFT action in the Boltzmann code \texttt{CAMB}~\cite{LewisCAMB:2000, HowlettCAMB:2012}. For the properties functions ($\alpha$-basis), we use the publicly available \texttt{mochiclass} \cite{Cataneo:2024uox}, a recently released branch of the code \texttt{hi\_class} \cite{Bellini_2020,Zumalacarregui:2016pph}.  We interfaced \texttt{EFTCAMB} and \texttt{mochiclass} with the MCMC sampler \texttt{cobaya} to perform Bayesian inference.

For the DESY3 3×2-pt analysis, we employ a likelihood that we specifically tailored for our modified gravity analyses.  The likelihood has been validated against the DESY3 modified-gravity results from \cite{DES:2022ccp} and it is noteworthy that we also made the same scales cuts as this paper limiting the data to linear scales where our theoretical modeling and parameterizations are valid. This likelihood has been integrated into our main pipeline using \texttt{desilike} and \texttt{cobaya}.

{It is worth noting that the perturbation theory employed for the full shape analysis here relies on the commonly used Einstein-de Sitter (EdS) kernels. This approach is valid when the growth rate and the matter density parameter are related by $f^2 = \Omega_m$. In $\Lambda$CDM and similar dark energy models, where $f \approx \Omega_m^{6/11}$ provides an excellent approximation, the use of EdS kernels is generally considered suitable. But, this assumption does not always hold in MG models.
However,  we note that our MG parameterizations are defined at the linearly perturbed Einstein Field Equations level, and that the scale cuts applied in the full-shape analysis of DESI (FS+BAO) \cite{DESI2024.VII.KP7B} ensure that nonlinear terms are small, specifically the one-loop terms in the effective field theory expansion. Moreover, the \texttt{velocileptors} method used in the underlying DESI full-shape analysis, and that we use here as well, has been compared with the MG non-linear code \texttt{fkpt} \cite{Rodriguez-Meza:2023rga}, demonstrating good agreement in loop corrections for small departures from GR. Such correction are expected to be small in such a case since the correction due to MG in nonlinearities is small (i.e one loop)  $\times$ small (i.e. deviation from GR).  Therefore, this validates in this work our results where our mean values for modified gravity parameters are found to be close to their GR values as it is the case of our most constraining data combinations where for example CMB and galaxy lensing  are added to DESI data as well as supernova data. These strongly constraining data combinations are found to indeed force us to be in this vicinity of GR where our use of  \texttt{velocileptors} and EdS assumptions are valid.} 


{Furthermore, we tested our approximation given by the EdS kernels when we fit the model given by Eq.\eqref{eq:muEvolution} below, with $c_1=\lambda=0$, that is, its scale-independent version. We compared power spectrum outputs of the code \texttt{fkpt}\footnote{\href{https://github.com/alejandroaviles/fkpt}{https://github.com/alejandroaviles/fkpt}} \cite{Rodriguez-Meza:2023rga}, that in the case of scale-independent MG, utilizes the full, correct kernels in MG, against the code utilizing EdS kernels. We estimated  the ratio between the power spectra with the full kernels versus the EdS ones. We considered the first three multipoles, $\ell=0$, 2 and 4 (although we do not use the latter in the fits), for several values of $\mu_0$ and redshift. 
For example, we provide in the appendix \cref{fig:EdSvsMG} that illustrate such a power ratio for the particular case where the modified gravity parameter $\mu_0=0.5$ at redshift $z=0.3$. We found that for larger redshifts, or smaller values of the parameter $\mu_0$, the differences are always smaller so this should be considered as an extreme case. 
As shown in the figure, the difference in the monopole is smaller   than the 1\% for scales $k< 0.12\,h\,\text{Mpc}^{-1}$ and smaller than the 2\% for scales $k<0.25\,h\,\text{Mpc}^{-1}$. 
For the quadrupole, 
 the difference is smaller than the 2.4\% for scales $k< 0.10\,h\,\text{Mpc}^{-1}$ and remains bellow 3.6\% when it gets to  $k=0.20\,h\,\text{Mpc}^{-1}$.    
Although we never fit the hexadecapole in this work, we just show it here for future references in the figure.   
Again, this is for value of $\mu_0=0.50$ which are far from the GR value of zero, and this is consistent with our discussion in the previous paragraph where the mean values obtained from our most constraining combinations of data sets are found to be $\sim 0.05$ for $\mu_0$. This analysis shows that the use of EdS is safe within DESI DR1, provided that the deviations from GR give kernel differences with EdS that remain small when using  Eq.\eqref{eq:muEvolution} and we work within such an assumption.} 

{Additionally, knowing how much the 2-loop corrections can be more important in MG than in the more commonly used GR-case is worth exploring. While our best fit MG parameters are close to GR values with very little deviations from it and in that case, one would expect similar effects in the MG case as in the GR case. However, to look into this point further, we performed comparisons of the linear, 1-loop and 2-loop spectra for the extreme case of $\mu_0=0.50$. We found that up to $k< 0.20\,h\,\text{Mpc}^{-1}$, the MG 1-loop and 2-loop estimations always remain subdominant, just like in GR, and that the respective differences between MG and GR are nearly constant over k and only constitute a multiplicative constant factor that ultimately can be absorbed by the linear bias factors, which are different in values in MG and GR. Such contributions from beyond 1-loop should be analyzed in detail for both GR and MG in future analyses}. 

In the case of scale-dependent MG the growth rate also becomes scale-dependent. Although, in principle, the perturbative technique should be modified to include the factors  $f(k, t)$  \cite{Rodriguez-Meza:2023rga}, for simplicity, we choose to use the same pipeline as in the rest of the paper since this approach is also expected to be accurate for small deviations from GR.

Last, we also note that our method relies on linear parameterizations only, in that sense our approach is not complete at quasi-linear scales. However, there are multiple ways to extend beyond linear order, even while enforcing physically motivated symmetries, see e.g.\ \cite{DAmico:2021rdb,Euclid:2023bgs}.  It is worth noticing that these additional terms only influence the theoretical power spectrum through loop contributions, which are likely to be minimal for theories that are close to GR. Hence,  in this work we are assuming that such (unspecified) non-linear terms can be neglected.

\begin{table}[t] 
    \begin{center}
    \resizebox{\textwidth}{!}{
     \renewcommand{\arraystretch}{1} 
    \begin{tabular}{|lllll|}
    \hline
    data or model & parameter & default & prior & comment\\  
    \hline 
    \textbf{DESI (\lcdm)}
     & $H_{0} \; (\kmsMpc)$ &---& $\mathcal{U}[20, 100]$ &---  \\
     & $\ob$ &---& $\mathcal{N}(0.02218, 0.00055^2)$ &BBN prior  \\
    & $n_{s}$ &---& $\mathcal{N}(0.9649, 0.042^2)$ &Planck 10$\sigma$\\  
    & $\ocdm$ &---& $\mathcal{U}[0.001, 0.99]$ &---\\
    & $\ln(10^{10} A_{s})$ &---& $\mathcal{U}[1.61, 3.91]$ &---\\    
    \hline     
    \textbf{CMB (\lcdm)} 
    & $100 \theta_{\mathrm{MC}}$ &---& $\mathcal{U}[0.5, 10]$ &---\\
    & $\tau$ & 0.0544 & $\mathcal{U}[0.01, 0.8]$ &---\\
    & $\ob$ &---& $\mathcal{U}[0.005, 0.1]$ & \mbox{no BBN prior}\\ 
    & $n_s$ &---& $\mathcal{U}[0.8, 1.2]$ & \mbox{no 10$\sigma$ prior}\\      
    \hline 
    \textbf{Beyond \lcdm} 
    & $w_0$ & $-1$ & $\mathcal{U}[-3, 1]$ &---\\
    (dynamical DE) & $w_{a}$ & $0$ & $\mathcal{U}[-3, 2]$ &---\\
    \hdashline    

    (redshift dependence) & $\mu_0$ & $0$ & $\mathcal{U}[-3, 3]$ & ---\\
    & $\Sigma_0$ & $0$ & $\mathcal{U}[-3, 3]$ &--- \\
    & $\eta_0$ & $0$ & $\mathcal{U}[-3, 3]$ &--- \\

    (for scale dependence) & $\lambda$ & $0$ & $\mathcal{U}[-5, 5]$ & ---\\
    & $c_1$ & $1$ & $\mathcal{U}[-5, 5]$ &--- \\
    & $c_2$ & $1$ & $\mathcal{U}[-5, 5]$ &--- \\

    (redshift-only binning) 
     & $\mu_1,\, \mu_2$       & $1$ & $\mathcal{U}[-3, 3]$ &--- \\
     & $\Sigma_1,\, \Sigma_2$ & $1$ & $\mathcal{U}[-3, 3]$ &--- \\
     (redshift \& scale binning) 
     & $\mu_1,\, \mu_2,\, \mu_3,\, \mu_4$ & $1$ & $\mathcal{U}[-3, 3]$ &--- \\
     & $\Sigma_1,\, \Sigma_2,\, \Sigma_3,\, \Sigma_4$ & $1$ & $\mathcal{U}[-3, 3]$ &--- \\
    \hdashline    
    ($\alpha$-basis) & $c_M$ & $0$ & $\mathcal{U}[-10, 10]$ &--- \\
    &$c_B$ & $0$ & $\mathcal{U}[-10, 10]$ &--- \\
    \hdashline    
    (EFT-basis) & $\Omega_0$ & $0$ & $\mathcal{U}[-1, 1]$ &--- \\
    &$s_0$ & $0$ & $\mathcal{U}[-5, 5]$ &--- \\
    \hline
    \textbf{nuisance (DESI)} 
    & $(1 + b_1) \sigma_8$ &  & $\mathcal{U}[0, 3]$ & each $z$-bin\\    
    & $b_2 \sigma_8^2$ &  & $\mathcal{N}[0, 5^2]$ & each $z$-bin\\    
    & $b_s \sigma_8^2$ &  & $\mathcal{N}[0, 5^2]$ & each $z$-bin\\ 
    & $\alpha_0$ &  & $\mathcal{N}[0, 12.5^2]$ & analytic\\
    & $\alpha_2$ &  & $\mathcal{N}[0, 12.5^2]$ & analytic\\
    & $\mathrm{SN}_0$ &  & $\mathcal{N}[0, 2^2]$ & analytic\\
    & $\mathrm{SN}_2$ &  & $\mathcal{N}[0, 5^2]$ & analytic\\
   \hline 
    \end{tabular}}
    \end{center}
    \caption{Parameters and priors used in the analysis. All of the priors are flat in the ranges given. Here, $\mathcal{U}$ refers to a uniform prior in the range given, while  $\mathcal{N}(x, \sigma^2)$ refers to the Gaussian normal distribution with mean $x$ and standard deviation $\sigma$. In addition to the flat priors on $w_0$ and $w_a$ listed in the table, we also impose the requirement $w_0+w_a<0$ in order to enforce a period of high-redshift matter domination. Similarly, modified gravity parameters $\mu_0$ and $\Sigma_0$ are imposed an extra prior $\mu_0 < 2 \Sigma_0 + 1$ (see \cref{sec:functional}). 
    Bias parameters $(1 + b_1) \sigma_8$, $b_2 \sigma_8^2$ and $b_2 \sigma_8^2$ are independent within each tracer and redshift bin, as well as counter-terms $\alpha_0$, $\alpha_2$ and stochastic parameters $\mathrm{SN}_0$, $\mathrm{SN}_2$, which are marginalized over analytically. 
    Refer to Ref. \cite{DESI2024.V.KP5} for detailed discussions of DESI FS nuisance parameters. Note that the BBN and $n_s$ priors are added as a default in the DESI (FS+BAO) analysis and noted as DESI (FS+BAO)+BBN+$n_{s10}$, but dropped when DESI data is combined with the CMB. 
    }
    \label{tab:priors}
\end{table}


\section{Constraints on modified gravity functions \tpdf{$\mu(z,k)$},  \tpdf{$\Sigma(z,k)$} and \tpdf{$\eta(z,k)$}}
\label{sec:phenomenological_parameters}


\subsection{Perturbed Einstein's equations and MG parameter formalism}
\label{sec:MG_formalism}

Following standard practices in the field, we introduce physically motivated phenomenological parameters for modified gravity (MG) into the perturbed Einstein Field Equations (EFEs) and constrain them using observational data to test any deviation from the predictions of general relativity (GR). For further details, see, e.g., the reviews \cite{Clifton:2011jh, Koyama:2015vza, Ishak:2018his} and references therein.


We adopt the conformal Newtonian gauge for the flat Friedmann–Lemaître–Robertson–Walker metric with scalar perturbations. The line element in this gauge is:
\begin{equation}
ds^2=a(\tau)^2[-(1+2\Psi)d\tau^2+(1-2\Phi)\delta_{ij}dx^i dx^j],
\label{eq:line-element}
\end{equation}
where $\Psi$ and $\Phi$ are the two gravitational potentials and $\tau$ is the conformal time.


The EFEs for this metric yield two evolution equations that describe how the gravitational potentials couple to the matter-energy content of spacetime. The first equation is a relativistic version of the Poisson equation in Fourier space, which, in the late universe when anisotropic stresses are negligible, simplifies to:
\begin{equation}
k^2\Psi = -4\pi G a^2 \mu(a,k) \sum_i\rho_i\Delta_i,
\label{eq:definition_mu}
\end{equation}
%
where \(\rho_i\) is the density of the matter species \(i\), and \(\Delta_i\) is the gauge-invariant, rest-frame overdensity for species \(i\). This equation governs the growth of linear structures in the universe.


The phenomenological function \(\mu(a, k)\) introduces scale and redshift dependencies to the gravitational coupling strength, thus modifying the growth rate of structure. This parameter is associated with the clustering of massive particles, and DESI directly constrains it through clustering measurements.


The second equation relates the two gravitational potentials, \(\Phi\) and \(\Psi\). In GR, these potentials are expected to be nearly equal at late times, as anisotropic stresses become negligible. However, in MG models, they can differ even in the late universe. This deviation is typically parameterized using the gravitational slip parameter:
\begin{equation}
\eta(a,k) \approx \frac{\Phi}{\Psi}.
\label{eq:definition_eta}
\end{equation}

By combining the two equations above, we derive an expression particularly relevant to the motion of massless particles in a gravitational field. This equation is especially important for gravitational lensing surveys and takes the form:
\begin{equation}
k^2(\Phi+\Psi)=-8\pi G a^2 \Sigma (a,k) \sum_i\rho_i\Delta_i.
\label{eq:definition_Sigma}
\end{equation}
Here, in the left-hand side, \((\Phi + \Psi) \), represents twice the Weyl potential, which governs the motion of massless particles. The right-hand side introduces the MG parameter \(\Sigma(a, k)\), which modifies the equation from its GR-based form.

At low redshift, where the anisotropic stress induced by free-streaming particles can be safely neglected, these MG parameters are related by,
\begin{equation}
\Sigma(a,k)=\frac{\mu(a,k)}{2}(\eta(a,k)+1).
\label{eq:MG_param_relation}
\end{equation}
In GR, these MG functions $\mu(a,k)$, $\Sigma(a,k)$ and $\eta(a,k)$ are predicted to be just one, leaving the perturbed EFEs unchanged from their standard form.

Lastly,  we note that our parameterization using  $\mu$, $\Sigma$ and $\eta$  are defined at the linearly perturbed EFEs level.  For DESI (FS+BAO), the scale cuts applied in the full-shape analysis \cite{DESI2024.VII.KP7B} ensure that nonlinear terms are small with implications as discusses above in section \cref{sec:data_methodology}. Moreover, the external data we use also rely on linear scales or has been reduced to them: SNIa and the CMB data are inherently linear; CMB lensing is almost entirely within the linear regime; and for the DESY3 (3×2-pt) data, we apply the same conservative scale cuts as in the DES MG paper \cite{DES:2022ccp}, limiting the data to linear scales.  Accordingly, our analyses here on modified gravity are based on linear scales where these parameterizations are well-defined. 

\subsection{Functional forms of MG parameterizations}
\label{sec:functional}

We consider a functional form of the MG parameters that includes both time and scale dependencies following previous works and based on the same motivation to seek whether modified gravity can be associated with the observed late time cosmic acceleration. The time dependence is often parameterized using a proportionality to the time evolution of the dark energy density parameter  $\Omega_{\text{DE}}(a)$, e.g. \cite{Simpson:2012ra,DES:2018ufa}.   For scale, it was shown in  \cite{BakerEtAl2014b,2013PhRvD..87j4015S} that within the quasi-static approximation,  a specific scale dependence for the MG parameters in the form of ratios of polynomials in the wave number, $k$,  can be adequate to capture such a dependence.   We therefore use the following forms (that were also used in, e.g., \cite{Planck:2015bue,BakerEtAl2014b,Garcia-Quintero:2020bac,Denissenya:2022ens}) to represent ratios of polynomials in $k$
\begin{equation}
\mu(a,k)=1+\mu_{0}\frac{\Omega_{\text{DE}}(a)}{\Omega_\Lambda} \left[ \frac{1+c_1 \left( \lambda H(a)/k \right)^2}{1+\left( \lambda H(a)/k \right)^2} \right]
\label{eq:muEvolution}
\end{equation}
and
\begin{equation}
\Sigma(a,k)=1+\Sigma_{0}\frac{\Omega_{\text{DE}}(a)}{\Omega_\Lambda} \left[ \frac{1+c_2 \left( \lambda H(a)/k \right)^2}{1+\left( \lambda H(a)/k \right)^2} \right].
\label{eq:SigmaEvolution}
\end{equation}
where the MG parameters $\mu_0$, and $\Sigma_0$ take the value of zero in GR. 

It is noteworthy to mention that while these forms of time and scale dependencies have been widely used in the literature and can serve for constraints and comparison with other works, they may have limitations and, like any other phenomenological parameterizations, maybe not cover all MG models, see e.g. \cite{Linder:2020xza}. Further discussions of phenomenological parameterizations including functional and binning forms can be found in, e.g. \cite{Clifton:2011jh,Koyama:2015vza,Ishak:2018his}. 

We note that the scale-dependent parameterization satisfies the limiting case that at high-$k$ (small scales),   $\mu(a)-1 \rightarrow \mu_0 \Omega_{\text{DE}}(a)/\Omega_\Lambda$ and $\Sigma(a)-1 \rightarrow \Sigma_0 \Omega_{\text{DE}}(a)/\Omega_\Lambda$.  Whilst for low-$k$ (large scales),  $\mu(a)-1 \rightarrow \mu_0 c_1 \Omega_{\text{DE}}(a)/\Omega_\Lambda$ and $\Sigma(a)-1 \rightarrow \Sigma_0 c_2 \Omega_{\text{DE}}(a)/\Omega_\Lambda$. Thus, the parameters $c_1$ and $c_2$ represent, respectively, the behavior of $\mu$ and $\Sigma$ at large scales. Since we are using units where we set the speed of light as $c=1$, the factor $H(a)/k$ becomes dimensionless.  Finally, when one sets $\lambda=0$ and $c_1=c_2=1$, this recovers the redshift-only dependence.

To finalize, we note that for the $\mu_0-\Sigma_0$ functional parameterization, we  impose a hard prior $\mu_0 < 2 \Sigma_0 + 1$ when running our MCMC chain inference, as done in previous studies, see e.g. \cite{DES:2018ufa,DES:2022ccp}. This prior is necessary to circumvent the part of parameter space where MG software codes based on \texttt{CAMB} run randomly into numerical errors when integrating the evolution of perturbations. However, this prior does not affect the interpretation of results. This can be seen in our figures showing that when using DESI-only, a horizontal-band posterior is expected. Similarly, we see that the CMB-only contour do hit this prior. But for any other dataset combination, the contours are much smaller, and thus unaffected by this prior. 

We also employ the parameterization $\mu(a)-\eta(a)$ to run separate analyses for the case of redshift-only using the following functional form that is slightly different from the above one \cite{Planck:2015bue,Planck-2018-cosmology,Andrade:2023pws}:
\begin{equation}
\mu(a)=1+E_{11}\Omega_{\text{DE}}(a)
\label{eq:muEvolution2}
\end{equation}
and 
\begin{equation}
\eta(a)=1+E_{22}\Omega_{\text{DE}}(a),
\label{eq:etaEvolution2}
\end{equation}
where we assume again a time evolution of MG parameters to be proportional to dark energy density in the context of cosmic acceleration.  The functions $\mu(a)$ and $\eta(a)$ take the value of one in GR. 
We report our results for this parameterization in terms of $\mu_0 \equiv\mu(a=1)-1$ and $\eta_0 \equiv \eta(a=1)-1$ which are determined from $E_{11}$, $E_{22}$ and the dark energy density today.  

\begin{table}
\centering
\resizebox{\columnwidth}{!}{%
    \small 
\setcellgapes{3pt}\makegapedcells  
\renewcommand{\arraystretch}{2.5} 
    \begin{tabular}{lccccc}
    \toprule
   {{\bf Flat} $\boldsymbol{\mu_0\Sigma_0}$ {\bf $\Lambda$CDM} }& {$\Om$} & {$\sigma_8$}  & $H_0  [${\rm km/s/Mpc}$] $ & {$\mu_0$} &{$\Sigma_0$}\\[-0.1cm]
    \midrule
    
    \makecell[l]{DESI (FS+BAO)\\\quad+BBN+$n_{s10}$} & $0.2956\pm 0.0096$ & $0.838\pm 0.034$ & $68.53\pm 0.75$ & $0.11^{+0.44}_{-0.54}$ \\
    \hdashline    

    CMB (PR3)-nl          & $0.3041\pm 0.0093$ & $0.742^{+0.13}_{-0.092}$ & $68.21\pm 0.71$ & $-0.66^{+1.5}_{-0.83}$ & $0.47^{+0.16}_{-0.22}$ \\
    
    CMB (CamSpec)-nl      & $0.3083\pm 0.0079$ & $0.743^{+0.13}_{-0.086}$ & $67.77\pm 0.59$ & $-0.64^{+1.4}_{-0.80}$ & $0.32^{+0.14}_{-0.20}$ \\
    
    CMB ($\texttt{LoLLiPoP}$-$\texttt{HiLLiPoP}$)-nl   & $0.3060\pm 0.0076$ & $0.737^{+0.13}_{-0.084}$ & $67.93\pm 0.57$ & $-0.73^{+1.4}_{-0.79}$ & $0.23^{+0.13}_{-0.20}$ \\

    CMB (PR3)-l   & $0.3105\pm 0.0083$ & $0.732^{+0.12}_{-0.096}$ & $67.71\pm 0.61$ & $-0.80^{+1.4}_{-0.88}$ & $0.25^{+0.12}_{-0.18}$ \\
    
    CMB (CamSpec)-l  & $0.3128\pm 0.0074$ & $0.734^{+0.13}_{-0.085}$ & $67.44\pm 0.54$ & $-0.77^{+1.4}_{-0.80}$ & $0.23^{+0.11}_{-0.18}$ \\
    
    CMB ($\texttt{LoLLiPoP}$-$\texttt{HiLLiPoP}$)-l   & $0.3093\pm 0.0073$ & $0.730^{+0.12}_{-0.094}$ & $67.69\pm 0.54$ & $-0.84^{+1.4}_{-0.85}$ & $0.22^{+0.12}_{-0.18}$ \\
    \hdashline    
 
    DESI+CMB (PR3)-nl   & $0.2985\pm 0.0055$ & $0.822\pm 0.024$ & $68.63\pm 0.44$ & $0.23\pm 0.24$ & $0.388^{+0.11}_{-0.086}$ \\
    
    DESI+CMB (CamSpec)-nl      & $0.3013\pm 0.0053$ & $0.822\pm 0.024$ & $68.29\pm 0.41$ & $0.24\pm 0.24$ & $0.26^{+0.12}_{-0.11}$ \\
    
    DESI+CMB ($\texttt{LoLLiPoP}$-$\texttt{HiLLiPoP}$)-nl    & $0.3006\pm 0.0051$ & $0.824\pm 0.024$ & $68.33\pm 0.40$ & $0.22\pm 0.24$ & $0.148^{+0.097}_{-0.12}$ \\
     
    DESI+CMB (PR3)-l        & $0.3023\pm 0.0053$ & $0.824\pm 0.024$ & $68.32\pm 0.41$ & $0.21\pm 0.24$ & $0.166\pm 0.074$ \\
    
    DESI+CMB (CamSpec)-l      & $0.3044\pm 0.0050$ & $0.823\pm 0.024$ & $68.05\pm 0.38$ & $0.21\pm 0.24$ & $0.144\pm 0.071$ \\
     
    DESI+CMB ($\texttt{LoLLiPoP}$-$\texttt{HiLLiPoP}$)-l     & $0.3028\pm 0.0050$ & $0.825\pm 0.024$ & $68.18\pm 0.38$ & $0.18\pm 0.24$ & $0.119^{+0.068}_{-0.076}$ \\
      \hdashline    

    DESI+CMB (PR3)-nl+DESY3  & $0.3027\pm 0.0051$ & $0.808\pm 0.023$ & $68.28\pm 0.40$ & $0.04\pm 0.22$ & $0.044\pm 0.047$ \\

    CMB-nl+DESY3   & $0.3074\pm 0.0081$ & $0.690^{+0.047}_{-0.064}$ & $67.93\pm 0.61$ & $-1.21^{+0.53}_{-0.65}$ & $0.19\pm 0.10$ \\

    \makecell[l]{DESI+CMB (PR3)-nl\\\quad+DESY3+DESSNY5}  & $0.3073\pm 0.0049$ & $0.810\pm 0.023$ & $67.93\pm 0.37$ & $0.04\pm 0.23$ & $0.028\pm 0.046$  \\

    DESI+CMB ($\texttt{LoLLiPoP}$-$\texttt{HiLLiPoP}$)-nl+DESY3    
  & $0.3027\pm 0.0048$ & $0.808\pm 0.023$ & $68.16\pm 0.37$ & $0.04\pm 0.23$ & $0.024\pm 0.046$ \\

    \makecell[l]{DESI+CMB ($\texttt{LoLLiPoP}$-$\texttt{HiLLiPoP}$)-nl\\\quad+DESY3+DESSNY5} 
    & $0.3068\pm 0.0047$ & $0.811\pm 0.022$ & $67.86\pm 0.35$ & $0.05\pm 0.22$ & $0.008\pm 0.045$ \\
    \midrule       
      {{\bf Flat} $\boldsymbol{\mu_0\Sigma_0 w_0w_a}$}& {$w_0$} &  {$w_a$} & $H_0  [${\rm km/s/Mpc}$] $ & {$\mu_0$} &{$\Sigma_0$}\\[-0.1cm]
    
  \midrule
       
    \makecell[l]{DESI+CMB ($\texttt{LoLLiPoP}$-$\texttt{HiLLiPoP}$)-nl\\\quad+DESY3+DESSNY5}  

& $-0.784\pm 0.061$ & $-0.82^{+0.28}_{-0.24}$ & $67.33\pm 0.62$ & $-0.24^{+0.32}_{-0.28}$ & $0.006\pm 0.043$ \\
    
   
    \bottomrule
    \end{tabular}
}
\caption{ 
     {Constraints on modified-gravity parameters $\mu_0$ and $\Sigma_0$ from DESI (FS+BAO) data alone, and in combination with external datasets. We show results in the flat \lcdm\ background expansion model and in the $(w_0, w_a)$ dark energy parameterization for the background. Constraints are quoted for the marginalized means and 68\% credible intervals in each case. In this and other tables, the shorthand notation ``CMB-l" is used to denote the addition of temperature and polarization data from \Planck\ and CMB lensing data from the combination of \Planck\ and ACT, while ``CMB-nl" means CMB lensing is not used. 
    }} 
    \vspace{0.1em}
    \label{tab:functional_forms}
    
\end{table}

\subsubsection{Results for redshift-dependent MG functions in \lcdm\ and \wowacdm\ backgrounds 
}
\label{sec:functional_a_dependence}

\begin{figure*}
\begin{tabular}{c c}
{\includegraphics[width=7.3cm]{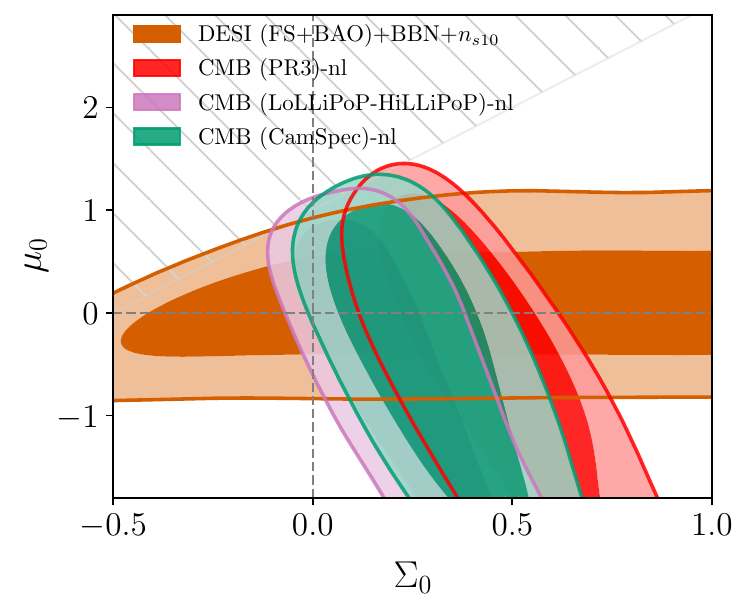}} &
{\includegraphics[width=7.3cm]{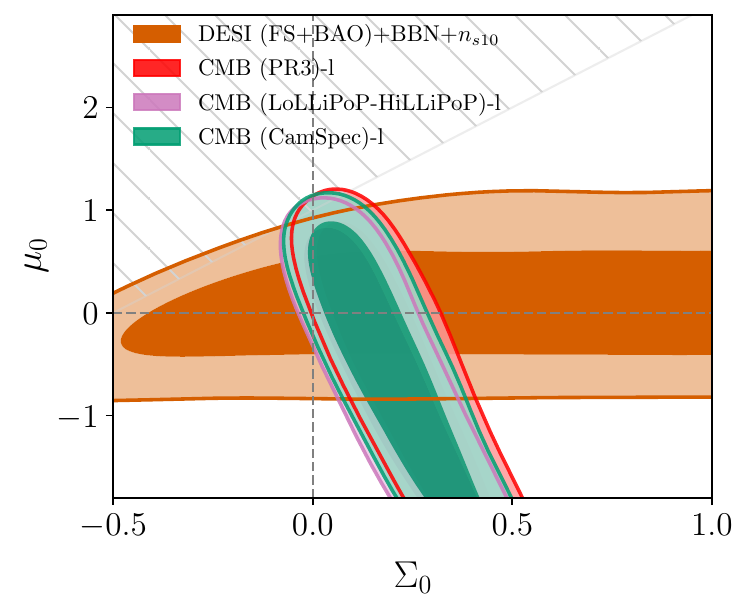}} \\
{\includegraphics[width=7.3cm]{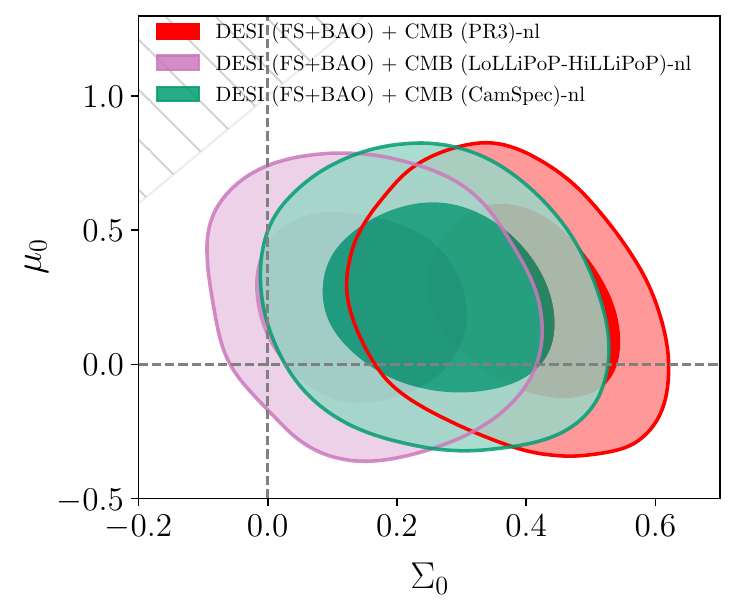}} &
{\includegraphics[width=7.3cm]{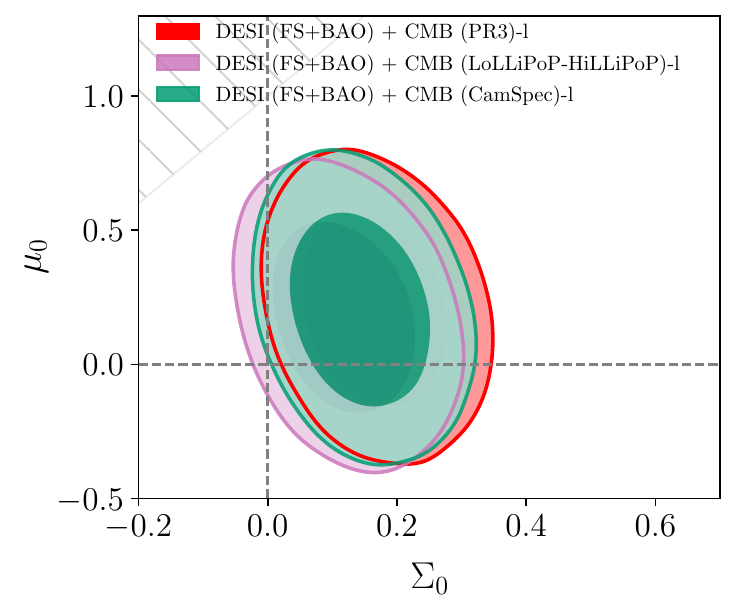}}
\end{tabular}
\caption{
 {MG parameterization $\mu$--$\Sigma$ with time-dependence only. 68\% and 95\% credible-interval contours in a $\Lambda$CDM background cosmology plus scalar perturbations to GR. 
{Top-Left:} DESI in the horizontal band and CMB no-lensing for the 3 different likelihoods. 
{Top-Right:} Similar constraints but adding CMB Lensing data. 
{Bottom-Left:} DESI combined with CMB no-lensing for the three likelihoods. 
{Bottom-Right:} Similar constraints as on bottom left, but adding CMB Lensing data.  See section \cref{sec:functional_a_dependence} for discussion. 
We note that the shaded area on the top left of figures shows the hard prior $\mu_0 < 2 \Sigma_0 + 1$ that is added due to a numerical computational limitation of MG software codes based on \texttt{CAMB} CMB code. As we explain in 
\cref{sec:functional}, this prior does not affect our main results from combinations of datasets.}
}
\label{Fig:functional_mu_Sigma_a_lcdm}
\end{figure*}

\begin{figure}[t]
\centering
\begin{subfigure}
    \centering
    \includegraphics[width=\textwidth]{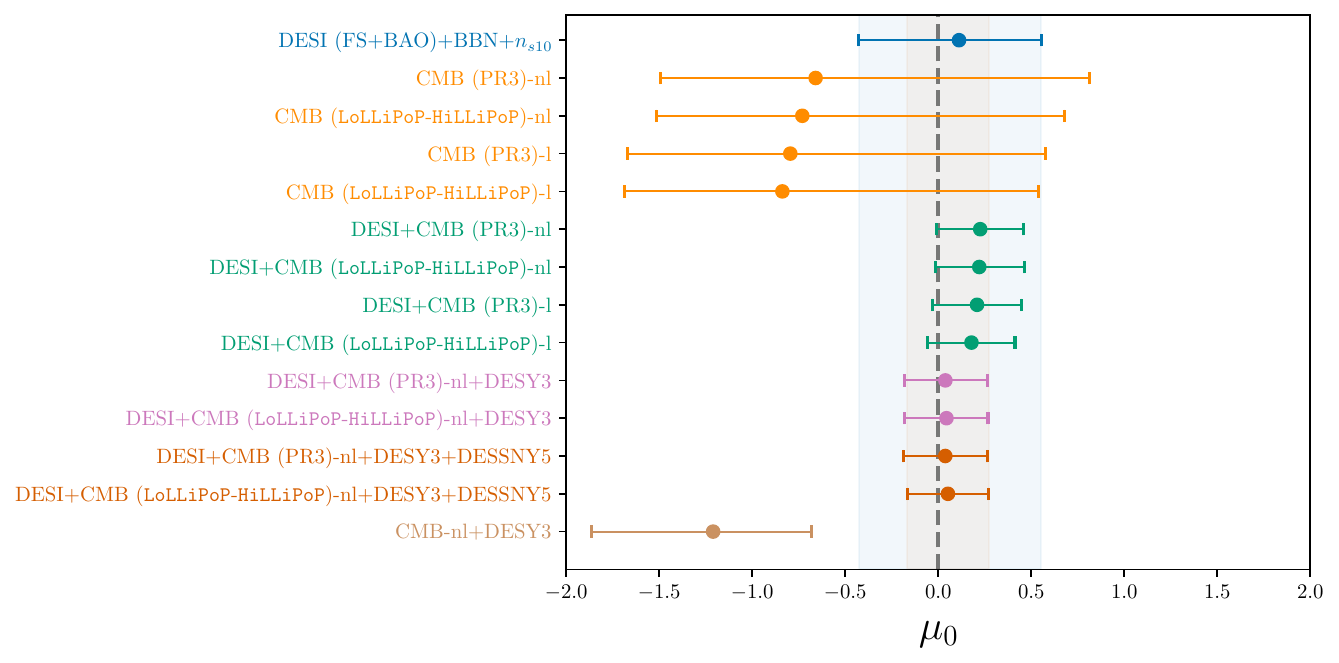}
\end{subfigure}
\vspace{0.5cm}
\begin{subfigure}
    \centering
    \includegraphics[width=\textwidth]{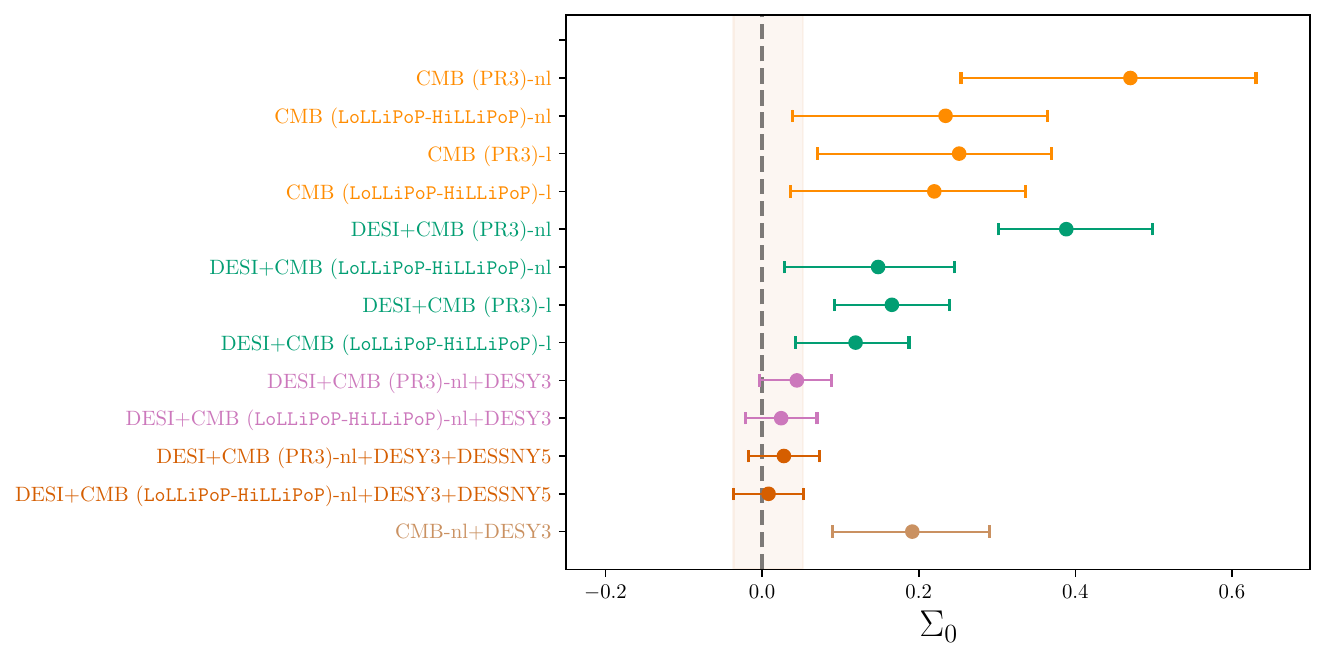}
\end{subfigure}
\caption{
MG parameterization of $\mu$--$\Sigma$ with time-dependence only. Marginalized means and 68\% credible intervals on $\mu_0$ and $\Sigma_0$ in a $\Lambda$CDM background cosmology plus scalar perturbations to GR. Note that DESI alone does not constrain $\Sigma_0$. 
}
\label{fig:wiskerplots_muSigma}
\end{figure}

\begin{figure*}
\centering
\begin{tabular}{c c}
{\includegraphics[width=7.0cm]{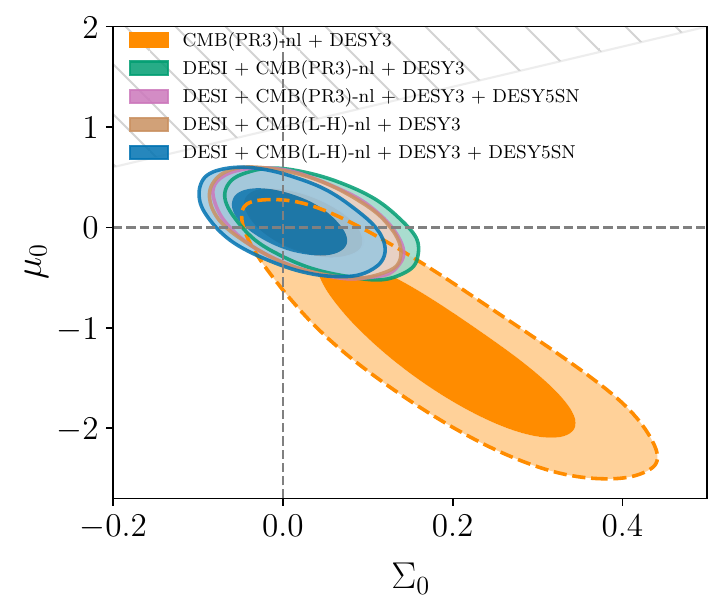}} 
{\includegraphics[width=7.0cm]{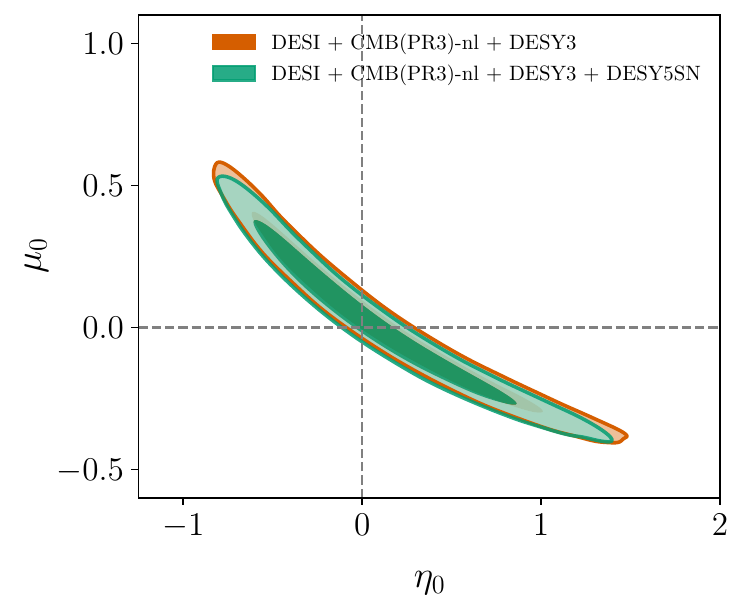}} 
\end{tabular}
\caption{
 Left: MG parameterization of $\mu$--$\Sigma$ with time-dependence only. 68\% and 95\% credible intervals in a $\Lambda \rm CDM$ background cosmology plus scalar perturbations to GR for the datasets indicated. 
Right: MG Functional parameterization for  $\mu$--$\eta$ with time-dependence only. 68\% and 95\% credible intervals in a $\Lambda \rm CDM$ background cosmology plus scalar perturbations to GR for the datasets indicated. See section \cref{sec:functional_a_dependence} for discussion for both panels. 
}
\label{Fig:functional_mu_Sigma_eta_a_lcdm}
\end{figure*}

Our results for the  $\mu$--$\Sigma$ parameterization with time-dependence only (i.e. fixing $c_1=1$, $c_2=1$ and/or $\lambda=0$ into \cref{eq:muEvolution} and \cref{eq:SigmaEvolution}) in a \lcdm\ background are presented in the four panels of \cref{Fig:functional_mu_Sigma_a_lcdm}, \cref{fig:wiskerplots_muSigma}, the left panel of 
\cref{Fig:functional_mu_Sigma_eta_a_lcdm} and summarized for the various dataset combinations in \cref{tab:functional_forms}.     

The top-left panel of \cref{Fig:functional_mu_Sigma_a_lcdm} shows the constraint from DESI (FS+BAO)+BBN+$n_{s10}$ on the MG parameter $\mu_0 = 0.11^{+0.44}_{-0.54}$ . DESI full shape power spectra are able to constrain this parameter via its embedded growth of large scale structure function associated with the clustering of massive particles. The DESI constraint and its credible-interval contours are centered around the value of zero predicted by general relativity and is fully consistent with it. However, the 68\% credible intervals still allow for significant departures from general relativity.  Moreover, the same figure confirms the expectation that DESI does not constrain the parameter $\Sigma_0$ that is associated with the dynamics of massless particles and, for example lensing, as shown by the horizontal ``band''.  

The other three credible-interval areas that appear nearly vertically in \cref{Fig:functional_mu_Sigma_a_lcdm} show the constraints from CMB with no-lensing from three Planck likelihoods, namely from PR3 \cite{Planck-2018-likelihoods}, Camspec \cite{Efstathiou:2021,Rosenberg:2022} and \texttt{LoLLiPoP}-\texttt{HiLLiPoP} \cite{Tristram:2021,Tristram:2023}. As explained at the end of \cref{sec:functional}, these CMB constraints are hitting the necessary computational prior $\mu_0 < 2 \Sigma_0 + 1$, but are nearly orthogonal to the DESI ``band'' and very complementary to it. 

We see in the same top-left panel that the constraint from Planck PR3 on the parameter $\Sigma_0$ is in tension with the zero value of general relativity. Moreover, the bottom-left panel shows that when DESI is added to Planck PR3, this tension reaches well-above the 3-$\Sigma$ level. It is worth noting that although DESI is not driving this tension, its addition to \emph{Planck} breaks parameter degeneracies and exacerbates the tension. 
This discordance of $\Sigma_0$ with GR when using CMB data was noted in Planck 2015 analysis \cite{Planck:2015bue} and confirmed in Planck 2018 \cite{Planck-2018-cosmology}. It was attributed to the CMB lensing anomaly or the $A_\text{lens}$  problem \cite{Calabrese08,Renzi18,Mokeddem23}. This anomaly and the corresponding $A_\text{lens}$ nonphysical parameter are associated with a systematic effect in Planck data that manifests as an excess lensing effect that smooths the peaks and troughs of the power spectra \cite{Planck-2018-cosmology}. This non physical parameter can enter as a multiplicative scaling factor in the lensing of the CMB power spectra. By construction, it should be equal to unity for the \lcdm\ model. 
However, it was found repetitively, first in WMAP data  \cite{Calabrese08} and then by Planck collaboration and PR3 data, e.g. \cite{Planck-2013-cosmology, Planck:2015bue,Planck-2018-cosmology} that the fit of a \lcdm\ model, plus the $A_\text{lens}$ parameter allowed to vary, gives a better fit to the data with an $A_\text{lens}$ value that departs from the unity value expected in the \lcdm\ model. 
The parameter $A_\text{lens}$ is known to be degenerate with other physical parameters and affects their accuracy, including the sum of Neutrino masses, spatial curvature and modified gravity parameters, see e.g. \cite{Planck:2015bue,Planck-2018-cosmology} for further general discussion. 

Relevant to our analysis, the nonphysical $A_\text{lens}$ parameter is degenerate with the parameter $\Sigma_0$ and, if not mitigated, provides a value of $\Sigma_0$ that departs from the GR zero value as shown for PR3 in the let-top panel of \cref{Fig:functional_mu_Sigma_a_lcdm}. 
Recently, the $A_\text{lens}$ anomaly was partly fixed with the Camspec Planck analysis \cite{Efstathiou:2021,Rosenberg:2022} and completely resolved  with the \texttt{LoLLiPoP} and \texttt{HiLLiPoP} likelihoods \cite{Tristram:2023}.  Interestingly, we find in our analysis that the departure of the $\Sigma_0$ parameter from the zero GR-value gets alleviated when using Planck PR4 Camspec, as the GR value is within the 95\% credible-interval contours, and then gets even closer to the GR value when using the \texttt{LoLLiPoP}-\texttt{HiLLiPoP} likelihoods.  This indicates that the found departure of $\Sigma_0$ from the GR value is rather related to the CMB lensing anomaly in the Planck PR3 data than to any new physics.\footnote{We note that while our papers were in DESI internal collaboration wide review, the paper \cite{Specogna:2024euz} appeared on the arXiv showing a similar finding about the $\Sigma_0$ tension being resolved when using \texttt{LoLLiPoP} and high-$\ell$ \texttt{HiLLiPoP}, and using a different modified gravity software (\texttt{MGCAMB} than the one (\texttt{ISiTGR}) the we used in our analysis. It is also worth mentioning as well that the very recent paper \cite{SPT-3G:2024atg} reports some more complex findings concerning the Planck lensing anomaly.}

As in previous works, we also find that this discordance of $\Sigma_0$ with GR becomes insignificant when the reconstructed CMB lensing data is added to the CMB power spectra as we show in the top-right panel of \cref{Fig:functional_mu_Sigma_a_lcdm} and \cref{tab:functional_forms} for all three Planck likelihoods, where we have added in the present analysis the specific results for Camspec and \texttt{LoLLiPoP}-\texttt{HiLLiPoP} likelihoods. 

In the next step, we combine DESI (FS+BAO) with CMB constraints with and without CMB lensing. We see in the bottom-left and bottom-right panels of \cref{Fig:functional_mu_Sigma_a_lcdm} as well as \cref{fig:wiskerplots_muSigma} and \cref{tab:functional_forms} that regardless of lensing, combining DESI and CMB breaks degeneracies among parameters and allows to improve the constraints on the parameter $\mu$ by roughly a factor of 5 compared to CMB-only and a factor of 2 compared to DESI-only. Moreover, adding DESI to CMB with lensing improves the constraints on $\Sigma_0$ by a factor of $1.5-2.0$ and adding DESI to CMB without lensing also tightens the constraint on this parameter by at least a factor of two. 

\begin{figure*}[t]
\centering
\includegraphics[width=7.3cm]{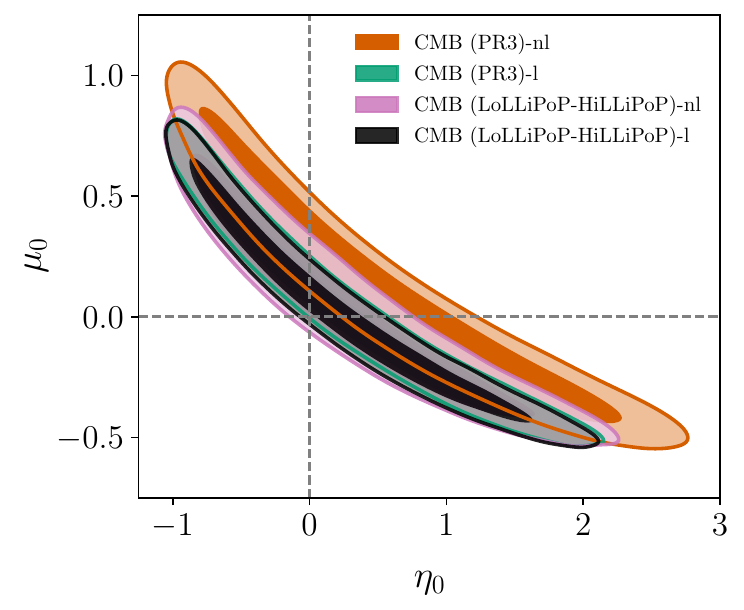} 
\includegraphics[width=7.3cm]{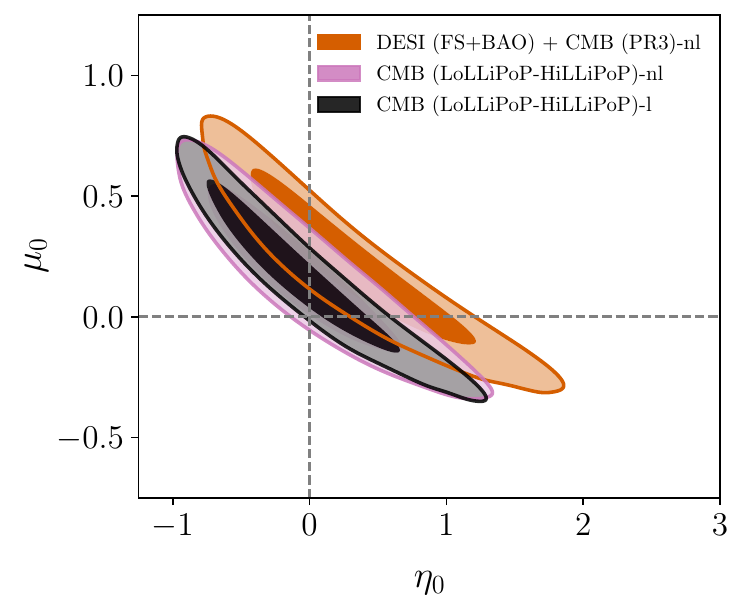} 
\caption{
MG parameterization of $\mu$--$\eta$ with time-dependence only. 68\% and 95\% credible intervals in a $\Lambda$CDM background cosmology plus scalar perturbations to GR. 
Left:  CMB constraints with and without lensing for the two likelihoods PR3 and \texttt{LoLLiPoP}-\texttt{HiLLiPoP} for Planck data. Right: DESI (FS+ BAO) + CMB combinations using the same two likelihoods, respectively. 
}
\label{fig:functional_mu_eta_a_lcdm}
\end{figure*}

As expected, the addition of the DES Y3 3$\times$2-pt data to the combination of DESI and CMB without lensing improves the constraints on the lensing-sensitive MG parameter $\Sigma_0$ by roughly a factor of two, while providing only marginal improvements on the parameter $\mu_0$, see \cref{tab:functional_forms} and left-panel of \cref{Fig:functional_mu_Sigma_eta_a_lcdm}.  We do not add CMB lensing to these combinations because of covariances with DES Y3 3$\times$2-pt data.  

It is worth noting as well that when adding DESI data to the combination CMB-without-lensing and DESY3, we obtain roughly a factor of 2.5 improvement on $\mu_0$ and roughly a factor of 2 improvement on $\Sigma_0$, as shown on \cref{Fig:functional_mu_Sigma_a_lcdm} and \cref{tab:functional_forms}. 

Finally, the addition of type Ia supernova data (e.g.\ the DES-YR5 SN  that  we use here as an example) adds practically no further improvement to the DESI+CMB-nl+DES Y3 combination in a \lcdm\ background cosmology. We provide the combinations with SN~Ia here for comparison, see \cref{tab:functional_forms} and left panel of \cref{Fig:functional_mu_Sigma_eta_a_lcdm}.  SN~Ia data will play a more important role when we adopt the \wowacdm\ cosmology background which we will discuss in \cref{sec:functional_ak_results}.  Moreover, we also use this full combination when we analyze other demanding cases such as binning with multiple parameters or when we include both redshift and scale. Again, even in some of these cases, the gain is very small when we add supernovae but we keep them in the combination to have consistent comparisons between our own different cases but also other previous studies that kept supernovae in their external datasets.

In sum, we find that all our results for the $\mu-\Sigma$ parameterization are consistent with GR for all dataset combinations. The tightest constraints we obtain on both parameters, and free from the $A_\text{lens}$ anomaly mentioned above, come from the combination DESI+CMB (\texttt{LoLLiPoP}+\texttt{HiLLiPoP})-nl+DES Y3 or DESI+CMB (\texttt{LoLLiPoP}+\texttt{HiLLiPoP})-nl+DES Y3+DESY5 SN where the latter provides only minute improvements but we quote them for comparison with other cases:  
\twoonesig[6.5cm]
{\mu_0    &= 0.05\pm 0.22,}
{\Sigma_0 &= 0.008\pm 0.045,} 
{DESI (FS+BAO)+CMB ($\texttt{LoLLiPoP}$-$\texttt{HiLLiPoP}$)+ DESY3+ DESY5 SN . \label{eq:mu_sigma_DESI_CMB_DES3x2}}

The constraints on MG parameters in \cref{eq:mu_sigma_DESI_CMB_DES3x2} are comparable in precision to the ones from \cite{DES:2022ccp} using two decades of BAO+RSD from SDSS + CMB (PR3) + DESY3 ($3\times 2$-pt)+ PantheonPlus SN, but we note that 1-year only of data from DESI can provide comparable constraining power on specifically $\mu_0$ as two decades of BAO+RSD data from SDSS 
 \cite{2020ApJ...901..153D} and the entire BAO from 6dFGS \cite{2009MNRAS.399..683J}.  We also observe that constraints on $\mu_0$ including DESI with or without other datasets are more centered around the GR value than those from SDSS which show a mild shift of slightly above 1$\sigma$ from the GR zero value. This shows the constraining power of DESI and the promise of the four years of data to come from the DESI program.

Next, we now consider constraints in the \lcdm\ background for the  $\mu-\eta$ parameterization with time-only dependence as shown in our  \cref{fig:functional_mu_eta_a_lcdm}, \cref{Fig:functional_mu_Sigma_eta_a_lcdm} and \cref{tab:functional_forms_mu_eta}.  
The results for  $\mu_0-\eta_0$ are very comparable to the ones for the $\mu_0-\Sigma_0$. Specifically, DESI (FS+BAO)+BBN+$n_{s}$  gives $\mu_0 = 0.17^{+0.45}_{-0.56}$ with similar error bars.  Again, like in previous studies, CMB (PR3) no-lensing gives results on $\eta_0$ that are in tension with GR due to the Planck PR3 lensing anomaly indicated above and manifest in the $A_\text{lens}$ parameter. But when using the \texttt{LoLLiPoP}-\texttt{HiLLiPoP} likelihood for Planck, such a tension goes away for $A_\text{lens}$ and we find that the tension for $\eta_0$ also goes away as shown in \cref{fig:functional_mu_eta_a_lcdm} and \cref{tab:functional_forms_mu_eta}. When adding CMB lensing, the contours are shifted to the GR values in both cases, as in the $\mu_0-\Sigma_0$ case.  As further above, adding CMB with or without lensing to DESI improves the constraints on $\mu_0$ by roughly a factor of 2 and, likewise, adding DESI to CMB improves constraints on $\eta$ by roughly a factor of two as well. Finally, using the combination DESI+CMB (PR3)-nl+DESY3+DESY5SN gives us the best constraints as

\twoonesig[6.5cm]
{\mu_0    &=  0.02^{+0.19}_{-0.24},}
{\eta_0 &= 0.09^{+0.36}_{-0.60}.} 
{DESI (FS+BAO)+CMB (PR3)+ DESY3+ DESY5 SN. \label{eq:mu_eta_DESI_CMB_DES3x2_DESY5SN}}

We finish this section with results for the $\mu$--$\Sigma$ parameterization but in a \wowacdm\ cosmological background.  We assume a dynamical dark energy model with an equation 
of state that takes the commonly-used form \cite{Chevallier:2001,Linder2003}:  

\begin{equation}
w_{\text{DE}}(a) = w_0 + w_a(1-a).
\end{equation}

\begin{figure*}
\centering
\begin{tabular}{c c}
{\includegraphics[width=7.4cm]{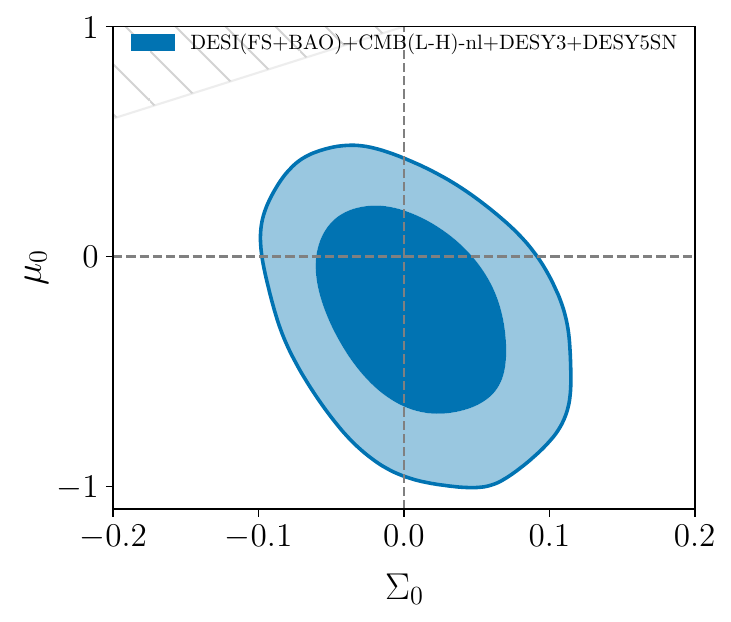}} 
{\includegraphics[width=7.6cm]{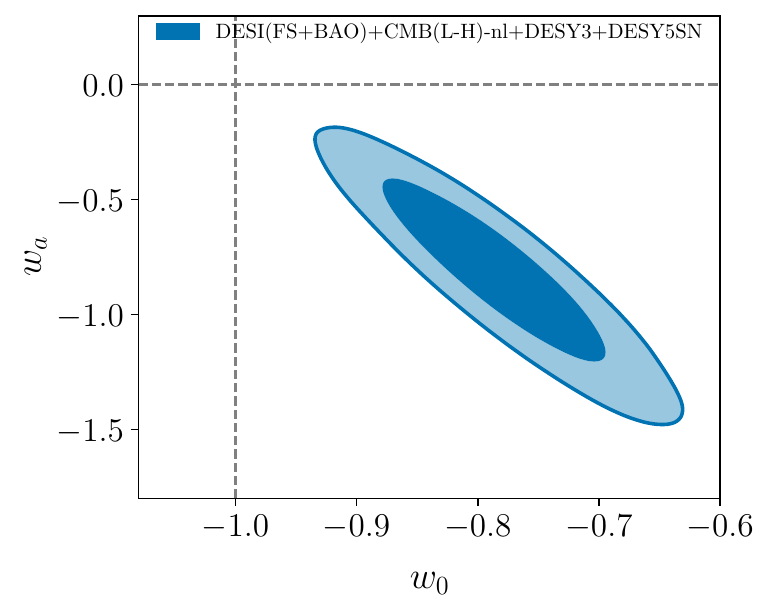}} 
\end{tabular}
\caption{
 Left: MG Functional parameterization for  $\mu$--$\Sigma$ with time-dependence only. 68\% and 95\% credible intervals in a \wowacdm\ background cosmology plus scalar perturbations to GR for the dataset indicated. 
Right: 68\% and 95\% credible intervals for the dark energy equation of state parameters $(w_0,w_a)$. See section \cref{sec:functional_a_dependence} for discussion for both panels. 
}
\label{fig:functional_mu_Sigma_w0wa}
\end{figure*}

Our results for $\mu_0$ and $\Sigma_0$ MG parameters, as well as the equations of state parameters ($w_0,w_a$), are given in the last two rows of \cref{tab:functional_forms} and \cref{fig:functional_mu_Sigma_w0wa} for the our  constraining combination of our datasets, i.e.\,  DESI+CMB (\texttt{LoLLiPoP}-\texttt{HiLLiPoP})-nl +DESY3+DESY5SN.  

The first observation is that, unlike the case of the \lcdm\ background, here in the \wowacdm\ background the addition of a supernovae dataset to the DESI+CMB-nl+DESY3 combination does have a significant effect in providing substantially tighter constraints on both the dark energy equation of state parameters as well as the MG parameters. For such a full combination of datasets, we obtain for the following constraints for w(a): 

\twoonesig[5.5cm]
{w_0 &= -0.784\pm 0.061,}
{w_a &=  -0.82^{+0.28}_{-0.24},}
{DESI+CMB ($\texttt{LoLLiPoP}$-$\texttt{HiLLiPoP}$)-nl + DESY3 + DESY5 SN. \label{eq:w0_wa_mu0_Sigma0}}
and 
\twoonesig[5.5cm]
{\mu_0 &   = -0.24^{+0.32}_{-0.28},}
{\Sigma_0 &=0.006\pm 0.043,}
{DESI+CMB ($\texttt{LoLLiPoP}$-$\texttt{HiLLiPoP}$)-nl + DESY3 + DESY5 SN. \label{eq:mu0_Sigma0_w0_wa}}
for MG parameters. 

We note here the interesting result that despite adding two MG parameters to the model, the constraints on  ($w_0,w_a$) still show a well-above 
 3-$\sigma$ preference for a dynamical dark energy with MG parameter constraints being consistent with GR values.

\begin{table}
\centering
\resizebox{\columnwidth}{!}{%
    \small 
\setcellgapes{3pt}\makegapedcells  
\renewcommand{\arraystretch}{2.0} 
    \begin{tabular}{lccccc}
    \toprule

 {{\bf Flat} $\boldsymbol{\mu_0\eta_0}$ {\bf CDM} }& {$\Om$} & {$\sigma_8$}  & $H_0  [${\rm km/s/Mpc}$] $ & {$\mu_0$} &{$\eta_0$}\\[-0.1cm]
     \midrule    
     
     \makecell[l]{DESI (FS+BAO)\\\quad+BBN+$n_{s10}$}  
     & $0.2959\pm 0.0097$ & $<0.853$ & $68.52\pm 0.75$ & $0.17^{+0.45}_{-0.56}$ \\

    CMB (PR3)-nl           
    & $0.3041\pm 0.0092$ & $0.815^{+0.030}_{-0.054}$ & $68.21\pm 0.70$ & $0.12^{+0.26}_{-0.54}$ & $0.62^{+0.69}_{-1.3}$ \\
 
   CMB (\texttt{LoLLiPoP}-\texttt{HiLLiPoP})-nl   
   & $0.3059\pm 0.0078$ & $0.810^{+0.029}_{-0.042}$ & $67.94\pm 0.58$ & $0.06^{+0.25}_{-0.43}$ & $0.29^{+0.53}_{-1.0}$ \\

   CMB (PR3)-l            & $0.3094\pm 0.0083$ & $0.810^{+0.030}_{-0.044}$ & $67.79\pm 0.62$ & $0.04^{+0.25}_{-0.45}$ & $0.31^{+0.58}_{-1.1}$ \\
 
    CMB ($\texttt{LoLLiPoP}$-$\texttt{HiLLiPoP}$)-l    & $0.3093\pm 0.0074$ & $0.812^{+0.028}_{-0.044}$ & $67.69\pm 0.55$ & $0.04^{+0.24}_{-0.45}$ & $0.28^{+0.55}_{-1.1}$ \\

    DESI+CMB (PR3)-nl           & $0.2987\pm 0.0055$ & $0.822\pm 0.024$ & $68.62\pm 0.43$ & $0.22\pm 0.24$ & $0.33^{+0.44}_{-0.62}$ \\

    DESI+CMB ($\texttt{LoLLiPoP}$-$\texttt{HiLLiPoP}$)-nl      
  & $0.3006\pm 0.0051$ & $0.820\pm 0.023$ & $68.34\pm 0.39$ & $0.17\pm 0.23$ & $-0.03^{+0.35}_{-0.57}$ \\

   DESI+CMB ($\texttt{LoLLiPoP}$-$\texttt{HiLLiPoP}$)-l  & $0.3012\pm 0.0049$ & $0.822\pm 0.023$ & $68.29\pm 0.38$ & $0.18\pm 0.23$ & $-0.06^{+0.34}_{-0.57}$ \\

    DESI+CMB (PR3)-nl+DESY3

    & $0.3021\pm 0.0051$ & $0.806^{+0.022}_{-0.026}$ & $68.32\pm 0.39$ & $0.03^{+0.20}_{-0.27}$ & $0.13^{+0.40}_{-0.66}$ \\
    
    \makecell[l]{DESI+CMB (PR3)-nl+DESY3\\\quad+DESSNY5}  

& $0.3069\pm 0.0049$ & $0.808\pm 0.021$ & $67.96\pm 0.37$ & $0.02^{+0.19}_{-0.24}$ & $0.09^{+0.36}_{-0.60}$ \\

    \bottomrule
    \end{tabular}
}
\caption{ 
     {Constraints on modified-gravity parameters $\mu_0$ and $\eta_0$ from DESI (FS+BAO) data alone, and in combination with external datasets. We show results in the flat \lcdm\ background expansion model. Constraints are quoted for the marginalized means and 68\% credible intervals in each case.}
    }
    \vspace{0.1em}
    \label{tab:functional_forms_mu_eta}
    
\end{table}

\subsubsection{Results for redshift and scale dependent MG functions in \lcdm\ and \wowacdm\ backgrounds 
}
\label{sec:functional_ak_results}


Our results for the  $\mu$--$\Sigma$ parameterization with both time and scale dependence in a \lcdm\ background  are presented in 
\cref{tab:functional_forms_ak}.  We use the time and scale dependencies as expressed in the functional form in  \cref{eq:muEvolution} and \cref{eq:SigmaEvolution}.  This adds the three parameters we discussed in section \cref{sec:functional} noted as $\lambda$, $c_1$ and $c_2$. 

 Clearly, this model requires more constraining power for MG parameters from the data and we use here the combination DESI + CMB (PR3)\footnote{We use here PR3 instead of \texttt{LoLLiPoP}-\texttt{HiLLiPoP} as our MCMC chains were taking a much larger time to converge in this case}-nl + DES Y3 + DESY5 SN.  While we find that the constraints on $\mu_0$ and $\Sigma_0$ are easily obtained with comparable precision to those form redshift-only dependence, the scale dependence parameters are harder to constrain. Typically, the $\lambda$ parameter is difficult to constrain, and it only sets the scale below which the MG parameters start to be sensitive to the scale-dependent effects. We investigate two choices for $\lambda$. We first set $\lambda=10$, which allows the function $\mu(a=1,k)$ to begin evolving asymptotically from $1+\mu_0$ to $1+\mu_0 c_1$ at scales $k<0.01$Mpc$/h$. This somehow matches the transition scale that we use in the binning methods. Alternatively, we also set $\lambda=100$, where the scale dependence on $\mu$ is induced starting from $k<0.1$Mpc$h^{-1}$, as an extreme case. The results of these constraints can be found in \cref{tab:functional_forms_ak}.
 
 It is found that in both cases, the constraints on $\mu_0$ and $\Sigma_0$ do not deteriorate substantially from the time-only dependence cases and are consistent with GR. The 68\% error bars on the parameters $c_1$ and $c_2$ while still too wide are also consistent with the GR value of 1. As we will see further below, the binned parameterization in both redshift and scale is found to provide much better constraints on all MG parameters than the functional form here. It remains an open question whether other scale functional parameterization are able to provide better constraints using current data or not and we leave such open question (and beyond the scope of our paper) to be explored in other future analyses but refer the readership to our results in the binning method in  \cref{sec:binning-method}.     

\begin{table}
\centering
\resizebox{\columnwidth}{!}{%
    \small 
\setcellgapes{3pt}\makegapedcells  
\renewcommand{\arraystretch}{3.0} 
    \begin{tabular}{lccccc}
    \toprule
 {{\bf Flat $\Lambda$CDM background} }& {$\mu_0$} & {$\Sigma_0$}  & {$\lambda$ fixed-value}  & {$c_1$} &{$c_2$}\\[-0.1cm]
     \midrule    
    \makecell[l]{DESI+CMB (PR3)-nl\\\quad+DESY3+DESY5SN}  
      & $0.03^{+0.13}_{-0.14}$ & $0.027\pm 0.043$ &   10                  & $0.9^{+1.4}_{-1.6}$           & ---                  \\
    \makecell[l]{DESI+CMB (PR3)-nl\\\quad+DESY3+DESY5SN}  
 
    & $0.12^{+0.14}_{-0.22}$ & $0.015\pm 0.045$ &  100 &$0.7^{+1.1}_{-2.3}$    & $0.52^{+0.94}_{-1.4}$ \\   
       
    \bottomrule
    \end{tabular}
}
\caption{ 
     {Constraints on modified-gravity parameters $\mu_0$ and $\Sigma_0$ with time (redshift) and scale evolution. We show results in the flat \lcdm\ background expansion. Constraints are quoted for the marginalised means and 68\% credible intervals in each case, see discussion in \cref{sec:functional_ak_results}}.
    } 
    \vspace{0.1em}
    \label{tab:functional_forms_ak}
\end{table}

\subsection{Binned MG parameterizations}
 \label{sec:binning-method}

\begin{table}[t!]
\begin{center}
 \begin{tabular} {| c | c | c | c |}
\hline
  & \multicolumn{3}{c|}{Redshift bins} \\ \hline
Scale bins & $0\leq z< 1$ & $1\leq z< 2$ & $z\geq 2$ \\ \hline
$0\leq k < 0.01$ & $\mu_1$, $\Sigma_1$ & $\mu_3$, $\Sigma_3$ & GR is assumed\\ \hline
$k\geq 0.01$ & $\mu_2$, $\Sigma_2$ & $\mu_4$, $\Sigma_4$ & GR is assumed \\
\hline
\end{tabular}
\end{center}
\caption{
{Redshift and scale (i.e.\ wavenumber) bins used in this work, with the corresponding MG parameters. In the context of cosmic acceleration, deviations from GR are tested in the range $0 \leq z < 2$ (more discussion in the text). }}
\label{Table:Bins}
\end{table}

As mentioned earlier, we extend our analysis to binning methods that do not assume a specific analytical functional form for the MG parameters. This will complement the functional methods and also validate them. We first consider an analysis that employs binning in redshift (time) only, and then an analysis that includes binning in both redshift and scale. Using bins in redshift-only requires less constraining power and has been done  before in, for example \cite{Mueller:2016kpu}, 
where results were found consistent with GR and with no significant improvement in the fit over the $\Lambda$CDM model, but, again, with error bars that leave room for a lot of improvement.  On the other hand, other works that used bins in both redshift and scale observed some mild deviations from GR using previous survey datasets, e.g. \cite{Zhao:2010dz,Johnson:2015aaa,Daniel2010MG2,Joudaki:2016kym,Joudaki2018,Garcia-Quintero:2020bac}, and that is worth investigating here. 

Indeed, with the additional constraining power available to us here from DESI and DESY3, we will explore such a dual binning in redshift and scale.  For that, we set four bins consisting of two bins in redshift and two bins in scale that are implemented in ISiTGR. We consider the redshift bins to be fit in the ranges $0<z\leq z_{\text{div}}$ and $z_{\text{div}}<z\leq z_{\text{TGR}}$ where $z_{\text{div}}=1$ and $z_{\text{TGR}}=2$. $z_{\text{div}}$ is the redshift that divides the two bins and $z_{\text{TGR}}$ is the redshift above which we assume that GR is the correct theory. This has been designed with the idea of cosmic acceleration in mind where we seek for any modification to GR at relatively late times and assume GR at earlier times, if  $z>z_{\text{TGR}}=2$. For the binning in scale, we use a bin with $k\leq k_\text{c}$ and another one with $k>k_\text{c}$, where $k_\text{c}=0.01$ Mpc$^{-1}$ is the scale dividing them. Such a dividing scale roughly represents the scale at which the non-CMB probes start to play a role for $k>k_\text{c}$ as well as the scale of matter-radiation equality horizon specifying the matter power spectrum turnover. To encapsulate this, we note each MG parameter by $X(z,k)$ and write 
\begin{equation}
\begin{split}
X(z,k) & = \frac{1+X_{z_1}(k)}{2}+\frac{X_{z_2}(k)-X_{z_1}(k)}{2}\tanh\left(\frac{z-z_{\text{div}}}{z_{\text{tw}}}\right) \\
& + \frac{1-X_{z_2}(k)}{2}\tanh\left(\frac{z-z_{\text{TGR}}}{z_{\text{tw}}}\right).
\end{split}
\label{eq:binning_ak1}
\end{equation}
We note that we have constructed the binning parameterizations here to be  centered around the value of 1 which will be considered as the GR expected value. This is summarized in \cref{Table:Bins}.  By design, (\cref{eq:binning_ak1}) gives a smooth and continuous transition of the MG parameters between the redshift bins. The transition width is controlled by the parameter $z_{\text{tw}}$ that sets how rapidly the transition from one bin to another happens in time. Obviously, a very small value of $z_{\text{tw}}$ could lead to numerical errors and a rejection of such parameter values so we have chosen a moderate value for such a transition with $z_{\text{tw}}=0.05$.  

\begin{figure*}
\begin{tabular}{c}
{\includegraphics[width=15.0cm]{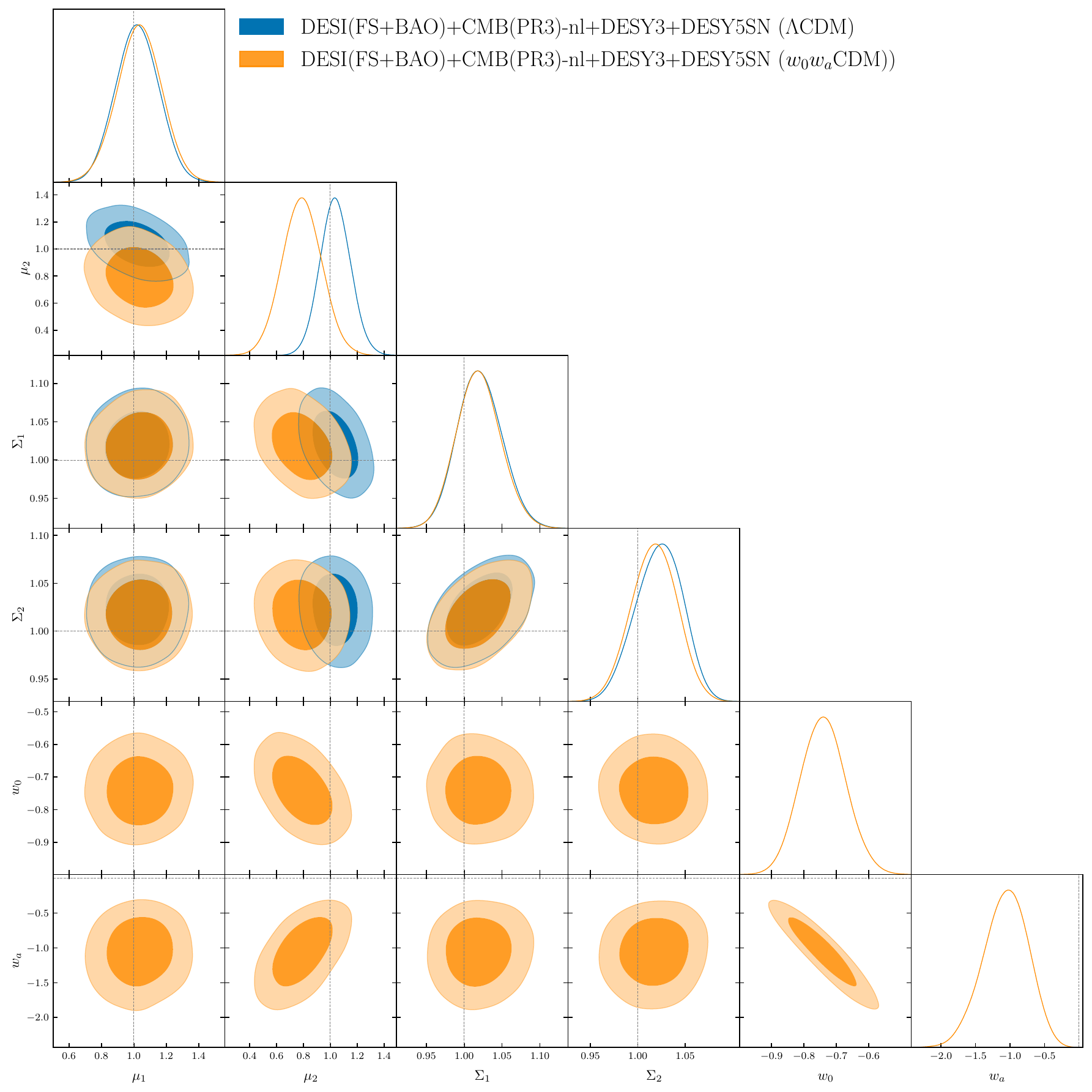}} 
\end{tabular}
\caption{
 {Constraints on redshift-binned form for the $\mu-\Sigma$ MG functions in \lcdm\ and \wowacdm\ cosmological backgrounds, respectively, plus scalar perturbations to GR. The binning in redshift results are indicated in the first part of \cref{Table:Bins}. The contours represent the 68\% and 95\% credible intervals for the combination DESI(FS+BAO)+CMB(PR3)-nl+DESY3+DESY5SN. The results are listed in the first part of \cref{tab:binning_ak}.
 }
}
\label{fig:constraints_binning_a}
\end{figure*}

\begin{figure*}
\begin{tabular}{c}
{\includegraphics[width=15.5cm]{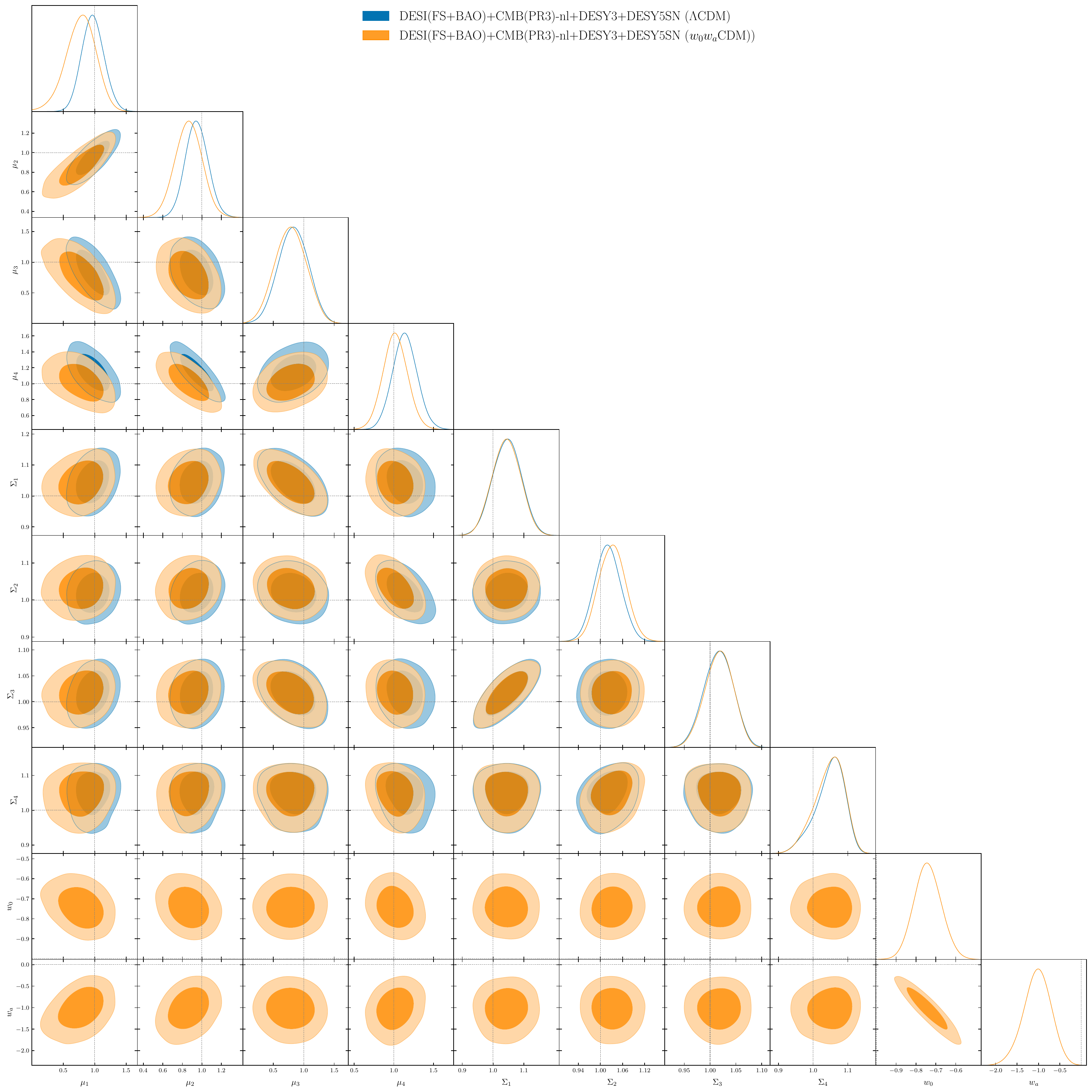}} 
\end{tabular}
\caption{
 {Constraints for redshift and scale binned form for the $\mu-\Sigma$ MG functions in \lcdm\ and \wowacdm\ cosmological backgrounds, respectively, plus scalar perturbations to GR. The binning in time and scale are as indicated in \cref{Table:Bins}. The contours represent the 68\% and 95\% credible intervals for the DESI(FS+BAO)+CMB(PR3)-nl+DESY3+DESY5SN combination. The results are listed in the second part of \cref{tab:binning_ak}.
}
}
\label{fig:constraints_binning_a_k}
\end{figure*}

Next, we set the functions $X_{z_1}(k)$ and $X_{z_2}(k)$ for scale binning using a hyperbolic tangent function for scale (as is done for the redshift bins) with a transition parameter $k_{\text{tw}}=k_{\text{c}}/10$. These parameters are thus given by
\begin{equation}
X_{z_1}(k) = \frac{X_2+X_1}{2}+\frac{X_2-X_1}{2}\tanh\left(\frac{k-k_c}{k_{\rm tw}}\right)
\label{eq:binning_ak2}
\end{equation}
and
\begin{equation}
X_{z_2}(k) = \frac{X_4+X_3}{2}+\frac{X_4-X_3}{2}\tanh\left(\frac{k-k_c}{k_{\rm tw}}\right).
\label{eq:binning_ak3}
\end{equation}
This formulation of the scale binning is labeled as traditional binning method in ISiTGR documentation \cite{Dossett:2011tn,Garcia-Quintero:2019xal} and we employ it here. Notice that with this configuration, the DESI data directly impacts the $X_1$ and $X_3$ parameters for scales $k>k_c$, while the parameters $X_2$ and $X_4$ are constrained by CMB. This by construction implies that we are using the whole DESI data to probe the MG parameters at scales below the matter power spectrum turnover, and the MG effects at larger scales are sensitive to the CMB data. Finally, for our binning in redshift only, we assume that $X_{z_1}(k)=X_1$ and $X_{z_2}(k)=X_2$ are constants and do not provide them with any scale dependence. This is equivalent to having two redshift bins, in the ranges $0<z<1$ and $1<z<2$, parameterized by $X_1$ and $X_2$, respectively.  In this case, all the MG parameters are constrained by DESI, as we are probing redshift ranges covered by the DESI tracers. 

\subsubsection{Results for binning in redshift  in \lcdm\ and \wowacdm\ backgrounds }

Our results for binning in redshift are given in \cref{fig:constraints_binning_a} and \cref{tab:binning_ak}.  We expect to obtain analogous results for the $\mu(a)$-$\Sigma(a)$  or  the $\mu(a)$-$\eta(a)$ space, so for a consistent presentation over the sub-sections, we show results for the former pair.  We derive constraints on MG parameters in both the flat $\Lambda$CDM expansion background and the flat $w_0w_a$CDM expansion background.

We fix the scale dependence in \cref{eq:binning_ak1} -- \cref{eq:binning_ak3}, which effectively gives two redshift bins with parameters $\mu_1$ and $\Sigma_1$ defined in the first bin with $0\leq z< 1$  and parameters $\mu_2$ and $\Sigma_2$ defined in the second bin with $1\leq z< 2$, while for $z\geq 2$ the parameters are set to take the GR value of 1 in the binning form convention we use. 

The combinations DESI+CMB(PR3)+DESY3+DESY5SN provides our best constraints on the 4 parameters as follows:

\fouronesig[5.5cm]
{\mu_1&    =1.02\pm 0.13,\\[-0.45cm]}
{\mu_2 &   = 1.04\pm 0.11,\\[-0.45cm]}
{\Sigma_1 &= 1.021\pm 0.029,\\[-0.45cm]}
{\Sigma_2 &= 1.022^{+0.027}_{-0.023} }
{DESI+CMB(PR3)+DESY3+DESY5SN. \label{eq:muSigmaDESI_LH_DESY3_DESY5SN}}

All our constraints on the MG parameters are consistent with GR with clear improvement compared to the redshift-binning results with the four parameters obtained in a previous study \cite{Mueller:2016kpu} using other datasets from SDSS, CMB, and CMB lensing data instead of 3$\times$2-pt weak-lensing along with clustering data.  Our constraints for $\mu_1$ and $\mu_2$ are comparable to the forecast made for DESI plus CMB and CMB Lensing in \cite{Toda:2024fgv} where similar overall binning ranges in redshift were employed, although their higher redshift bin goes up to redshift 3 while we assumed GR for redshift above 2. 

Finally, its worth noting that in the \wowacdm, while the MG parameters are found all consistent with GR, we still find values of the equations of state parameters that show preference for a dynamical dark energy, see \cref{tab:binning_ak}.

Finally, it is interesting to find that the results for our binning scheme give competitive constraints on MG parameters compared to functional forms. The results are not only independent from the functional form results but also consistent with them within the constraining power of current data which could address some concerns expressed about some limitations in functional forms. It would be good to see if this holds in future studies using a larger number of bins. 

In sum, our results for binning in redshift show that all the constraints on the 4 MG parameters are consistent with GR and in agreement with the functional method.

\subsubsection{Results for binning in redshift and scale  in \lcdm\ and \wowacdm\ backgrounds }

\begin{table}[t!]
\begin{center}
\resizebox{\columnwidth}{!}{
    \small 
\renewcommand{\arraystretch}{2.0} 
\begin{tabular}{p{4.0cm} c c c c c c c}
\hline \hline
\multicolumn{6}{c}{\textbf{Redshift binning}}\\\hline         Datasets  & & $\mu_1$ & $\mu_2$ & $\Sigma_1$ & $\Sigma_2$ & $w_0$  & $w_a$\\ \hline \\[-0.5cm]
 {{\bf $\Lambda$CDM background}}\\
\makecell[l]{DESI+Planck (PR3)\\  
             +DES Y3 + DES5YSN}     
        &  & $1.02\pm 0.13$ & $1.04\pm 0.11$ & $1.021\pm 0.029$ & $1.022^{+0.027}_{-0.023}$ & ---& ---\\[0.2cm] 
 {{\bf $w_0w_a$CDM background}}\\  
\makecell[l]{DESI+Planck (PR3)\\  
        +DES Y3 + DES5YSN} 
        
       & & $1.03\pm 0.14$ & $0.79\pm 0.15$ & $1.019\pm 0.029$ & $1.017\pm 0.024$ & $-0.740\pm 0.070$ & $-1.06^{+0.35}_{-0.30}$ \\[0.2cm]

\hline \hline
\multicolumn{6}{c}{\textbf{Redshift and scale (wavenumber) binning}}\\  \hline
Datasets   &    & $i=1$ & $i=2$ & $i=3$ & $i=4$ & $w_0$  & $w_a$\\ \hline \\[-0.5cm]
 {{\bf $\Lambda$CDM background}}\\
\multirow{2}{*}{\makecell[l]{DESI+Planck (PR3)\\   
 +DES Y3 + DESY5SN}}
 & $\mu_i$ & $0.97\pm 0.18$ & $0.95\pm 0.11$ & $0.83\pm 0.24$ & $1.14\pm 0.15$  & ---& --- \\

& $\Sigma_i$
& $1.045\pm 0.046$ & $1.020\pm 0.035$ & $1.017\pm 0.028$ & $1.048^{+0.050}_{-0.031}$   & ---& ---\\

{{\bf $w_0 w_a$CDM background}}\\ 
\multirow{2}{*}{\makecell[l]{DESI+Planck (PR3)\\   
 +DES Y3 + DESY5SN}}
 & $\mu_i$
& $0.78^{+0.25}_{-0.22}$ & $0.86\pm 0.14$ & $0.78\pm 0.25$ & $1.02\pm 0.15$ & $-0.741\pm 0.068$ & $-1.03^{+0.34}_{-0.30}$ \\

 & $\Sigma_i$
& $1.044\pm 0.045$ & $1.032\pm 0.036$ & $1.017^{+0.029}_{-0.026}$ & $1.047^{+0.050}_{-0.034}$  & $\shortparallel$ & $\shortparallel$ \\
                       
\hline  \hline
                                                            
\end{tabular}
}
\end{center}
\caption{
Constraints on the $\mu$ and $\Sigma$ binned in redshift (top part of the table), and in redshift and scale (bottom part). We show results in both the flat $\Lambda$CDM expansion background and the flat $w_0w_a$CDM expansion background. Constraints are quoted for the marginalized means and 68\% credible intervals in each case. 
}
\label{tab:binning_ak}
\end{table}

Our results for binning in redshift and scale are given in \cref{fig:constraints_binning_a_k} and \cref{tab:binning_ak}.
Results in the table are provided in both the flat $\Lambda$CDM and the flat $w_0w_a$CDM expansion backgrounds. 

The use of the full \cref{eq:binning_ak1} -- \cref{eq:binning_ak3} gives 4 $\mu_i$ and 4 $\Sigma_i$ MG parameters to be constrained by the data.
Specifically, we have two bins in redshift and two bins in scale that are combined as shown in \cref{Table:Bins}. Our results from \cref{fig:constraints_binning_a_k} can thus be categorized as crossing ``low''-$z$ and ``high''-$z$ versus small scales and large scales. 

We find that all 8 MG parameters are around the GR values of 1 (as designed in the binning scheme) and consistent with Einstein's theory.  The 68\% credible intervals for $\mu_i$ range from 11\% to 25\% and those on $\Sigma_i$ range from 3\% to 5\%. So, interestingly, current combined datasets start giving tight and informative constraints when using binned forms including both redshift and scale which is very promising in testing modified gravity using cosmological data. 

Moreover, in the flat $w_0w_a$CDM expansion background, we still find that a dynamical dark energy is preferred by the data.   


\section{Constraints on MG EFT/alpha parameterization}
\label{sec:EFT_alpha_parameterization}

The {Effective Field Theory (EFT) of dark energy} \cite{Creminelli_2009,Park_2010,Gubitosi:2012hu,Gleyzes:2014rba} is a powerful framework to study general modifications of gravity\footnote{In this work, we focus on the Horndeski class of theories \cite{Horndeski:1974wa}. The Horndeski Lagrangian encompasses most dark energy and modified gravity models with a scalar degree of freedom and second-order equations of motion. } in a flexible and unified manner. In this section, we present the constraints using both the EFT-basis \cite{2013JCAP...08..025G,Gubitosi:2012hu} and the $\alpha$-basis \cite{Bellini_2014}. The EFT-basis and $\alpha$-basis are inter-convertible with redefinitions of variables in the \textit{effective} Lagrangian\footnote{For the complete equations, see equations (55) and (56) of \cite{2014arXiv1405.3590H}}. The EFT-basis is advantageous because it closely reflects the underlying structures in the \textit{effective} Lagrangian through changing operator coefficients, whereas the $\alpha$-basis directly characterizes the properties of the linearized scalar field perturbations, offering a more direct connection to observational data. In particular, in the $\alpha$-basis, the background evolution is clearly separated from dynamics of linearized perturbations, controlled by the functions $\alpha_i(t)$. However, it is more convenient to work on EFT-basis if extending beyond second derivatives in the Einstein equations. In this study, the two bases are equivalent frameworks. One also needs to be careful about the Boltzmann solver precision to ensure a fair comparison of observables computed using both the EFT- and $\alpha$-basis.

In the absence of a compelling microscopic theory for dark energy, it is nevertheless possible to constrain its phenomenology from observations in a model-agnostic way using a few free parameters. This ``bottom-up'' approach does not specify the functional form of the Lagrangian; instead, it parameterizes the time evolution of the EFT functions characterizing departures from \lcdm/General Relativity, while remaining agnostic about the field theory content of the model.

\subsection{EFT-basis  }\label{sec:EFT-basis} 
The action of the EFT of dark energy in unitary gauge is
\begin{equation}
  \begin{aligned}
S_{\text{DE}} &= \int d^4 x \sqrt{-g} \bigg[ M_{\text{Pl}}^2 [1+\Omega(t)] \frac{R}{2} - \Lambda(t) - c(t) g^{00} \\[0.2cm]
&+\frac{M_{2}^{4}(t)}{2}(\delta g^{00})^2 - \bar{M_{1}}^3(t) \frac{1}{2} \delta g^{00} \delta K - \bar{M_{2}}^2(t) \frac{1}{2} (\delta K)^2  \\[0.2cm]
&- \bar{M_3}^2(t) \frac{1}{2} \delta \tensor{K}{_\nu^\mu} \delta \tensor{K}{_\mu^\nu}  + \hat{M}^2(t) \frac{1}{2} \delta g^{00} \delta R^{(3)}\\[0.2cm]
&+m_2(t) \partial_i g^{00}\partial^i g^{00} \bigg]+ S_{\text{m}}(g_{\mu\nu}, \Psi_{\text{m}}), 
  \label{eq:EFTDEaction}
  \end{aligned}
\end{equation}
where $M_{\rm Pl}$ is the Planck mass, $R$ is the Ricci scalar, $\delta R^{(3)}$ is the perturbation of the spatial component of the Ricci scalar, $\delta g^{00}$ is defined as $g^{00}+1$, $\delta  \tensor{K}{_\mu^\nu}$ is the perturbation of the extrinsic curvature, $\delta K$ is its trace, and $S_{\text{m}}(g_{\mu\nu}, \Psi_{\text{m}})$ is the action of matter field except dark energy. There are nine time dependent functions in the action modeling the dark energy \{$\Omega(t)$, $\Lambda(t)$, $c(t)$, $M_2(t)$, $\bar{M_{1}}(t)$, $\bar{M_{2}}(t)$, $\bar{M_{3}}(t)$, $\hat{M}(t)$, $m_2(t)$\}. The functions \{$\Lambda(t)$, $c(t)$\} affect the background evolution. After specifying the expansion history, these two functions are determined from the Friedman equations. The rest of free functions only change the perturbation evolution. We study the EFT of dark energy models using \texttt{EFTCAMB} \cite{Hu:2013twa,Raveri2014}.

In this basis, the second-order EFT functions are defined in a dimensionless form.
\begin{equation}
    \begin{aligned}
          \gamma_1 &= \frac{M_{2}(t)^4}{m^2_0 H^2_0}, &
    \gamma_2 &= \frac{\bar{M}_{1}(t)^{3}}{m^2_0 H_0}, &
    \gamma_3 &= \frac{\bar{M}_{2}(t)^{2}}{m^2_0}, \\
    \gamma_4 &= \frac{\bar{M}_{3}(t)^2}{m^2_0}, &
    \gamma_5 &= \frac{\hat{M}(t)^{2}}{m^2_0}, &
    \gamma_6 &= \frac{m_{2}(t)^{2}}{m^2_0}.
    \end{aligned}
    \label{eq:gamma}
\end{equation}

We demand the following constraints to avoid higher-order spatial derivatives \cite{2013JCAP...08..025G,2015JCAP...02..018G}: \begin{equation}
    \begin{aligned}
    m_2 &= 0, \\
    \hat{M}^2 &= \frac{\bar{M_2}^2}{2} = -\frac{\bar{M_3}^2}{2}.
    \end{aligned}
    \label{eq:hornpri}
\end{equation} 

The constraint from Eq.~\ref{eq:hornpri} is equivalent to $2\gamma_5 = \gamma_3 = -\gamma_4$ and $\gamma_6 = 0$. Additionally, the EFT-basis can be converted into the $\mu(z,k)$ and $\Sigma(z,k)$ (see \cite{2016PhRvD..94j4014P} for details).

For background evolution, we consider both $
\Lambda$CDM and $w_0w_a$CDM background cosmology.
For the $w_0w_a$CDM background evolution, the dark energy equation of motion follows the
Chevallier-Polarski-Linder (CPL) parametrization \cite{Chevallier:2001,Linder2003} as used previously:  \begin{equation}
     w_{\text{DE}}(a) = w_0 + w_a(1-a).
 \end{equation} The prior is $w_0\in[-3,1]$ and $w_a\in[-3,2]$. 

We assume the following time-dependence parametrization for the EFT model:
\begin{equation}\label{eq:param_EFT_basis}
    \Omega(a) = \Omega_0 a^{s_0}, \quad \gamma_i(a) = 0,
\end{equation}
where $\Omega_0$ and $s_0$ are free parameters to be constrained and $i$ = 1, 2, 3. The parameter $s_0$ models how early $\Omega(a)$ returns to GR prediction. $\Omega(a)$ controls non-minimal coupling to gravity. As mentioned above, the EFT parameters can be converted into parameters in the $\alpha$-basis $\{ \alpha_M, \alpha_B, \alpha_K, \alpha_T\}$ discussed below in \cref{sec:alpha_basis} (see \cite{Bellini_2014} for full definitions). The EFT parameter $\Omega(a)$ is related to the $\alpha_M$ through
\begin{equation}
    \alpha_M = \frac{a}{\Omega + 1}\frac{d\Omega}{da}.
\end{equation} The $\gamma_1$ affects kineticity in the EFT of dark energy Lagrangian and $\gamma_2$ relates to the kinetic braiding. Both are set to zero. 

The parameter $\gamma_3$ relates the speed of gravitational waves to the speed of light through 
\begin{equation}
\frac{c_T^2}{c^2} = 1 - \frac{\gamma_3(a)}{1 + \Omega(a) + \gamma_3(a)},
\end{equation}we choose $\gamma_{3}(a) =0$ to avoid non-luminal gravitational-wave speed at low redshifts given the constraint from gravitational-wave event GW170817. \cref{tab:priors} shows the priors on the EFT parameters. 

Additionally, we require both ghost stability and gradient stability. The former requires there is no wrong sign of the kinetic term. The latter avoids negative speed of the sound propagation, $c_s^2 < 0$ in the equations of motion of perturbations.

\subsubsection{Results in \tpdf{$\Lambda$}CDM background}\label{subsec:LCDM-model}
First, we report the constraints on the EFT of DE model assuming the $\Lambda$CDM background. \cref{fig:Omega_contours} shows the constraints on \{$\Omega_0$, $s_0$\} in the $\Lambda$CDM background with $\gamma_{1,2} = 0$.
The combination of DESI (FS + BAO), CMB with lensing, and five-year SN~Ia sample from DES gives the following constraints on background properties:

\begin{figure*}
\centering
\includegraphics[width=0.46\columnwidth]{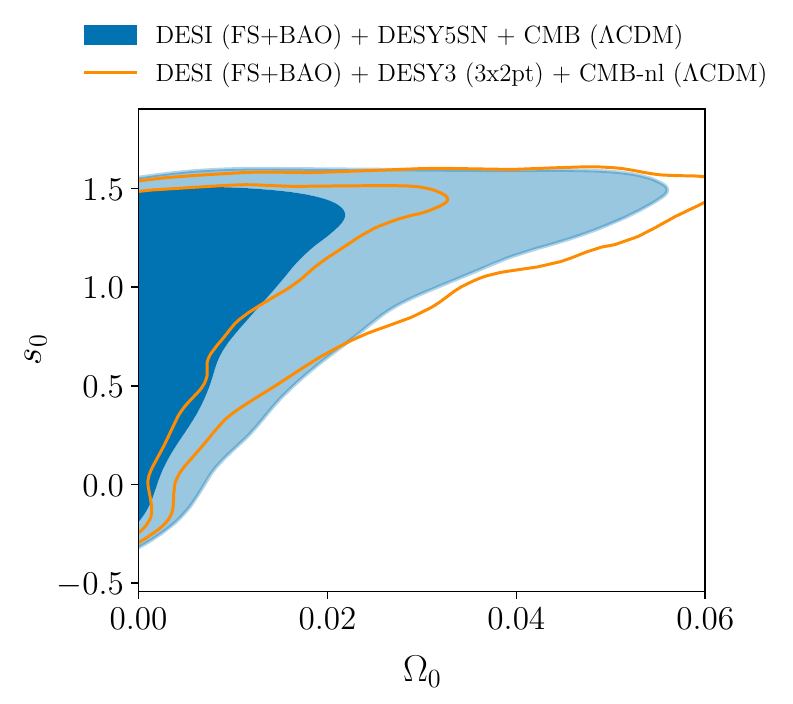}
 \includegraphics[width=0.47\columnwidth]{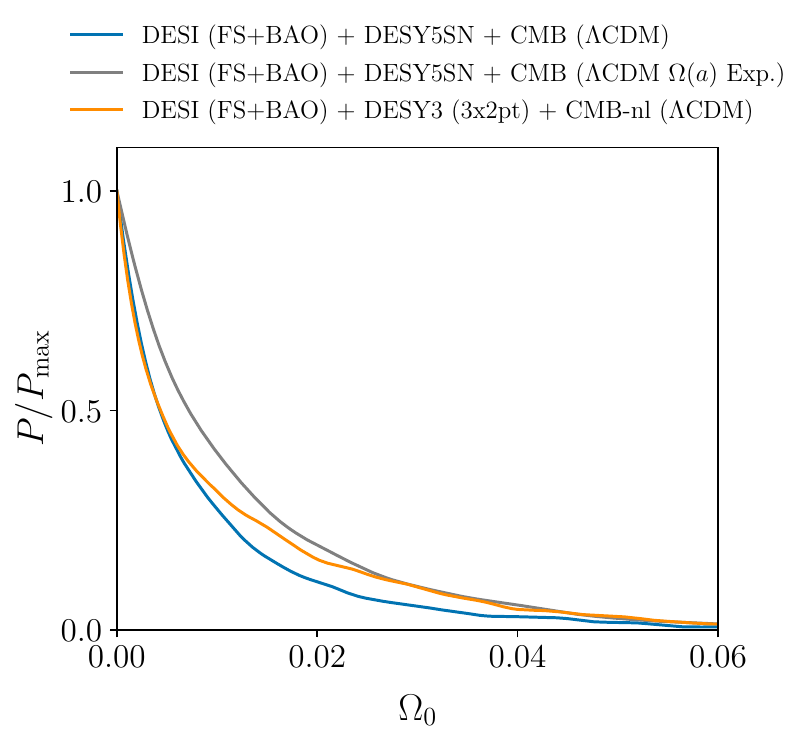}
\caption{
Left: The 68\% and 95\% credible contours on the EFT parameters $\{\Omega_0, s_0\}$ using DESI (FS+BAO), CMB (with lensing), and DESY5SN data in the $\Lambda$CDM expansion history.
We present the constraints on the EFT parameter $\Omega_0$ for values greater than zero, as the region where $\Omega_0 < 0$ is unphysical due to stability conditions. Right: The marginalized posterior distribution for the EFT parameter $\Omega_0$ under power law and exponential parameterizations in $\Lambda$CDM background. Both parameterizations yield consistent constraints on $\Omega_0(a)$, with no deviation from GR predictions.
}
\label{fig:Omega_contours}
\end{figure*}

\threeonesig[4cm]
{\Om &= 0.3220\pm 0.0047,\\[-0.45cm]}
{\sigma_8 &= 0.8152\pm 0.0052,\\[-0.45cm]}
{H_0 &= (66.87\pm 0.34)\,\kmsMpc.}
{DESI (FS+BAO) + DESY5SN + CMB. \label{eq:DESI_DESY5SN_CMB_LCDM}}
We note that a higher value of $\Omega_m$ relative to the best-fit $\Lambda$CDM values is preferred, similar to the constraints in the $w_0w_a$CDM background without modified gravity \cite{DESI2024.VII.KP7B}.

The constraints on the EFT parameters are the following:
\twoonesig[4.5cm]
{\Omega_0 = 0.01189^{+0.00099}_{-0.012},\\[-0.45cm]}
{s_0 = 0.996^{+0.54}_{-0.20},}
{DESI (FS+BAO) + DESY5SN + CMB. \label{eq:DESI_DESY5SN_CMB_EFT}}
The parameter $\Omega(a)$ controls the non-minimal coupling through the effective Planck mass term $M_{\text{Pl}}^2 (1+\Omega(t))$. The constraints on the EFT parameters are consistent predictions of GR in this model. Note that the no-ghost and no-gradient conditions already imply that $\Omega(a) > 0$ (see \cite{2020Frusciante} for more details). Using combinations of DESI (FS+BAO), DESY5SN, and CMB measurement, we find the 95\% C.L. constraint to be $\Omega_0 < 0.0412$. The $s_0$ value is consistent with one, implying a linear evolution of $\Omega(a)$.

Furthermore, combining DESI (FS + BAO) with CMB with no-lensing, weak lensing and galaxy clustering datasets from DESY3 ($3\times 2$-pt), we obtain the following constraints for the EFT parameters:
\twoonesig[4.5cm]
{\Omega_0 =  0.0150^{+0.0041}_{-0.016},\\[-0.45cm]}
{s_0 = 1.06^{+0.49}_{-0.15},}
{DESI (FS+BAO) + DESY3 (3$\times$2-pt) + CMB-nl. \label{eq:DESI_DESY3_CMBnl_EFT}}
The 95\% C.L. constraint is $\Omega_0 < 0.0476$ in this combination of data sets. This result is consistent with the prediction of GR, i.e., $\Omega(a) = 0$. This constraint yields a tighter constraint on the parameter $s_0$ compared to previous constraint that utilized five-year SN Ia data.

The right panel of \cref{fig:Omega_contours} shows the marginalized posterior distribution on $\Omega_0$ in $\Lambda$CDM expansion history. We also include the marginalized posterior distributions when modeling $\Omega(a)$ as an exponential evolution in the $\Lambda$CDM background: $\Omega(a) = \exp\left( \frac{\alpha_{\text{M0}}}{\beta} a^{\beta} \right) - 1.0$, where $\alpha_{\text{M0}}$ and $\beta$ are parameters we aim to constrain (see \cref{fig:Omega_contours}). We find a tight constraint of $\alpha_{\text{M0}}< 0.0445$ in the 95\% C.L., and that $\beta$ is $1.22^{+0.71}_{-0.23}$. Given the small value of $\alpha_{\text{M0}}$ and $\beta$ close to unity, the exponential evolution effectively approximates a linear evolution of $\Omega(a)$, aligning with our constraints assuming a power-law evolution. We also note that we have tighter constraints on EFT parameter $\Omega(a)$ compared to similar constraints in the literature, e.g., \cite{Planck:2015bue,2019PhRvD..99f3538F}. 

\subsubsection{Results in \tpdf{$w_0w_a$}CDM background}
\label{sec:DE}

\begin{figure*}
\centering
\includegraphics[width=0.492\columnwidth]{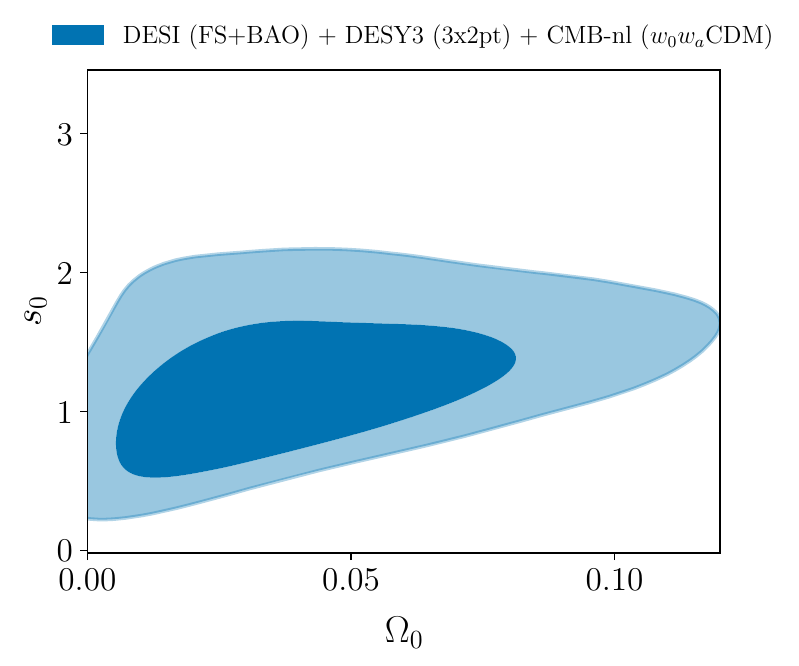}
 \includegraphics[width=0.46\columnwidth]{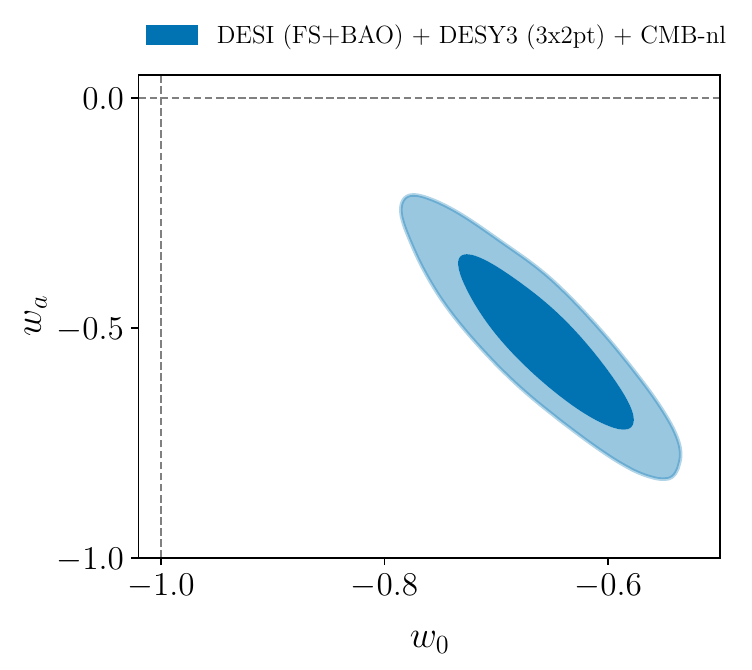}
\caption{
Left: The marginalized posterior distribution for the EFT parameter $\Omega_0$ under power law parametrization in $w_0w_a$CDM background. The constraints on EFT $\Omega_0(a)$ are consistent with GR predictions within 2$\sigma$. Again, we present the constraints on the EFT parameter $\Omega_0$ for values greater than zero. Right: The 68\% and 95\% credible contours on the $\{w_0, w_a\}$ using DESI (FS+BAO), DESY3 ($3\times 2$-pt), and CMB measurement without lensing in the $w_0w_a$CDM expansion history. The tight constraint on $\{w_0, w_a\}$ is caused by simple assumption in~\cref{eq:param_EFT_basis} of the parametrization and setting  $\gamma_{1,2} = 0$.
}
\label{fig:Omega_w0wa_contours}
\end{figure*}

In this section, we discuss the constraints on the EFT parameters when fixing background to $w_0w_a$CDM cosmology. \cref{fig:Omega_w0wa_contours} shows the constraints shows on the EFT parameters \{$\Omega_0$, $s_0$\} with $\gamma_{1,2} = 0$ in the $w_0w_a$CDM background. In this model, the constraints on the dark energy equation of state from DESI (FS+BAO), DESY3 ($3\times 2$-pt), and CMB measurement without lensing are the following:
\twoonesig[4.5cm]
{w_0 &= -0.657\pm 0.051,\\[-0.45cm]}
{w_a &= -0.53\pm 0.12,}
{DESI (FS+BAO) + DESY3 (3$\times$2-pt) + CMB-nl . \label{eq:w0_wa_DESY5SN_CMB}}

We note that the small error bars on $w_0$ and $w_a$ here do not come from the constraining power of the data. It is a result of assuming a simple parametrization in~\cref{eq:param_EFT_basis} for EFT parameter \( \Omega(a) \), which restricts the exploration of further possible EFT models. Subsequently, this imposes a tighter constraint on the \( w_0 \)-\( w_a \) plane compared to when, for example, binned non-parametric reconstruction \cite{Genye2024} and allowing more freedom on EFT parameters (e.g., varying \{$\Omega(a)$, $\gamma_1$(a), $\gamma_2$(a)\} simultaneously)\footnote{We tried to constrain EFT parameters with \{$\Omega(a)$, $\gamma_1$(a), $\gamma_2$(a)\} varying simultaneously using combination of DESI (FS+BAO), DESY5SN, and CMB; however, the mcmc chains were difficult to converge in this case.}. See~\cite{Pan:2025psn} for results using an improved parametrization for \(\Omega(a)\) inferred from non-parametric reconstruction methods, which avoids the issue described above.

Combining DESI (FS + BAO), DESY3 ($3\times 2$-pt), and CMB with no-lensing, we obtain the following constraints for the EFT parameters:
\twoonesig[4.5cm]
{\Omega_0 = 0.043^{+0.016}_{-0.031},\\[-0.45cm]}
{s_0 = 1.23^{+0.31}_{-0.42},}
{DESI (FS+BAO) + DESY3 (3$\times$2-pt) + CMB-nl. \label{eq:DESI_DESY3_CMBnl_EFT}}

The constraints on EFT parameters are compatible with GR predictions in this dataset combinations. At the 95\% C.L., we find the constraint of EFT $\Omega_0$ to be $\Omega_0 = 0.043^{+0.052}_{-0.043}$. Despite consistent with GR in the EFT parameters, indication of dynamics dark energy still persists in the $w_0$ - $w_a$ plane. Past work constraining $\Omega_0$ and $s_0$ in the $w_0w_a$CDM background did not find such signal using BOSS BAO measurement, Supernovae, CMB, and weak lensing from KiDS \cite{2019PhRvD..99f3538F}. 

We also observe that our results yield a lower value of $H_0$, which worsens the Hubble tension, and higher $\Omega_m$ values. This 
may be caused by the parametrization of EFT \( \Omega(a) \) in ~\cref{eq:param_EFT_basis} is not a good fit to the data, which gives larger $w_0$ - $w_a$ deviation. However, further work is needed to understand what is driving these constraints.

The full-shape analysis of these EFT parameters may be subject to projection effects and other systematics, considering the full shape information includes many more nuisance parameters and are thus more sensitive to modified gravity models compared to BAO measurement\footnote{Ref. \cite{Pan2024JCAP} shows that the BAO measurement is robust to Horndeski models considering EFT parameters with power law evolution.}. Nevertheless, we expect projection effects to be small with the inclusion of the DESY3 ($3\times 2$-pt) and CMB measurements with lensing. Table~\ref{tab:EFT_omega_gamma} summarizes all the constraints we have obtained for the EFT parameters.

\begin{table}
    \centering
    \resizebox{\textwidth}{!}{%
    \small
    \setcellgapes{2pt}\makegapedcells  
    \renewcommand{\arraystretch}{2.0} 
    \begin{tabular}{lccccccc}
    \toprule
    \midrule
    \textbf{Model/Dataset}& $\Omega_{\mathrm{m}}$ & $\sigma_8$ & $H_0$ & $w_{0}$ & $w_{a}$ & $\Omega_0$ & $s_0$ \\
    \midrule
    \makecell[l]{\textbf{Flat} $\boldsymbol{\Lambda}$\textbf{CDM} \\[0.1cm] \quad ($\boldsymbol{\Omega(a)}$ free, $\boldsymbol{\gamma_{1,2}(a)=0}$)} &  &  &  &  &  &  &  \\
    \midrule
    \makecell[l]{DESI(FS+BAO)+DESY5SN \\[0.1cm] \quad+CMB}& $0.3220\pm 0.0047$ & $0.8152\pm 0.0052$ & $66.87\pm 0.34$ & -1 & 0 & $0.01189^{+0.00099}_{-0.012}$ & $0.996^{+0.54}_{-0.20}$  \\
     \makecell[l]{DESI(FS+BAO)+DESY3(3$\times$2-pt) \\[0.1cm] \quad+CMB-nl}& $0.3187\pm 0.0046$ & $0.8135\pm 0.0066$ & $67.11\pm 0.34$ & -1 & 0 & $0.0150^{+0.0041}_{-0.016}$ & $1.06^{+0.49}_{-0.15}$  \\
    \midrule
    \makecell[l]{\textbf{Flat} $\boldsymbol{w_0w_a}$\textbf{CDM} \\[0.1cm] \quad ($\boldsymbol{\Omega(a)}$ free, $\boldsymbol{\gamma_{1,2}(a)=0}$)} &  &  &  &  &  &  &  \\
    \midrule
    \makecell[l]{DESI(FS+BAO)+DESY3(3$\times$2-pt) \\[0.1cm] \quad+CMB}& $0.3647\pm 0.0087$ & $0.756\pm 0.010$ & $62.34\pm 0.70$ & $-0.657\pm 0.051$ & $-0.53\pm 0.12$ & $0.043^{+0.016}_{-0.031}$ & $1.23^{+0.31}_{-0.42}$  \\
    \midrule
    \end{tabular}
    } 

    \caption{
       Mean values and 68\% credible intervals on the cosmological parameters for a subclass of Horndeski model in the EFT-basis. 
       }
    \label{tab:EFT_omega_gamma}
\end{table}

\subsection{\tpdf{$\alpha$}-basis} \label{sec:alpha_basis}

Next, we present constraints on the EFT of DE using an alternative basis, the so-called $\alpha$-basis \cite{Bellini_2014}.
In this formalism, the dynamics of the linear perturbations associated with the scalar degree of freedom $\phi$ are fully specified by four free functions of time.
Namely, $\alpha_M(t)\equiv d \ln{M_*^2}/d\ln{a}$ characterizing the running of the \textit{effective} Planck mass $M_*^2(t)=[1+\Omega(t)]M_\text{Pl}^2$, $\alpha_B(t)$ controlling the mixing between the kinetic terms of the scalar and the metric, $\alpha_K(t)$ related to the scalar field's kinetic term, and finally, the tensor speed excess $\alpha_T(t)=c_T^2-c^2$.
The so-called $\alpha$-functions have a direct mapping\footnote{see e.g. Appendix A of \cite{Bellini_2014} for the exact definitions} to the $G_i(X,\phi)$ functions appearing in the Horndeski Lagrangian \cite{Horndeski:1974wa} and have the advantage of providing a closer link with observations. In our baseline analysis, we adopt the commonly used parameterization
\begin{equation}\label{eq:alphas}
    \alpha_i(a) = c_i~\Omega_\mathrm{DE}(a)~,
\end{equation}
where $i = \{M, B, K, T\}$, $\Omega_\mathrm{DE}(a) \equiv \frac{8\pi G}{3H^2(a)}\rho_{\rm DE}(a)$, and $c_i$ is a constant free parameter. While this choice is not unique, \cref{eq:alphas} provides a good approximation for certain subclasses of Horndeski models \cite{Pujolas:2011he,Barreira:2014jha,Bellini:2015oua}. This approximation is further supported by the expectation that dark energy (DE) significantly influences the dynamics only at late times, implying $\alpha_i \to 0$ as $z \to \infty$. 
Different parameterizations, such as the EFT basis ($\Omega, \gamma_i$) in \cref{eq:param_EFT_basis} and the $\alpha$-basis in \cref{eq:alphas}, span distinct functional spaces and may lead to subtle differences in the derived constraints \cite{Bellini_2014,Bellini:2015xja,Noller:2018wyv}. Thus, testing multiple parameterizations is essential for obtaining robust, unbiased results. However, it is important to emphasize that, as with \textit{any} parametric analysis, the results should be interpreted with caution. For a detailed discussion on these and other parameterizations commonly used in the literature, we refer the reader to Refs. \cite{Linder:2015rcz,Linder:2016wqw,Gleyzes:2017kpi,Denissenya:2018mqs,Lombriser:2018olq,Noller:2018wyv}.

Under the quasi-static approximation (QSA) \cite{2013PhRvD..87j4015S,DeFelice:2011hq,Sawicki:2015zya}, the $\alpha$-functions can be related to the phenomenological functions $\mu(z, k)$ and $\Sigma(z, k)$ described previously. Assuming $\alpha_T = 0$ at all times, they can be expressed as \cite{Bellini_2014, Ishak:2019aay}
\begin{subequations}\label{eq:MuSigma-alphas}
    \begin{align}
        \mu(z) &= \frac{M^2_\text{Pl}}{M^2_*}\left[1 + \frac{2(\alpha_M + \frac{1}{2}\alpha_B)^2}{c_s^2(\alpha_K + \frac{3}{2}\alpha_B^2)}\right],\\
        \Sigma(z) &= \frac{M^2_\text{Pl}}{M^2_*}\left[1 + \frac{(\alpha_M + \frac{1}{2}\alpha_B)(\alpha_M + \alpha_B)}{c_s^2(\alpha_K + \frac{3}{2}\alpha_B^2)}\right]\;,
    \end{align}
\end{subequations}
where $M^2_\text{Pl}$ and $M^2_*$ are the bare and effective Planck masses.
The stability conditions discussed in \cref{sec:EFT-basis} require $(\alpha_K+\frac32\alpha_B^2)>0$ to avoid ghosts and $c_s^2>0$ to prevent gradient instabilities, where $c_s^2$ is given by Eq. (3.13) in \cite{Bellini_2014}.

Motivated by the simultaneous detection of GW170817 and its electromagnetic ($\gamma$-ray) counterpart GRB170817A \cite{LIGOScientific:2017zic}, which constrains $\alpha_T \lesssim 10^{-15}$ \cite{Ezquiaga:2017ekz,PhysRevLett.119.251301,Kreisch_2018,Kase_2019,Creminelli:2017sry,Copeland_2019}, we focus here on the subclass of models satisfying $\alpha_T = 0$.\footnote{Note that this constraint applies only at $z \simeq 0$, and in principle, $\alpha_T(z) \neq 0$ could be allowed in the past. See also \cite{PhysRevLett.121.221101} for a discussion on the validity of the EFT of DE at LIGO scales.} In what follows, we also fix $c_K = 10^{-2}$ since observations are generally insensitive to $\alpha_K$ \cite{Bellini:2015xja,Reischke:2018ooh}.  Thus, the remaining functions are the running $\alpha_M$ and the braiding $\alpha_B$. To derive constraints on cosmological parameters, we use the publicly available Boltzmann solver \texttt{mochi\_class} \cite{Cataneo:2024uox,Zumalacarregui:2016pph,Bellini_2020} interfaced with the MCMC sampler \texttt{cobaya} \cite{Torrado:2021}. In addition to the four (time-dependent) $\alpha_i(t)$'s, we need to specify the evolution of the effective energy density, $\rho_\phi(z)$. In this work, following the structure in the previous subsection, we consider both a $\Lambda\rm CDM$ expansion history ($w_0=-1,w_a=0$) and a $w_0w_a\rm CDM$ expansion history, where $w_0\in[-3,1]$ and $w_a\in[-3,2]$ are free to vary. In addition to the usual $\Lambda\rm CDM$ parameters, we also vary the coefficients $c_i$ by imposing flat uninformative priors $c_i\in[-10,10]$. Let us note that under such assumptions, some parameter combinations might lead to ghosts or gradient/tachyonic instabilities \cite{Bellini_2014}. To avoid ill-defined (pathological) theories, we reject those points in parameter space violating the stability conditions tested within \texttt{hi\_class} \cite{Zumalacarregui:2016pph}. 

We will present constraints for three different subclasses of models, which translates into ``activating'' certain properties of the linear perturbations. The first class of interest is the one with maximal freedom, allowing both the running $\alpha_M$ and the braiding $\alpha_B$ to vary following \cref{eq:alphas}. The second one, closely related to the first model presented in \cref{sec:EFT-basis}, is the subclass of models with no braiding, $\alpha_B=0$.
Finally, we focus on a third subclass of models satisfying $\alpha_B=-2\alpha_M$, dubbed ``no-slip'' gravity \cite{Linder:2018jil} (i.e. $\Phi=\Psi$), for which $\Sigma=\mu=M_\mathrm{Pl}^2/M_*^2$, as is obvious from \cref{eq:MuSigma-alphas}. Note that the subclass of Horndeski theories satisfying $\alpha_M=-\alpha_B$ corresponds to the well-known case of $f(R)$ gravity, which might be the subject of future work. 

\subsubsection{Results in \tpdf{$\Lambda\rm CDM$} background}\label{sec:LCDM-expansion}

\begin{figure}[ht]
    \centering
\includegraphics[width=0.45\columnwidth]{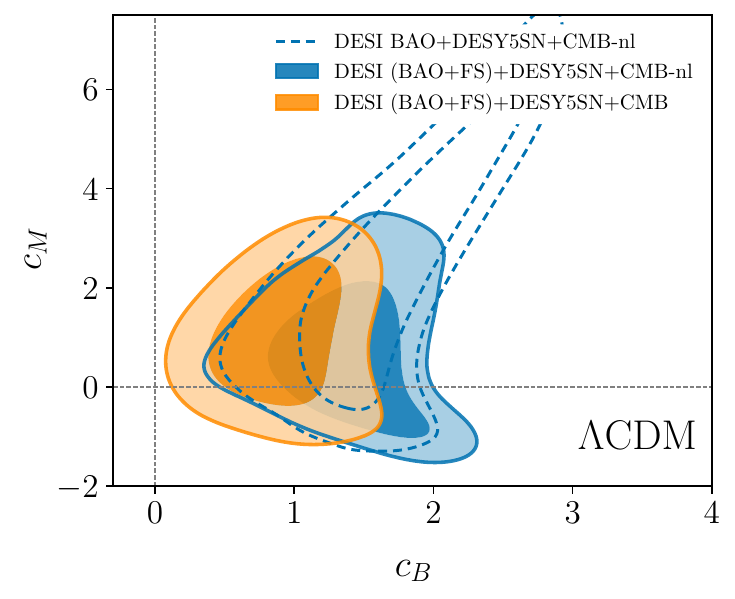}
\includegraphics[width=0.44\columnwidth]{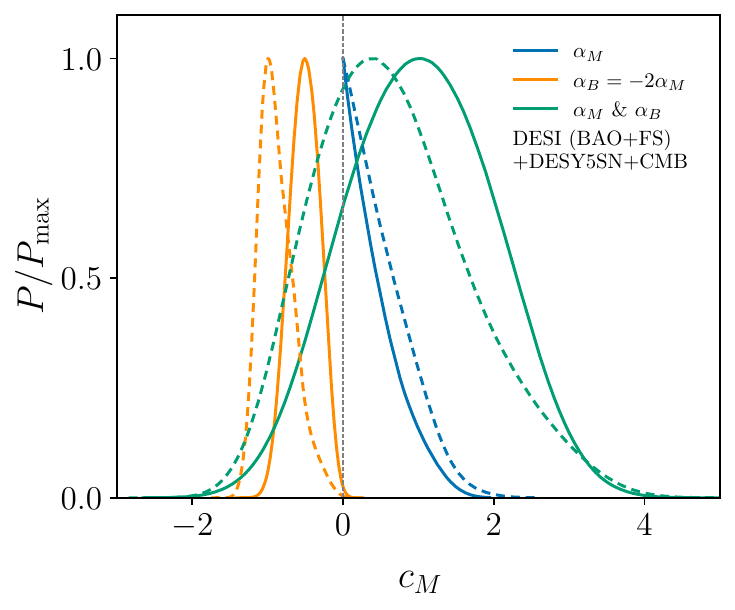}
    \caption{
    \textit{Left panel}: 2D posterior distributions for the EFT coefficients $c_M$ and $c_B$ and for various data combinations, where CMB-nl refers to the CMB anisotropies with no lensing. Note that the region $c_M<0$ and $c_B>0$ is plagued by gradient instabilities, which explain the sharp ``cut-off'' of the posterior distributions in the bottom-left region of the plot. \textit{Right panel}: Marginalized 1D posterior distribution for $c_M$ and the three classes of models considered. These constraints are derived using the DESI (BAO+FS)+DESY5SN+CMB combination, with (solid) and without (dashed) CMB lensing, assuming a \lcdm\ expansion history. Note that stability conditions exclude the region $c_M<0$ for theories with $\alpha_B=0$ (blue).}
    \label{fig:LCDM_cM_cB}
\end{figure}

We start by constraining the cosmological and EFT parameters assuming a $\Lambda\rm CDM$ expansion history. In what follows, we report the constraints when allowing for both $\alpha_B$ and $\alpha_M$ to vary in time, according to \cref{eq:alphas}. The constraints on the background quantities are 
\threeonesig[4cm]
{\Om &= 0.3054\pm 0.0050,\\[-0.45cm]}
{\sigma_8 &=  0.837\pm 0.017,\\[-0.45cm]}
{H_0 &= (68.08\pm 0.38)\,\kmsMpc.}
{DESI (FS+BAO) + DESY5SN + CMB~. \label{eq:Om0_sigma8_H0_DESI_DESY3_LCDM}}
We find that the constraints on the (background) cosmological parameters are relatively stable across the three sub-classes of models studied here, as reported in \cref{tab:EFT_alphas}. For the coefficients describing the evolution of the $\alpha_i$'s, we get
\twoonesig[4.5cm]
{c_M    &= 1.05\pm 0.96,}
{c_B &= 0.92\pm 0.33,} 
{DESI (FS+BAO) + DESY5SN + CMB~. \label{eq:cM_cB_DESI_CMB_DESy5_LCDM}}
The marginalized posterior distribution is shown in \cref{fig:LCDM_cM_cB}. The data indicates a mild preference for $\alpha_B \neq 0$, while remaining consistent with no running of the Planck mass ($\alpha_M = 0$). Stability bounds, in particular, due to gradient instabilities, exclude significant regions in parameter space, such as models with $c_M \ll 0$. When growth measurements from DESI are not included, the constraints on the $\alpha_i$'s are primarily driven by the late Integrated Sachs-Wolfe (ISW) effect on the CMB \cite{Bellini:2015xja,Kreisch_2018}. Large values of $\alpha_M$ (or $\alpha_B$)  modify the late-time evolution of the gravitational potentials ($\dot\Phi$ and $\dot\Psi$), resulting in excess power at large angular scales (low-$\ell$) \cite{Zumalacarregui:2016pph,Noller:2018wyv}. 

When both $\alpha_B$ and $\alpha_M$ are allowed to vary, they can interfere destructively, suppressing the low-$\ell$ ISW tail. This interaction can lead to significant deviations from GR while still maintaining a satisfactory fit to the data.
This degeneracy is broken when full-shape measurements of the power spectrum multipoles are included, as they tightly constrain the running of the Planck mass, $\alpha_M$, by probing the growth of structures at late times. The combined data favor the region $c_B \gtrsim 0$ and $c_M \lesssim 2$. Let us note that at this stage that such a region can be efficiently probed by cross-correlating galaxies with the CMB  \cite{Stolzner:2017ged,Seraille:2024beb}. Including such cross-correlation would result in even tighter constraints on the $\alpha_i$'s through a more sensitive probe of the ISW effect.

A notable subclass of theories, which falls nicely in the region currently allowed by observations, is ``no-slip'' gravity \cite{Linder:2018jil}. This subclass of theories is characterized by $\alpha_B = -2\alpha_M$, which ensures $\Phi = \Psi$ and a slip parameter of $\eta \equiv 1$. In such theories, the mild preference for $c_B\neq0$ is reflected in the 1d marginalized posterior distribution for $c_M$, shown in the right panel of \cref{fig:LCDM_cM_cB}. 


For models with no braiding ($\alpha_B = 0$), known as ``only-run'' gravity \cite{Linder:2020xza}, stability conditions impose $\alpha_M \ge 0$, as shown in  \cref{fig:LCDM_cM_cB}. In such theories, although dark energy does not cluster on subhorizon scales, the growth of matter perturbations is still affected by the non-minimal coupling ($\alpha_M \neq 0$). Consequently, the inclusion of full-shape measurements results in an upper bound on $c_M < 1.14$ 
at $95\%$ C.L., consistent with GR. These results can be seen as complementary to the ones presented in \cref{sec:EFT-basis}, for the first model, where $\Omega(a)$ is free and $\gamma_{1}(a)=\gamma_2(a)=0$.

\subsubsection{Results in \tpdf{$w_0w_a\rm CDM$} background}\label{sec:w0waalpha-expansion}
\begin{figure}[ht]
    \centering
    \includegraphics[width=0.6\textwidth]{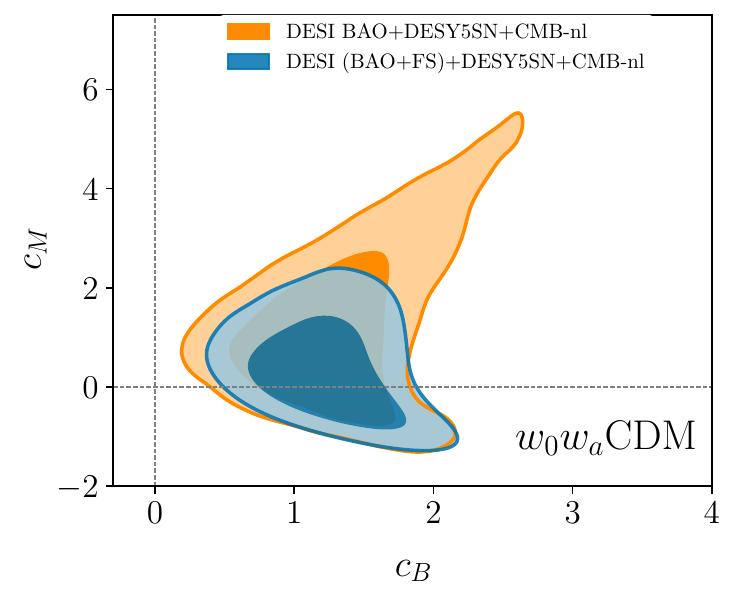}
    \caption{2D posterior distributions for the EFT coefficients $c_M$ and $c_B$ and for various data combinations.}
    \label{fig:w0waCDMc_M_C_B}
\end{figure}
Next, we let the effective dark energy density  $\rho_\phi$ vary with time according to the canonical $w(a)=w_0
+w_a(1-a)$ parametrisation \cite{Chevallier:2001,Linder2003}. 
Our background parameters estimates are
\threeonesig[4cm]
{\Om &= 0.3131\pm 0.0063,\\[-0.45cm]}
{\sigma_8 &= 0.832\pm 0.017,\\[-0.45cm]}
{H_0 &= (67.36\pm 0.62)\,\kmsMpc.}
{DESI (FS+BAO) + DESY5SN + CMB-nl  . \label{eq:Om0_sigma8_H0_DESI_DESY3_w0waCDM}}
Despite a small increase in $\Omega_m$, these are in good agreement with those reported in the previous \cref{sec:LCDM-expansion}, where we assumed a \lcdm -expansion history. The equation of state parameters are constrained to be
\twoonesig[4.5cm]
{w_0 &= -0.801\pm 0.65, }
{w_a &=  -0.70\pm 0.29,}
{DESI (FS+BAO) \dataplus DESY5SN + CMB-nl. \label{eq:DESI_CMB_DESY5_w0wa}}
We note that, while the statistical significance decreases due to an extended parameter space, the tantalizing hints for $w(z)\neq-1$ previously reported in \cite{DESI2024.VI.KP7A,DESI2024.VII.KP7B,DESI:2024aqx,DESI:2024kob} remain.
For the coefficients describing the evolution of the $\alpha$-functions, we obtain
\twoonesig[4.5cm]
{c_M    &= 0.33^{+0.63}_{-0.91},}
{c_B &= 1.25\pm 0.33,} 
{DESI (FS+BAO) + DESY5SN +CMB-nl. \label{eq:cM_cB_DESI_CMB_DESy5_wowaCDM}}
Our results (see also \cref{fig:w0waCDMc_M_C_B}) are consistent with no running of the Planck mass, $\alpha_M = 0$, as predicted by General Relativity. Interestingly, when allowing the dark energy equation of state to vary with time\textemdash under the assumption of $\alpha_i\propto\Omega_{\rm DE}$\textemdash the combined data continues to favor a non-zero braiding parameter, $c_B \neq 0$ \cite{Deffayet:2010qz,Pujolas:2011he,DAmico:2016ntq}. These findings, summarized in \cref{tab:EFT_alphas}, align with recent literature \cite{Raveri:2019mxg,Seraille:2024beb,Chudaykin:2024gol,Taule2024,Lu:2025gki}
however, the inclusion of DESI's full-shape measurements in this work places strong constraints on $\alpha_M$ and consistently shows the possibility of $\alpha_B \neq 0$. {From a theoretical standpoint, we expect at least one of the $\alpha_i\neq0$ to stabilize the phantom-crossing in $w(z)$, as suggested by parametric and non-parametric techniques~\cite{DESI:2025fii}.
Given the importance of such results, more work is needed to clarify the source of such deviations from GR, be it due to a physical or systematic origin, and we leave that for future work.} For example, as discussed in \cref{sec:functional_a_dependence}\textemdash and shown in \cref{Fig:functional_mu_Sigma_a_lcdm}\textemdash the derived modified gravity constraints are moderately sensitive to the choice of CMB likelihood. Notably, the statistical significance of deviations from GR decreases when moving from Planck PR3 to newer CamSpec \cite{Efstathiou:2021,Rosenberg:2022}
or \texttt{HiLLiPoP}/\texttt{LoLLiPoP} \cite{Tristram:2021,Tristram:2023} likelihoods based on Planck PR4, which lack the $A_\mathrm{lens}$ and $\Omega_k$ anomalies that often correlate with modifications to gravity \cite{Specogna:2024euz}. Due to the high dimensionality of our parameter space, we did not repeat the analysis with these alternative CMB likelihoods. However, we anticipate that the results would trend towards GR-predicted values ($\alpha_B = \alpha_M = 0$) when using Planck PR4 data, especially with the addition of DESY3 ($3 \times 2$pt) measurements.

An important point worth mentioning is that full-shape analyses based on the EFTofLSS may be subject to prior and projection effects (thoroughly discussed in \cite{DESI2024.V.KP5}), particularly in extended parameter spaces \cite{DESI2024.VII.KP7B}, as considered here. We assume that the combination of DESI (BAO+FS), DESY5SN, and CMB data effectively mitigates such projection effects, though further work is needed to clarify whether these effects or other systematics in the data could contribute to the preference for $\alpha_B \neq 0$. Lastly, we expect that including the ISW effect could be crucial in constraining the $\alpha$-functions. We leave this for future work.
\vspace{0.5cm}

\begin{table}[h]
    \centering
    \resizebox{\columnwidth}{!}{%
        \small 
    \setcellgapes{2pt}\makegapedcells  
    \renewcommand{\arraystretch}{3.0} 
            \begin{tabular}{cccccccc}
            \toprule
            \midrule
            model/dataset & $H_0$ &$\Omega_{\mathrm{m}}$ & $\sigma_8$ &  $w_{0}$ & $w_{a}$ & $c_M$ & $c_B$ \\
            \midrule
            \makecell[c]{\textbf{Flat} \boldsymbol{$\Lambda$}\textbf{CDM}\\
            [0.1cm] 
            \quad (\boldsymbol{$\alpha_M$} \& \boldsymbol{$\alpha_B$} free)} &  &  &  &  &  &  &  \\
            \makecell[l]{DESI BAO+DESY5SN+CMB-nl}& $68.20\pm 0.43$ & $0.3042\pm 0.0056$ & $0.871^{+0.025}_{-0.048}$ & $-1$ &$0$ & $2.1^{+1.4}_{-2.6}$ & $ 1.66^{+0.47}_{-0.55}$\\
            \makecell[l]{DESI(FS+BAO)+DESY5SN\\[0.1cm] \quad+CMB-nl} & $68.14\pm 0.41$
            & $0.3049\pm 0.0054$  
            & $0.840^{+0.011}_{-0.021}$  
            & $-1$ &$0$
            &$0.43^{+0.80}_{-1.3}$  
            & $1.43\pm 0.40$  
            \\
            \makecell[l]{DESI(FS+BAO)+DESY5SN\\[0.1cm] \quad+CMB}& $68.06\pm 0.37$ 
            & $0.3057\pm 0.0048$  
            & $ 0.835\pm 0.015$  
            & $-1$ &$0$
            & $0.98\pm 0.89$  
            & $0.91\pm 0.31$ 
            \\
            \midrule
            (\boldsymbol{$\alpha_M$} free \& \boldsymbol{$\alpha_B=0$}) &&&&&&&\\
            \makecell[l]{DESI BAO+DESY5SN+CMB-nl}& $67.82\pm 0.41$ & $ 0.3088\pm 0.0055$ & $0.8180^{+0.0076}_{-0.010}$ & $-1$ & $0$ & $ 0.44^{+0.12}_{-0.43}$ & $0$\\
            \makecell[l]{DESI(FS+BAO)+DESY5SN\\[0.1cm] \quad+CMB-nl} & $ 67.78\pm 0.37$ 
            & $0.3093\pm 0.0050$ 
            & $0.8190^{+0.0079}_{-0.0098}$ 
            & $-1$ & $0$ &
            $<0.636$ & $0$  \\
            \makecell[l]{DESI(FS+BAO)+DESY5SN\\[0.1cm] \quad+CMB}& $67.71\pm 0.35$ 
            & $0.3103^{+0.0043}_{-0.0049}$ 
            & $0.8219^{+0.0072}_{-0.0095}$ 
            & $-1$ &$0$ &
            $0.54^{+0.12}_{-0.54}$ 
            & $0$  \\
            \midrule
            (\boldsymbol{$\alpha_B=-2\alpha_M$}) &&&&&&&\\
            \makecell[l]{DESI BAO+DESY5SN+CMB-nl}& $68.14\pm 0.42$ & $0.3049\pm 0.0055$ & $0.8306^{+0.0092}_{-0.0077}$ & $-1$ & $0$ & $-0.85^{+0.18}_{-0.31}$ & $-2c_M$\\
            \makecell[l]{DESI(FS+BAO)+DESY5SN\\[0.1cm] \quad+CMB-nl} & $68.14\pm 0.43$
            & $0.3048\pm 0.0056$ 
            & $0.8287^{+0.0098}_{-0.0076}$ 
            & $-1$ &$0$
            & $-0.89^{+0.17}_{-0.26}$ 
            & $-2c_M$  \\
            \makecell[l]{DESI(FS+BAO)+DESY5SN\\[0.1cm] \quad+CMB} & $ 68.04\pm 0.41$ 
            & $0.3062\pm 0.0053$  
            & $0.8161\pm 0.0053$  
            & $-1$ &$0$
            &$-0.51\pm 0.21$  
            & $-2c_M$  \\
            \midrule
            \midrule
            \makecell[c]{\textbf{Flat} \boldsymbol{$w_0w_a$}\textbf{CDM}\\
            [0.1cm] 
            \quad (\boldsymbol{$\alpha_M$} \& \boldsymbol{$\alpha_B$} free)} &  &  &  &  &  &  &  \\
            \makecell[l]{DESI BAO+DESY5SN+CMB-nl}& $67.21^{+0.66}_{-0.75}$ & $0.3136\pm 0.0067$ & $0.843^{+0.017}_{-0.028}$ & $-0.775\pm 0.073$ & $-0.75\pm 0.33$ & $0.93^{+0.70}_{-1.4}$ &$1.23^{+0.41}_{-0.47}$\\
            
            \makecell[l]{DESI(FS+BAO)+DESY5SN\\[0.1cm] \quad +CMB-nl} & $67.36\pm 0.62 
            $  & $0.3131\pm 0.0062$ 
            & $0.832^{+0.013}_{-0.018}$ 
            & $-0.801\pm 0.065$ 
            & $-0.70\pm 0.29$ 
            & $0.33^{+0.63}_{-0.91}$ 
            & $1.25\pm 0.33$ 
            \\
            \makecell[l]{DESI(FS+BAO)+DESY5SN\\[0.1cm] \quad +CMB} & $67.35\pm 0.65$ 
            & $0.3127\pm 0.0063$ 
            & $0.823\pm0.017$ 
            & $-0.803^{+0.062}_{-0.072}$ 
            & $-0.68^{+0.31}_{-0.25}$ 
            & $0.39^{+0.56}_{-1.1}$ 
            & $0.95^{+0.24}_{-0.37}$ 
            \\
            \midrule
            \midrule
            \end{tabular}
    }

    \caption{
        Mean values and 68\% credible intervals on MG and cosmological parameters for various subclasses of Horndeski models, using the $\alpha$-basis and various dataset combinations. 
        }
    \label{tab:EFT_alphas}
    \end{table}

\pagebreak
\section{Conclusions}
\label{sec:conclusions}
We derive constraints on modified gravity parameters using data from the full-shape (FS) modeling of the power spectrum, including the effects of redshift-space distortions from the first year of DESI Data Release 1 (DR1). This clustering data is very sensitive to the growth rate of large scale structure and is very effective at constraining gravity theory at cosmological scales. 

We present results for DESI in combination with other available datasets including: the CMB temperature and polarization data from Planck as well as CMB lensing from Planck and ACT, BBN constraints on the physical baryon density, the galaxy weak lensing and clustering as well as their cross-correlation referred to as the DESY3,  and supernova data from DES Y5. We avoid using CMB lensing and the DESY3 (3$\times$2-pt) data at the same time in any combination due to their covariance. 

We first consider the often-used $\mu(a,k)$-$\Sigma(a,k)$ phenomenological parameterization (as well as $\eta(a,k)$) in order to test deviations from general relativity. In this approach, one aims to test whether the data shows any departures from the values predicted by GR (zero in this parameterization) without assuming a specific model of modified gravity.  By construction, $\mu(a,k)$ is featured in the equation that governs the dynamics of massive particles, while $\Sigma(a,k)$ appears in the equation that governs the dynamics of massless particles that can be constrained by gravitational lensing as an example.  

We start by deriving constraints for $\mu(a)$--$\Sigma(a)$ using a functional form to express the dependence on time (or redshift) of such parameters in a \lcdm\ cosmological background.  We find that DESI (FS+BAO)+BBN+$n_{s10}$ gives $\mu_0 = 0.11^{+0.44}_{-0.54}$ which is consistent with the GR value of zero for this scheme. DESI produces no direct constraints on the parameter $\Sigma_0$; however, when combined with other datasets, it breaks degeneracies in cosmological parameters and allows to significantly improve the constraints on $\Sigma_0$.   

We next derive constraints on these parameters using the Planck CMB data with and without lensing. We find that using Planck CMB PR3 without lensing gives results on $\Sigma_0$ that are in some tension with the GR value of zero. This was associated, in previous studies, with the CMB lensing anomaly (a systematic effect) that was usually expressed in terms of non-unity of the non-physical parameter $A_\text{lens}$.  In fact when CMB (PR3) is combined with DESI, this tension raises well above the 3-$\sigma$ level (see bottom-left panel of \cref{Fig:functional_mu_Sigma_a_lcdm}) (but again, this is driven by Planck PR3 not DESI).  We also derive constraints on these MG parameters using the Planck CMB likelihood Camspec and the most recent \texttt{LoLLiPoP}-\texttt{HiLLiPoP}. The problem of $A_\text{lens}$ is partly alleviated with Camspec and resolved with \texttt{LoLLiPoP}-\texttt{HiLLiPoP}.  We find in our new results on MG that the tension in $\Sigma_0$ is alleviated with Camspec and goes away with \texttt{LoLLiPoP}-\texttt{HiLLiPoP}. We then combine DESI (FS+BAO) with the no-lensing CMB data using the three likelihoods and observe the same trend for this tension. This thus seems to demonstrates the connection of this tension to the lensing anomaly, and that it seems to be related to a possible systematic effect in Planck PR3.

We find the tightest constraints on the two MG parameters come from our combination DESI+CMB-no-lensing+DESY3+DESY5-SN  and are given by $\mu_0  =  0.05\pm 0.22$ and $\Sigma_0 =  0.008\pm 0.045$  and  similarly $\mu_0 = 0.01^{+0.19}_{-0.24}$ and $\eta_0 =  0.09^{+0.36}_{-0.60}$  (but noting that the DESY5 SN in this case is not adding any significant further constraints but this will not be the case for the \wowacdm\ background or other extended dependencies). All the constraints are consistent with the GR predicted values but the resultant constraint on $\Sigma_0$ is found to be nearly a factor of 5 better than that on $\mu_0$. 

We then consider the same parameterization and time evolution but in a \wowacdm\ cosmological background.  In view of the increase in the total number of cosmological parameters, we use the full combination DESI+CMB (\texttt{LoLLiPoP}-\texttt{HiLLiPoP})-nl + DESY3 + DESY5 SN and find that adding the supernova dataset does make a significant improvement on both the dark energy equation of state parameters and also the MG parameters. Interestingly, even in this extended case of parameters, the constraints on the dark energy parameters still indicate a preference for a time evolving equation of state with   $w_0 = -0.784\pm 0.061$ and $w_a = -0.82^{+0.28}_{-0.24}$ while the MG results $\mu_0  = -0.24^{+0.32}_{-0.28}$ and $\Sigma_0 =   0.006\pm 0.043$ are still all consistent with GR. 

Next, we finish for the functional parameterization by allowing for both time and scale dependence. That is done by adding three further MG parameters that control the scale functional dependence, i.e. $\lambda$, $c_1$ and $c_2$. We again use the full and most constraining combination of datasets, DESI+CMB (PR3)-nl+DESY3+DESY5SN and obtain,  $\mu_0= 0.03^{+0.13}_{-0.14}$ , $\Sigma_0=0.027\pm 0.043$, but the scale parameters remain difficult to constrain using this functional form. 
Interestingly, the binning method in redshift and scale does much better and is able to return meaningful constraints on all parameters. In all cases, the results are found to be consistent with GR. 

We now move to the binning parameterization where, instead of using analytical functions to express time or scale evolution, we rather design bins of redshifts and bins of scales with smoothed transitions. We start by binning in redshift-only and assume two bins with MG parameters $\mu_1$ and $\Sigma_1$ in a first bin with $0\leq z< 1$  and $\mu_2$ and $\Sigma_2$ in a second bin with $1\leq z< 2$. We assume that beyond $z\geq 2$, gravity is given by GR and all the parameters are set to take the GR value of 1 (note the convention for the binning form). Using again the combination DESI+CMB (PR3)-nl+DESY3+DESY5SN, we obtain  $\mu_1=1.02\pm 0.13$, $\mu_2=1.04\pm 0.11$, $\Sigma_1=1.021\pm 0.029$ and $\Sigma_2=1.022^{+0.027}_{-0.023}$, which are all consistent with GR. 

The next level in the binning parameterization is to allow for binning in both redshift and scale and we implement that. We use the two bins in redshift above but crossed with two other bins in scale, giving 4 parameters $\mu_i$  and 4 parameters $\Sigma_i$. Similarly, we find that the combination  DESI+CMB (PR3)-nl+DESY3+DESY5SN is able to give constraints on all 8 parameters, that are consistent with GR. 

It is worth noting that in both the functional forms and binned forms of MG parameterizations, when we use a \wowacdm\ expansion background, we still find that the combined data show preference for a dynamical dark energy with $w_0>-1$ and $w_a<0$.    

In sum for the above part, we find that all the constraints are consistent with the GR predicted values. We also find that current combined datasets provide a more precise measurement on the $\Sigma$ parameters than the $\mu$ parameters by up to a factor 5. This indicates that there is room for a lot of improvement on $\mu_0$ where DESI is expected to play a major role in reducing such uncertainties with its next four years of data.

We also note that the analysis including \textit{both} redshift and scale seems to start providing tight and meaningful constraints on the MG parameters for the binned parameterization which is an important step in testing modified gravity in cosmology. 

We next constrain the class of Horndeski theory in the effective field theory of dark energy approach. We assume both an EFT-basis and an $\alpha$-basis in the analysis. In the EFT-basis, we first assume non-minimal coupling with a parameterization $\Omega(a) = \Omega_0 a^{s_0}$ with $\gamma_1(a) = \gamma_2(a) = \gamma_3(a) = 0$. Specifically in a $\Lambda$CDM background, using data from DESI (FS+BAO), DESY5SN, and CMB data, we obtain $\Omega_0 = 0.01189^{+0.00099}_{-0.012}$ and $s_0 = 0.996^{+0.54}_{-0.20}$. Additionally, when combining DESI (FS+BAO), DES Y3 ($3\times 2$-pt), and CMB without lensing measurement, we obtain the constraints $\Omega_0 = 0.0150^{+0.0041}_{-0.016}$ and $s_0 = 1.06^{+0.49}_{-0.15}$ assuming a $\Lambda$CDM background. In the $w_0w_a$CDM background, we find that the EFT parameters $\Omega_0 = 0.043^{+0.016}_{-0.031}$ and $s_0 = 1.23^{+0.31}_{-0.42}$ are consistent with GR predictions within 2$\sigma$, and we still find an indication for dynamical dark energy. The constraints on the EFT parameters in this model in both the $\Lambda$CDM and the $w_0w_a$CDM background show consistency with the predictions of GR ($\Omega_0 = 0$), although these constraints were derived only for the power-law parametrization of the EFT functions adopted here.

Assuming $\alpha_i \propto \Omega_{\rm DE}(a)$, we investigated three distinct classes of Horndeski models, each characterized by different properties in the linear perturbations. Our results, analyzed using this basis, are broadly consistent with no running of the Planck mass ($c_M = 0$), as expected from general relativity (GR), {while consistently showing mild departures in $\alpha_B$, from its GR value of zero}. {This could indicate projection effects, potential systematics in the data, or new physics beyond $\Lambda\rm CDM$ and deserves further exploration.}
These findings underscore the critical role of growth and lensing measurements in constraining modifications to gravity. Below, we summarize the main results for each class: 

No Braiding ($\alpha_B = 0$): Assuming a \lcdm\ expansion history, the combined DESI (BAO+FS)+DESY5SN+CMB data constrains $c_M < 1.14$ at the $95\%$ confidence level, in agreement with the expectations from GR ($\alpha_M=0$). 
No-Slip Gravity ($\alpha_B = -2\alpha_M$): In this scenario, the relationship $\alpha_B = -2\alpha_M$ fixes the gravitational potentials such that $\Psi = \Phi$. Consequently, lensing is closely tied to the growth of structures. Our analysis yields $c_M = -0.51 \pm 0.21$, which is consistent with GR within 2-$\sigma$.  
Running \& Braiding ($\alpha_B$, $\alpha_M$): For this case, we find that DESI full-shape measurements are crucial for constraining $\alpha_M$. However, they are insufficient to fully break degeneracies with $\alpha_B$, leading to a mild deviation of $\alpha_B$ from 0.  Combining DESI full-shape constraints with CMB lensing and weak lensing measurements will prove essential for disentangling the effects of the running  $\alpha_M$ and the braiding $\alpha_B$ in upcoming analyses. {In summary, while the combined DESI (FS+BAO)+DESY5SN+CMB data yield results consistent with GR when $\alpha_B$ is set to zero, {there is a mild yet consistent preference for $c_B > 0$} when this braiding term $\alpha_B$ is allowed to vary. This preference is more pronounced in the absence of lensing constraints.} 
{More work is needed to explore the origin of such departure from GR  and whether it is due to unknown systematic effects or new physics, and we leave that for future studies.
}

{Our results are based on the DESI full shape analysis that employed perturbation theory that has been compared in previous studies to an MG non-linear code, e.g. \cite{Rodriguez-Meza:2023rga}, and found to be in good agreement in loop corrections for small departures from GR as we find here for MG parameter mean values from our most constraining dataset combinations. However, future work should be conducted to estimate more quantitatively any modeling systematics related to MG models and to determine if any additional systematic errors need to be added to the overall error budget of the full shape results and corresponding MG constraints.} 

In sum, we focused in this paper on an analysis dedicated to testing modified gravity at cosmological scales using data from DESI, a Stage IV dark energy experiment, in combination with other publicly available datasets. We find that one year of data from DESI is able to provide constraints on MG parameters that are as competitive as two decades of data from SDSS. Moreover, DESI provides direct effective constraints on the MG parameter(s) associated with the growth rate of large scale structure using the full shape analysis indicating that forthcoming DESI data will play a major role in constraining the nature of gravity theory at cosmological scales.    

\section{Data Availability}
The data used in this analysis will be made public along the Data Release 1 (details in \url{https://data.desi.lbl.gov/doc/releases/}).
\acknowledgments
MI acknowledges that this material is based upon work supported in part by the Department of Energy, Office of Science, under Award Number DE-SC0022184 and also in part by the U.S. National Science Foundation under grant AST2327245. RC is funded by the Czech Ministry of Education, Youth and Sports (MEYS) and European Structural and Investment Funds (ESIF) under project number CZ.02.01.01/00/22\_008/0004632.
GV acknowledges the support of the Eric and Wendy Schmidt AI in Science Postdoctoral Fellowship at the University of Chicago, a Schmidt Sciences, LLC program.
AA, GN, and HN acknowledge support by CONAHCyT CBF2023-2024-162 and JLCC CBF2023-2024-589. AA acknowledges support by PAPIIT IG102123. GN acknowledges the support of DAIP-UG and the computational resources of the DCI-UG DataLab.
CGQ acknowledges support provided by NASA through the NASA Hubble Fellowship grant HST-HF2-51554.001-A awarded by the Space Telescope Science Institute, which is operated by the Association of Universities for Research in Astronomy, Inc., for NASA, under contract NAS5-26555.

This material is based upon work supported by the U.S.\ Department of Energy (DOE), Office of Science, Office of High-Energy Physics, under Contract No.\ DE–AC02–05CH11231, and by the National Energy Research Scientific Computing Center, a DOE Office of Science User Facility under the same contract. Additional support for DESI was provided by the U.S. National Science Foundation (NSF), Division of Astronomical Sciences under Contract No.\ AST-0950945 to the NSF National Optical-Infrared Astronomy Research Laboratory; the Science and Technology Facilities Council of the United Kingdom; the Gordon and Betty Moore Foundation; the Heising-Simons Foundation; the French Alternative Energies and Atomic Energy Commission (CEA); the National Council of Humanities, Science and Technology of Mexico (CONAHCYT); the Ministry of Science and Innovation of Spain (MICINN), and by the DESI Member Institutions: \url{https://www.desi. lbl.gov/collaborating-institutions}. 

The DESI Legacy Imaging Surveys consist of three individual and complementary projects: the Dark Energy Camera Legacy Survey (DECaLS), the Beijing-Arizona Sky Survey (BASS), and the Mayall z-band Legacy Survey (MzLS). DECaLS, BASS and MzLS together include data obtained, respectively, at the Blanco telescope, Cerro Tololo Inter-American Observatory, NSF NOIRLab; the Bok telescope, Steward Observatory, University of Arizona; and the Mayall telescope, Kitt Peak National Observatory, NOIRLab. NOIRLab is operated by the Association of Universities for Research in Astronomy (AURA) under a cooperative agreement with the National Science Foundation. Pipeline processing and analyses of the data were supported by NOIRLab and the Lawrence Berkeley National Laboratory. Legacy Surveys also uses data products from the Near-Earth Object Wide-field Infrared Survey Explorer (NEOWISE), a project of the Jet Propulsion Laboratory/California Institute of Technology, funded by the National Aeronautics and Space Administration. Legacy Surveys was supported by: the Director, Office of Science, Office of High Energy Physics of the U.S. Department of Energy; the National Energy Research Scientific Computing Center, a DOE Office of Science User Facility; the U.S. National Science Foundation, Division of Astronomical Sciences; the National Astronomical Observatories of China, the Chinese Academy of Sciences and the Chinese National Natural Science Foundation. LBNL is managed by the Regents of the University of California under contract to the U.S. Department of Energy. The complete acknowledgments can be found at \url{https://www.legacysurvey.org/}.
Any opinions, findings, and conclusions or recommendations expressed in this material are those of the author(s) and do not necessarily reflect the views of the U.S.\ National Science Foundation, the U.S.\ Department of Energy, or any of the listed funding agencies.

The authors are honored to be permitted to conduct scientific research on Iolkam Du’ag (Kitt Peak), a mountain with particular significance to the Tohono O’odham Nation.

\newpage
\appendix


\section{Einstein de Sitter Kernels versus MG full kernels using }
\label{sec:EdSvsMG}

\begin{figure}
    \centering
    \includegraphics[width=0.6\linewidth]{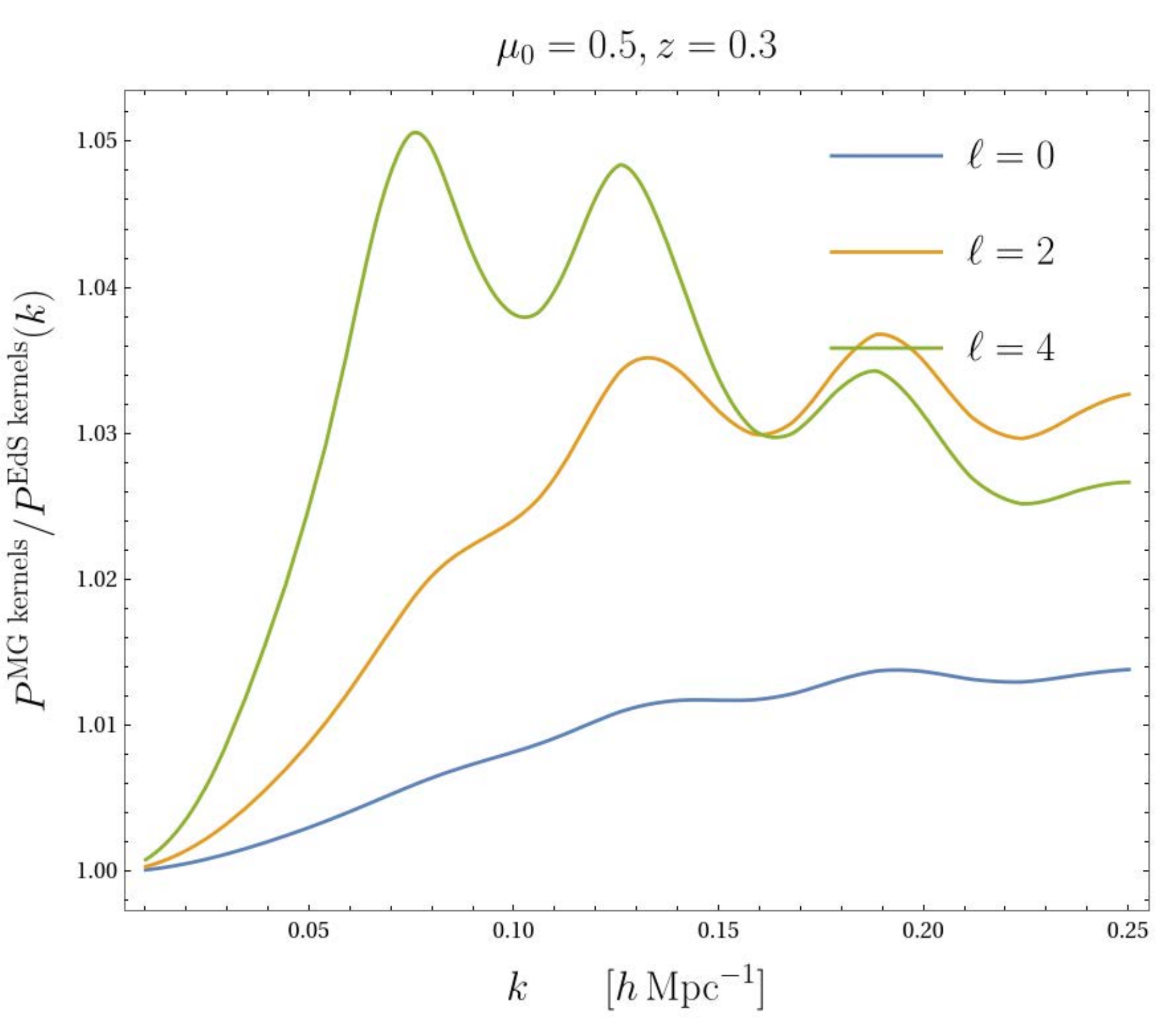}
    \caption{{Comparison between MG kernels using \texttt{fkpt} and EdS kernels to compute the 1-loop corrections of the power spectrum for the model given by Eq. \eqref{eq:muEvolution} with $c_1=\lambda=0$. The ratios are shown for a case that depart from GR with $\mu_0=0.5$. Shown are the power spectrum multipoles $\ell=0$, 2 and 4, although the latter is not used in the fits for this analysis.   }}
    \label{fig:EdSvsMG}
\end{figure}

{We added in this appendix \cref{fig:EdSvsMG} in support of the discussion of the EFT kernels in \cref{sec:data_methodology} and refer the reader to that subsection.}


\section{Constraints and comparison using ShapeFit results}
\label{sec:comparison_shapefit}

The cosmological information can be extracted from large-scale structures using two primary methods: the Full Shape method as used in \cref{sec:phenomenological_parameters} or compression techniques. Here, we perform a direct comparison of the constraints obtained from ShapeFit compression with those derived from the full-shape method in the context of a \lcdm\ background with time-only dependence of MG parameters. The compression method, such as ShapeFit (SF) \cite{Brieden:2021edu}, relies on fitting power-spectrum multipoles using a template {cosmology} and a set of {free} parameters that capture information about the underlying cosmology. The ShapeFit compression combines BAO scaling parameters, $\alpha_\parallel$ and $\alpha_\perp$, the growth of the structure, $df$, and two shape parameters, $m$ and $n$, which model the broadband shape of the linear power spectrum, pivoting at a specific scale, allowing the capture of information from both the matter-radiation equality epoch and the spectral index through:
\begin{equation}
    P^\prime_{\rm lin} =P^{\rm fid}_{\rm lin}\exp\left\{\frac{m}{a}\tanh\left[a\ln\left(\frac{k}{k_p}\right)\right]+n\ln\left(\frac{k}{k_p}\right) \right\},
\end{equation}
where $k_p$ and $a$ values has been kept fixed at $k_p=0.03 h^{\rm fid}{\rm Mpc}^{-1}$, $a=0.6$ and $n=0$. We set $n = 0$ due to the strong anti-correlation between the shape parameters $n$ and $m$ \cite{Brieden:2021edu, KP5s3-Noriega}. 

We closely follow the compression outlined in \cite{DESI2024.V.KP5} and utilize the ShapeFit measurements from six redshift bins combined with distance-scale information from the post-reconstruction correlation function, and BAO \lya likelihood \cite{DESI2024.IV.KP6} used in DESI DR1 BAO measurements. For more details of the Shapefit pipeline, we refer readers to section 4 of \cite{DESI2024.V.KP5}. The ShapeFit compression, similar to RSD $f \sigma_{s8}$ measurements, serves as an independent probe for the redshift dependence of the gravitational constant $\mu(z)$.  For the sake of brevity, we will denote the combined likelihood as DESI (SF + BAO).

\begin{figure*}
\centering

\includegraphics[width=0.6\textwidth]{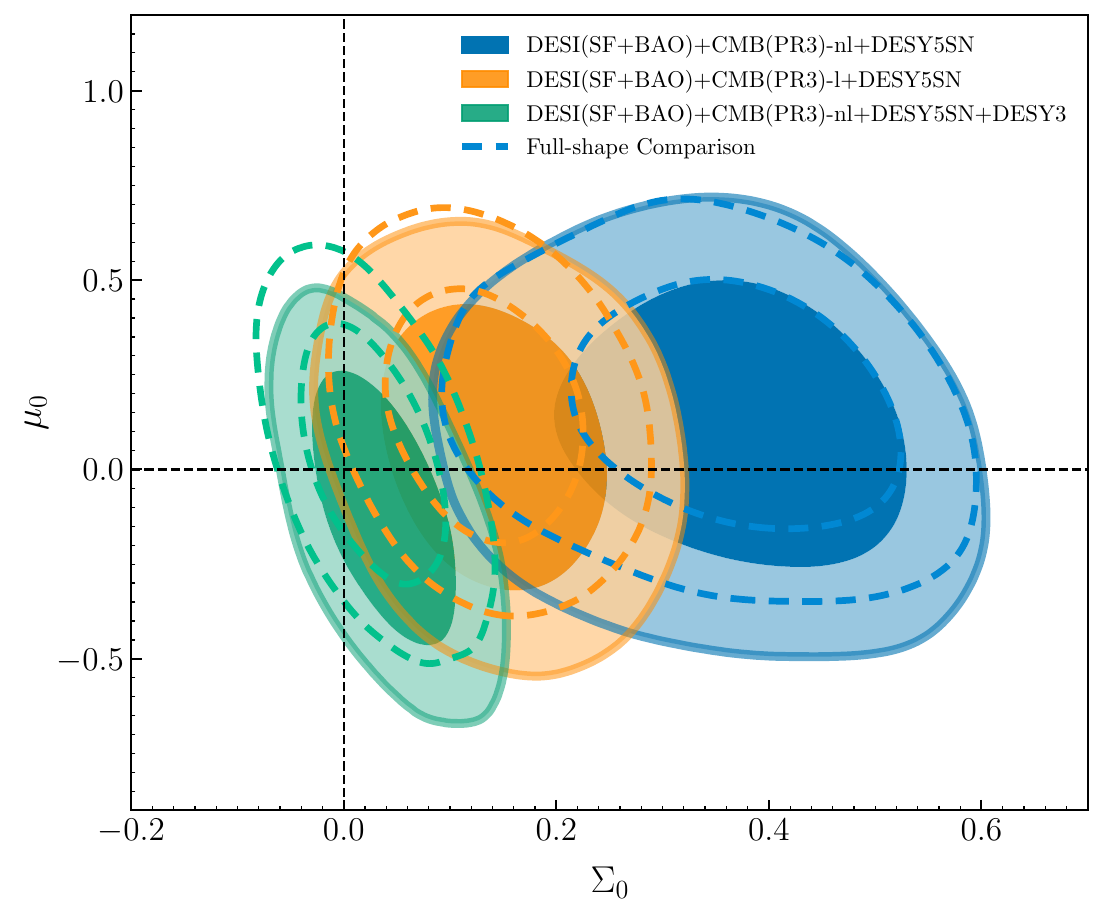}
\caption{
The data combination of DESI(SF + BAO), CMB (without lensing), and DESY5SN, depicted in blue contours, presents 68\% and 95\% credible limits on $\mu_0$ and $\Sigma_0$ parameters. Upon the addition of CMB lensing to the combination, represented in orange, the constraints on $\Sigma_0$ shift to a lower value, aligning with the predictions of GR. Lastly, the combination of DESI(SF + BAO) + CMB(PR3)-nl + DESY5SN + DESY3, shown in green, is completely consistent with GR ($\mu_0=0$, $\Sigma_0=0)$. We also compare these constraints with the DESI (FS + BAO) results, which are illustrated with dotted lines and discussed in the previous section.}
\label{Fig:SF_MG_constraints}
\end{figure*}

When considering the ($\mu_0$, $\Sigma_0$) anstaz discussed in \cref{sec:phenomenological_parameters}, we find DESI (SF + BAO) + CMB(PR3)-nl + DESY5SN gives the following constraints: 

\begin{eqnarray*}
\mu_0 &=& 0.10\pm 0.25,\\
\Sigma_0 &=& 0.363^{+0.12}_{-0.092}, 
\end{eqnarray*} 

These constraints tighten up when DESY3 is added to the mix. 
\begin{eqnarray*}
\mu_0 &=& -0.10\pm 0.23,\\
\Sigma_0 &=& 0.039^{+0.043}_{-0.048}, 
\end{eqnarray*} 

However, when CMB lensing is used instead of DESY3, we obtain the following constraints 
\begin{eqnarray*}
\mu_0 &=& -0.06\pm 0.25 ,\\
\Sigma_0 &=& 0.144\pm 0.071, 
\end{eqnarray*} 

In \cref{Fig:SF_MG_constraints}, we show the constraints obtained from the DESI (SF + BAO) in combination with the CMB (without lensing) and DESY5SN, depicted in blue. The incorporation of CMB lensing data lowers the constraints on the parameter $\Sigma_0$, making them more consistent with GR. Finally, the combination of DESI(SF + BAO) + CMB(PR3)-nl + DESY5SN + DESY3 further tightens the constraints on $\Sigma_0$, centering credible contours around prediction of GR ($\mu_0=0$, $\Sigma_0=0$). Additionally, we show corresponding DESI(FS+BAO) data combinations in dotted lines, which, as expected, provide tighter constraints compared to those from SF+BAO, indicating consistency across different dataset combinations.

Our results reveal that the constraints on the parameters $\mu_0$ and $\Sigma_0$ are overall consistent, although slightly weaker when using SF compression in comparison to the full-shape analysis and as expected. This difference is attributed to the fact that ShapeFit employs a single SF parameter ($m$) to capture information around the matter-radiation equality epoch. In contrast, the full-shape analysis leverages the complete shape of the power spectrum, allowing for efficient extraction of the cosmological information.

Despite the agreement between full-shape and Shapefit results, it's important to note that the ShapeFit compression is not designed and validated for MG scenarios and extra caution should be exercised when interpreting the results within the context of MG theories.

\section{Projection effects in the context of modified gravity}
\label{sec:projection}

In this appendix, we investigate projection effects by comparing the Maximum A Posteriori (MAP) values to the mean values of posteriors within the framework of the MG parameterization. In this context, projection effects refer to the shifts that may occur when nuisance parameters, such as bias, stochastic terms, and counter-terms, are partially degenerate with cosmological parameters, see 
e.g.\,Ref. \cite{DESI2024.V.KP5} for detailed discussions. When marginalizing over these nuisance parameters, the peak of the posterior distribution can shift away from the marginal posterior distribution. 

We determine the MAP by maximizing the log-posterior using the \texttt{prospect} \cite{prospect} package, initiated from the maximum log-posterior points identified within the MCMC chains. \cite{prospect} utilizes simulated annealing, a gradient-free stochastic optimization algorithm, which incorporates adaptive step-size tuning and covariance matrices derived from the MCMC to optimize the log-posterior effectively \footnote{We have validated the \texttt{prospect} results with the $w_0-w_a$ model results presented in the appendix of \cite{DESI2024.VII.KP7B} where another software was used, and the findings were very similar.}
\begin{figure}[h!]
    \centering
    \includegraphics[width=0.48\textwidth]{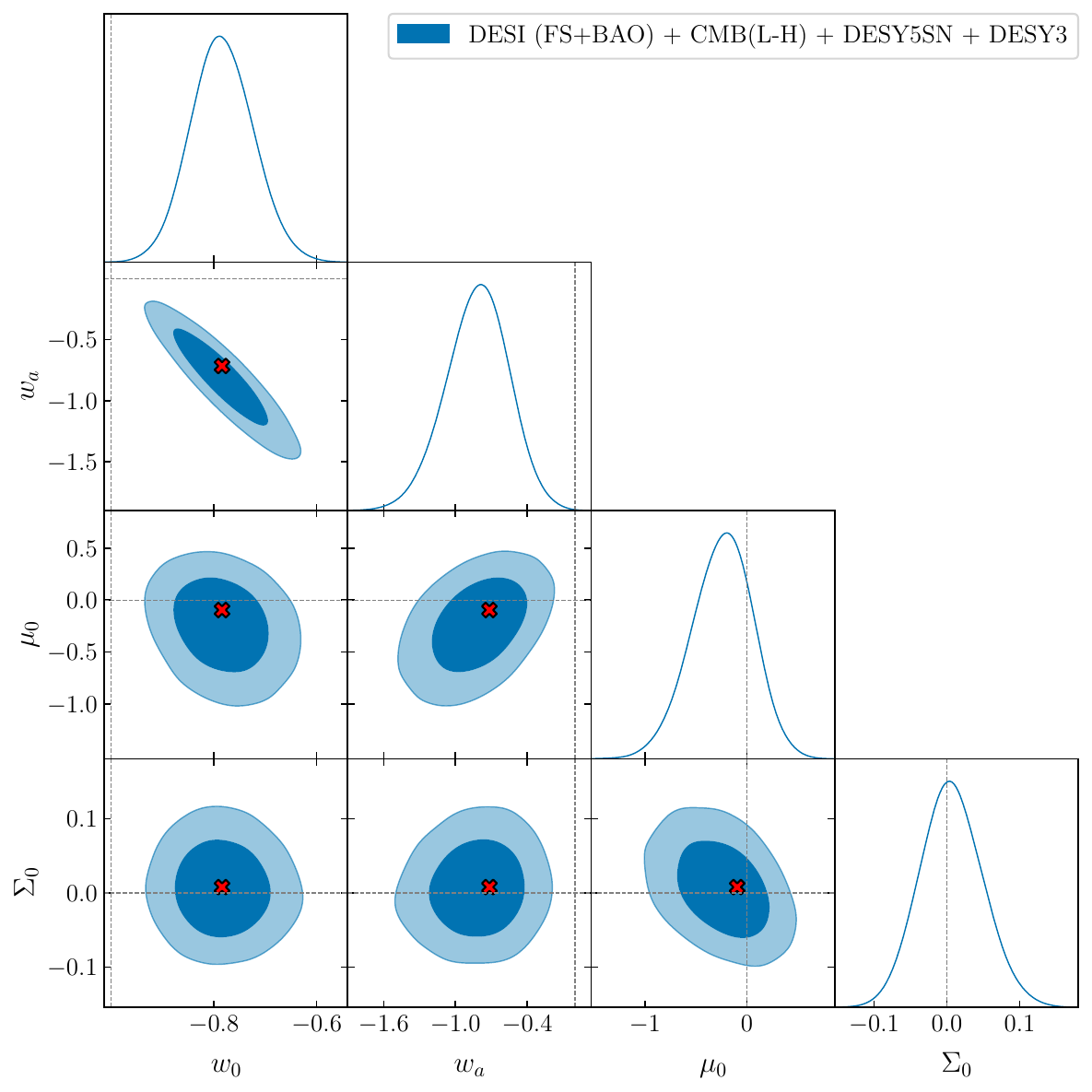}
    \includegraphics[width=0.48\textwidth]{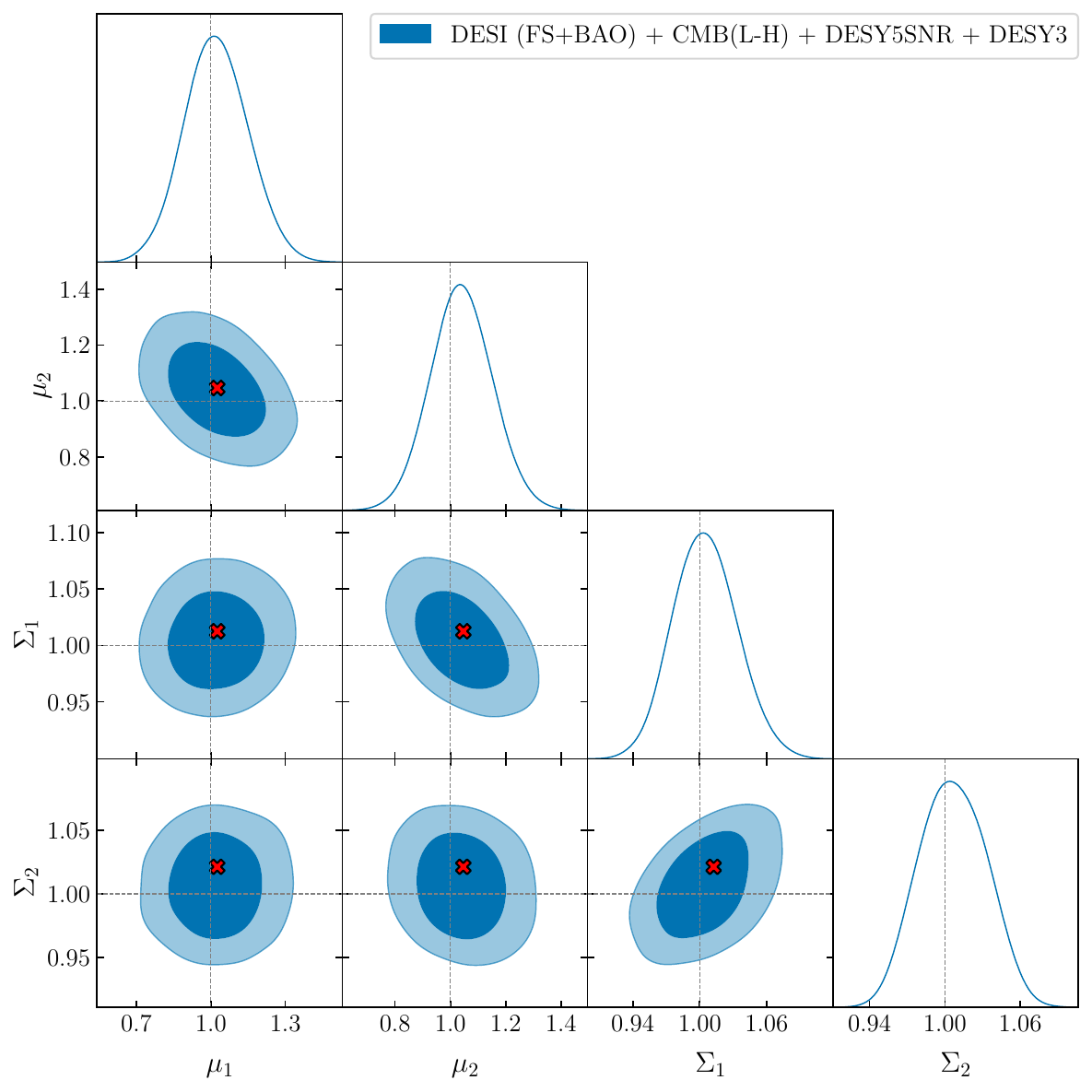}
    \caption{ The impact of projection effects on the modified gravity parameters. The left subplot illustrates the marginal posterior distribution of $\mu_0-\Sigma_0$ with a $w_0$-$w_a$ cosmological expansion background using DESI+CMB(L-H)+DESY5SN+DESY3, with blue contours representing  68\% and 95\% credible interval regions, and the red cross highlighting the maximum a posteriori (MAP) values obtained from \texttt{prospect}. The right subplot presents the corresponding constraints for binned $\mu(z)$ and $\Sigma(z)$ in the $\Lambda$CDM expansion background.}
    \label{fig:mg_projection}
\end{figure}

As illustrated in Figure~\ref{fig:mg_projection}, the maximum a posteriori and mean values for the MG parameters $\mu_0$ and $\Sigma_0$ fall well within 1$\sigma$, demonstrating that projection-induced biases are limited when all datasets are combined, supporting the robustness of the MG constraints presented in our study. 
We report in \cref{tab:mapvalues} the MAP values for MG and other cosmological parameters for various selected dataset combinations to allow comparison with the posterior results in the analysis.  
We leave a detailed characterization of these effects for various MG models and datasets for future work.

\begin{table}
\centering
\resizebox{\columnwidth}{!}{%
\begin{tabular}{lccccc}
\toprule
{{\bf Flat} $\boldsymbol{\mu_0\Sigma_0}$ {\bf $\Lambda$CDM} } & $\Omega_{\rm m}$ & $\sigma_8$ & $H_0$ [${\rm km/s/Mpc}$] & $\mu_0$ & $\Sigma_0$ \\

\hline

 \makecell[l]{DESI+CMB (PR3)-l} & $0.3016$ & $0.8287$ & $68.38$ & $0.2531$ & $0.1612$ \\
 \makecell[l]{DESI+CMB (PR3)-nl + DESY3} & $0.302$ & $0.8057$ & $68.34$ & $0.02158$ & $0.05945$ \\
 \makecell[l]{DESI+CMB ($\texttt{LoLLiPoP}$-$\texttt{HiLLiPoP}$)-l} & $0.3018$ & $0.8325$ & $68.25$ & $0.2536$ & $0.1027$ \\
 \makecell[l]{DESI+CMB ($\texttt{LoLLiPoP}$-$\texttt{HiLLiPoP}$)-nl\\\quad+DESY3} & $0.3019$ & $0.8123$ & $68.24$ & $0.08183$ & $0.03205$ \\
 \makecell[l]{DESI+CMB ($\texttt{LoLLiPoP}$-$\texttt{HiLLiPoP}$)-nl\\\quad+DESY3+DESSNY5}  & $0.3068$ & $0.8155$ & $67.86$ & $0.1013$ & $0.01112$ \\

\hline 

{{\bf Flat} $\boldsymbol{\mu_0\eta_0}$ {\bf $\Lambda$CDM} } & $\Omega_{\rm m}$ & $\sigma_8$ & $H_0$ [${\rm km/s/Mpc}$] & $\mu_0$ & $\eta_0$ \\

\hline
  \makecell[l]{DESI+CMB ($\texttt{LoLLiPoP}$-$\texttt{HiLLiPoP}$)-l}  & $0.3011$ & $0.8343$ & $68.31$ & $0.4148$ & $-0.4939$ \\
   \makecell[l]{DESI+CMB ($\texttt{LoLLiPoP}$-$\texttt{HiLLiPoP}$)-nl\\\quad+DESY3+DESSNY5} & $0.3075$ & $0.81$ & $67.82$ & $0.05567$ & $-0.06469$ \\

\midrule
      {{\bf Flat} $\boldsymbol{\mu_0\Sigma_0 w_0w_a}$}& {$w_0$} &  {$w_a$} & $H_0  [${\rm km/s/Mpc}$] $ & {$\mu_0$} &{$\Sigma_0$}\\[-0.1cm]
    
  \midrule
       
    \makecell[l]{DESI+CMB ($\texttt{LoLLiPoP}$-$\texttt{HiLLiPoP}$)-nl\\\quad+DESY3+DESSNY5}  

& $-0.7838$ & $-0.7161$ & $66.72$ & $-0.096$ & $0.0078$ \\
    
    \midrule     
      {\bf Flat $\Lambda$CDM Redshift Binning} & $\mu_1$ & $\mu_2$ & $\Sigma_1$ & $\Sigma_2$ & -  \\[-0.1cm]
   
    \midrule \makecell[l]{DESI+CMB (PR3)-nl + DESY3 + DES5YSN}
    & $1.026$ & $1.047$ & $1.012$ & $1.022$ & $-$ \\
    
    \bottomrule
\end{tabular}
}
\caption{\label{tab:mapvalues} {Maximum A Posteriori (MAP) estimates of cosmological parameters and modified gravity parameters obtained using the PROSPECT for selected models and combinations of datasets to inform comparison with posterior values from \cref{tab:functional_forms} and \cref{tab:functional_forms_mu_eta}.}}
\end{table}


\bibliographystyle{JHEP}
\bibliography{refs_key_paper,DESI2024}


\input{MG.authorlist.nov18.affiliations}

\end{document}

%% file: MG.authorlist.nov18.tex
\affiliation{Affiliations are in Appendix \ref{sec:affiliations}}

\author[1]{{M.~Ishak}\orcidlink{0000-0002-6024-466X},}
\author[2]{{J.~Pan}\orcidlink{0000-0001-9685-5756},}
\author[3]{{R.~Calderon}\orcidlink{0000-0002-8215-7292},}
\author[3,4]{{K.~Lodha}\orcidlink{0009-0004-2558-5655},}
\author[5,6]{{G.~Valogiannis}\orcidlink{0000-0003-0805-1470},}
\author[7,8]{{A.~Aviles}\orcidlink{0000-0001-5998-3986},}
\author[9,8]{{G.~Niz}\orcidlink{0000-0002-1544-8946},}
\author[10,11]{{L.~Yi},}
\author[12]{{C.~Zheng},}
\author[13,1,14]{{C.~Garcia-Quintero}\orcidlink{0000-0003-1481-4294},}
\author[15]{{A.~de~Mattia}\orcidlink{0000-0003-0920-2947},}
\author[1]{{L.~Medina-Varela},}
\author[16]{{J.~L.~Cervantes-Cota}\orcidlink{0000-0002-3057-6786},}
\author[17,2]{{U.~Andrade}\orcidlink{0000-0002-4118-8236},}
\author[18]{{D.~Huterer}\orcidlink{0000-0001-6558-0112},}
\author[7,19]{{H.~E.~Noriega}\orcidlink{0000-0002-3397-3998},}
\author[10,11,20,21]{{G.~Zhao},}
\author[3,4]{{A.~Shafieloo}\orcidlink{0000-0001-6815-0337},}
\author[12]{{W.~Fang},}
\author[22]{{S.~Ahlen}\orcidlink{0000-0001-6098-7247},}
\author[23]{{D.~Bianchi}\orcidlink{0000-0001-9712-0006},}
\author[24]{{D.~Brooks},}
\author[25]{{E.~Burtin},}
\author[26]{{E.~Chaussidon}\orcidlink{0000-0001-8996-4874},}
\author[26]{{T.~Claybaugh},}
\author[27]{{S.~Cole}\orcidlink{0000-0002-5954-7903},}
\author[19]{{A.~de la Macorra}\orcidlink{0000-0002-1769-1640},}
\author[28]{{Arjun~Dey}\orcidlink{0000-0002-4928-4003},}
\author[29,30]{{K.~Fanning}\orcidlink{0000-0003-2371-3356},}
\author[26,31]{{S.~Ferraro}\orcidlink{0000-0003-4992-7854},}
\author[24,32]{{A.~Font-Ribera}\orcidlink{0000-0002-3033-7312},}
\author[33,34]{{J.~E.~Forero-Romero}\orcidlink{0000-0002-2890-3725},}
\author[35,36,37]{{E.~Gaztañaga},}
\author[38,35,39]{{H.~Gil-Mar\'in}\orcidlink{0000-0003-0265-6217},}
\author[26]{{S.~{Gontcho A Gontcho}}\orcidlink{0000-0003-3142-233X},}
\author[5]{{G.~Gutierrez},}
\author[40]{{C.~Hahn}\orcidlink{0000-0003-1197-0902},}
\author[41,42,43]{{K.~Honscheid},}
\author[44]{{C.~Howlett}\orcidlink{0000-0002-1081-9410},}
\author[28]{{S.~Juneau},}
\author[45]{{D.~Kirkby}\orcidlink{0000-0002-8828-5463},}
\author[26]{{T.~Kisner}\orcidlink{0000-0003-3510-7134},}
\author[26]{{A.~Kremin}\orcidlink{0000-0001-6356-7424},}
\author[26]{{M.~Landriau}\orcidlink{0000-0003-1838-8528},}
\author[46]{{L.~Le~Guillou}\orcidlink{0000-0001-7178-8868},}
\author[47,48]{{A.~Leauthaud}\orcidlink{0000-0002-3677-3617},}
\author[26]{{M.~E.~Levi}\orcidlink{0000-0003-1887-1018},}
\author[28]{{A.~Meisner}\orcidlink{0000-0002-1125-7384},}
\author[49,32]{{R.~Miquel},}
\author[50]{{J.~Moustakas}\orcidlink{0000-0002-2733-4559},}
\author[51]{{J.~ A.~Newman}\orcidlink{0000-0001-8684-2222},}
\author[15,26]{{N.~Palanque-Delabrouille}\orcidlink{0000-0003-3188-784X},}
\author[52,53,54]{{W.~J.~Percival}\orcidlink{0000-0002-0644-5727},}
\author[26,55,31]{{C.~Poppett},}
\author[56]{{F.~Prada}\orcidlink{0000-0001-7145-8674},}
\author[57]{{I.~P\'erez-R\`afols}\orcidlink{0000-0001-6979-0125},}
\author[41,58,43]{{A.~J.~Ross}\orcidlink{0000-0002-7522-9083},}
\author[59]{{G.~Rossi},}
\author[60]{{E.~Sanchez}\orcidlink{0000-0002-9646-8198},}
\author[26]{{D.~Schlegel},}
\author[18,2]{{M.~Schubnell},}
\author[61]{{H.~Seo}\orcidlink{0000-0002-6588-3508},}
\author[28]{{D.~Sprayberry},}
\author[2]{{G.~Tarl\'{e}}\orcidlink{0000-0003-1704-0781},}
\author[19]{{M.~Vargas-Maga\~na}\orcidlink{0000-0003-3841-1836},}
\author[28]{{B.~A.~Weaver},}
\author[29,62,30]{{R.~H.~Wechsler}\orcidlink{0000-0003-2229-011X},}
\author[15]{{C.~Yèche}\orcidlink{0000-0001-5146-8533},}
\author[46]{{P.~Zarrouk}\orcidlink{0000-0002-7305-9578},}
\author[26]{{R.~Zhou}\orcidlink{0000-0001-5381-4372},}
\author[10]{{H.~Zou}\orcidlink{0000-0002-6684-3997},}

%% file: MG.authorlist.nov18.affiliations.tex

\section{Author Affiliations}
\label{sec:affiliations}

\noindent \hangindent=.5cm $^{1}${Department of Physics, The University of Texas at Dallas, Richardson, TX 75080, USA}

\noindent \hangindent=.5cm $^{2}${University of Michigan, Ann Arbor, MI 48109, USA}

\noindent \hangindent=.5cm $^{3}${Korea Astronomy and Space Science Institute, 776, Daedeokdae-ro, Yuseong-gu, Daejeon 34055, Republic of Korea}

\noindent \hangindent=.5cm $^{4}${University of Science and Technology, 217 Gajeong-ro, Yuseong-gu, Daejeon 34113, Republic of Korea}

\noindent \hangindent=.5cm $^{5}${Fermi National Accelerator Laboratory, PO Box 500, Batavia, IL 60510, USA}

\noindent \hangindent=.5cm $^{6}${Department of Astronomy and Astrophysics, University of Chicago, 5640 South Ellis Avenue, Chicago, IL 60637, USA}

\noindent \hangindent=.5cm $^{7}${Instituto de Ciencias F\'{\i}sicas, Universidad Aut\'onoma de M\'exico, Cuernavaca, Morelos, 62210, (M\'exico)}

\noindent \hangindent=.5cm $^{8}${Instituto Avanzado de Cosmolog\'{\i}a A.~C., San Marcos 11 - Atenas 202. Magdalena Contreras, 10720. Ciudad de M\'{e}xico, M\'{e}xico}

\noindent \hangindent=.5cm $^{9}${Departamento de F\'{i}sica, Universidad de Guanajuato - DCI, C.P. 37150, Leon, Guanajuato, M\'{e}xico}

\noindent \hangindent=.5cm $^{10}${National Astronomical Observatories, Chinese Academy of Sciences, A20 Datun Rd., Chaoyang District, Beijing, 100012, P.R. China}

\noindent \hangindent=.5cm $^{11}${School of Astronomy and Space Science, University of Chinese Academy of Sciences, Beijing, 100049, P.R.China}

\noindent \hangindent=.5cm $^{12}${Department of Astronomy, University of Science and Technology of China, Hefei, Anhui, 230026, People’s Republic of China}

\noindent \hangindent=.5cm $^{13}${Center for Astrophysics $|$ Harvard \& Smithsonian, 60 Garden Street, Cambridge, MA 02138, USA}

\noindent \hangindent=.5cm $^{14}${NASA Einstein Fellow}

\noindent \hangindent=.5cm $^{15}${IRFU, CEA, Universit\'{e} Paris-Saclay, F-91191 Gif-sur-Yvette, France}

\noindent \hangindent=.5cm $^{16}${Departamento de F\'{i}sica, Instituto Nacional de Investigaciones Nucleares, Carreterra M\'{e}xico-Toluca S/N, La Marquesa,  Ocoyoacac, Edo. de M\'{e}xico C.P. 52750,  M\'{e}xico}

\noindent \hangindent=.5cm $^{17}${Leinweber Center for Theoretical Physics, University of Michigan, 450 Church Street, Ann Arbor, Michigan 48109-1040, USA}

\noindent \hangindent=.5cm $^{18}${Department of Physics, University of Michigan, Ann Arbor, MI 48109, USA}

\noindent \hangindent=.5cm $^{19}${Instituto de F\'{\i}sica, Universidad Nacional Aut\'{o}noma de M\'{e}xico,  Cd. de M\'{e}xico  C.P. 04510,  M\'{e}xico}

\noindent \hangindent=.5cm $^{20}${Institute for Frontiers in Astronomy and Astrophysics, Beijing Normal University, Beijing, 102206,P.R.China}

\noindent \hangindent=.5cm $^{21}${Chinese Academy of Sciences South America Center for Astronomy (CASSACA), National Astronomical Observatories of China, Beijing, 100101, P.R. China}

\noindent \hangindent=.5cm $^{22}${Physics Dept., Boston University, 590 Commonwealth Avenue, Boston, MA 02215, USA}

\noindent \hangindent=.5cm $^{23}${Dipartimento di Fisica ``Aldo Pontremoli'', Universit\`a degli Studi di Milano, Via Celoria 16, I-20133 Milano, Italy}

\noindent \hangindent=.5cm $^{24}${Department of Physics \& Astronomy, University College London, Gower Street, London, WC1E 6BT, UK}

\noindent \hangindent=.5cm $^{25}${IRFU, CEA, Universit\'{e} Paris-Saclay, F-91191 Gif-sur-Yvette, France"}

\noindent \hangindent=.5cm $^{26}${Lawrence Berkeley National Laboratory, 1 Cyclotron Road, Berkeley, CA 94720, USA}

\noindent \hangindent=.5cm $^{27}${Institute for Computational Cosmology, Department of Physics, Durham University, South Road, Durham DH1 3LE, UK}

\noindent \hangindent=.5cm $^{28}${NSF NOIRLab, 950 N. Cherry Ave., Tucson, AZ 85719, USA}

\noindent \hangindent=.5cm $^{29}${Kavli Institute for Particle Astrophysics and Cosmology, Stanford University, Menlo Park, CA 94305, USA}

\noindent \hangindent=.5cm $^{30}${SLAC National Accelerator Laboratory, Menlo Park, CA 94305, USA}

\noindent \hangindent=.5cm $^{31}${University of California, Berkeley, 110 Sproul Hall \#5800 Berkeley, CA 94720, USA}

\noindent \hangindent=.5cm $^{32}${Institut de F\'{i}sica d’Altes Energies (IFAE), The Barcelona Institute of Science and Technology, Campus UAB, 08193 Bellaterra Barcelona, Spain}

\noindent \hangindent=.5cm $^{33}${Departamento de F\'isica, Universidad de los Andes, Cra. 1 No. 18A-10, Edificio Ip, CP 111711, Bogot\'a, Colombia}

\noindent \hangindent=.5cm $^{34}${Observatorio Astron\'omico, Universidad de los Andes, Cra. 1 No. 18A-10, Edificio H, CP 111711 Bogot\'a, Colombia}

\noindent \hangindent=.5cm $^{35}${Institut d'Estudis Espacials de Catalunya (IEEC), 08034 Barcelona, Spain}

\noindent \hangindent=.5cm $^{36}${Institute of Cosmology and Gravitation, University of Portsmouth, Dennis Sciama Building, Portsmouth, PO1 3FX, UK}

\noindent \hangindent=.5cm $^{37}${Institute of Space Sciences, ICE-CSIC, Campus UAB, Carrer de Can Magrans s/n, 08913 Bellaterra, Barcelona, Spain}

\noindent \hangindent=.5cm $^{38}${Departament de F\'{\i}sica Qu\`{a}ntica i Astrof\'{\i}sica, Universitat de Barcelona, Mart\'{\i} i Franqu\`{e}s 1, E08028 Barcelona, Spain}

\noindent \hangindent=.5cm $^{39}${Institut de Ci\`encies del Cosmos (ICCUB), Universitat de Barcelona (UB), c. Mart\'i i Franqu\`es, 1, 08028 Barcelona, Spain.}

\noindent \hangindent=.5cm $^{40}${Department of Astrophysical Sciences, Princeton University, Princeton NJ 08544, USA}

\noindent \hangindent=.5cm $^{41}${Center for Cosmology and AstroParticle Physics, The Ohio State University, 191 West Woodruff Avenue, Columbus, OH 43210, USA}

\noindent \hangindent=.5cm $^{42}${Department of Physics, The Ohio State University, 191 West Woodruff Avenue, Columbus, OH 43210, USA}

\noindent \hangindent=.5cm $^{43}${The Ohio State University, Columbus, 43210 OH, USA}

\noindent \hangindent=.5cm $^{44}${School of Mathematics and Physics, University of Queensland, 4072, Australia}

\noindent \hangindent=.5cm $^{45}${Department of Physics and Astronomy, University of California, Irvine, 92697, USA}

\noindent \hangindent=.5cm $^{46}${Sorbonne Universit\'{e}, CNRS/IN2P3, Laboratoire de Physique Nucl\'{e}aire et de Hautes Energies (LPNHE), FR-75005 Paris, France}

\noindent \hangindent=.5cm $^{47}${Department of Astronomy and Astrophysics, UCO/Lick Observatory, University of California, 1156 High Street, Santa Cruz, CA 95064, USA}

\noindent \hangindent=.5cm $^{48}${Department of Astronomy and Astrophysics, University of California, Santa Cruz, 1156 High Street, Santa Cruz, CA 95065, USA}

\noindent \hangindent=.5cm $^{49}${Instituci\'{o} Catalana de Recerca i Estudis Avan\c{c}ats, Passeig de Llu\'{\i}s Companys, 23, 08010 Barcelona, Spain}

\noindent \hangindent=.5cm $^{50}${Department of Physics and Astronomy, Siena College, 515 Loudon Road, Loudonville, NY 12211, USA}

\noindent \hangindent=.5cm $^{51}${Department of Physics \& Astronomy and Pittsburgh Particle Physics, Astrophysics, and Cosmology Center (PITT PACC), University of Pittsburgh, 3941 O'Hara Street, Pittsburgh, PA 15260, USA}

\noindent \hangindent=.5cm $^{52}${Department of Physics and Astronomy, University of Waterloo, 200 University Ave W, Waterloo, ON N2L 3G1, Canada}

\noindent \hangindent=.5cm $^{53}${Perimeter Institute for Theoretical Physics, 31 Caroline St. North, Waterloo, ON N2L 2Y5, Canada}

\noindent \hangindent=.5cm $^{54}${Waterloo Centre for Astrophysics, University of Waterloo, 200 University Ave W, Waterloo, ON N2L 3G1, Canada}

\noindent \hangindent=.5cm $^{55}${Space Sciences Laboratory, University of California, Berkeley, 7 Gauss Way, Berkeley, CA  94720, USA}

\noindent \hangindent=.5cm $^{56}${Instituto de Astrof\'{i}sica de Andaluc\'{i}a (CSIC), Glorieta de la Astronom\'{i}a, s/n, E-18008 Granada, Spain}

\noindent \hangindent=.5cm $^{57}${Departament de F\'isica, EEBE, Universitat Polit\`ecnica de Catalunya, c/Eduard Maristany 10, 08930 Barcelona, Spain}

\noindent \hangindent=.5cm $^{58}${Department of Astronomy, The Ohio State University, 4055 McPherson Laboratory, 140 W 18th Avenue, Columbus, OH 43210, USA}

\noindent \hangindent=.5cm $^{59}${Department of Physics and Astronomy, Sejong University, Seoul, 143-747, Korea}

\noindent \hangindent=.5cm $^{60}${CIEMAT, Avenida Complutense 40, E-28040 Madrid, Spain}

\noindent \hangindent=.5cm $^{61}${Department of Physics \& Astronomy, Ohio University, Athens, OH 45701, USA}

\noindent \hangindent=.5cm $^{62}${Physics Department, Stanford University, Stanford, CA 93405, USA}